# The SPARQL2XQuery Interoperability Framework

## Utilizing Schema Mapping, Schema Transformation and Query Translation to Integrate XML and the Semantic Web [1]

Nikos Bikakis [† ¥ 2]    Chrisa Tsinaraki [#]    Ioannis Stavrakantonakis [§ 2]

Nektarios Gioldasis [#]    Stavros Christodoulakis [#]

[†] National Technical University of Athens | Greece
[¥] IMIS | "Athena" Research Center | Greece
*bikakis@dblab.ntua.gr*

[#] Technical University of Crete | Greece
*{chrisa, nektarios, stavros}@ced.tuc.gr*

[§] STI | University of Innsbruck | Austria
*ioannis.stavrakantonakis@sti2.at*

## ABSTRACT

The *Web of Data* is an open environment consisting of a great number of large inter-linked RDF datasets from various domains. In this environment, organizations and companies adopt the Linked Data practices utilizing Semantic Web (*SW*) technologies, in order to publish their data and offer SPARQL endpoints (i.e., SPARQL-based search services). On the other hand, the dominant standard for information exchange in the Web today is XML. Additionally, many international standards (e.g., *Dublin Core, MPEG-7, METS, TEI, IEEE LOM*) in several domains (e.g., Digital Libraries, GIS, Multimedia, e-Learning) have been expressed in XML Schema. The aforementioned have led to an increasing emphasis on XML data, accessed using the XQuery query language. The SW and XML worlds and their developed infrastructures are based on different data models, semantics and query languages. Thus, it is crucial to develop interoperability mechanisms that allow the Web of Data users to access XML datasets, using SPARQL, from their own working environments. It is unrealistic to expect that all the existing legacy data (e.g., Relational, XML, etc.) will be transformed into SW data. Therefore, publishing legacy data as Linked Data and providing SPARQL endpoints over them has become a major research challenge. In this direction, we introduce the SPARQL2XQuery Framework which creates an interoperable environment, where SPARQL queries are automatically translated to XQuery queries, in order to access XML data across the Web. The SPARQL2XQuery Framework provides a mapping model for the expression of OWL–RDF/S to XML Schema mappings as well as a method for SPARQL to XQuery translation. To this end, our Framework supports both manual and automatic mapping specification between ontologies and XML Schemas. In the automatic mapping specification scenario, the SPARQL2XQuery exploits the Xs2Owl component which transforms XML Schemas into OWL ontologies. Finally, extensive experiments have been conducted in order to evaluate the schema transformation, mapping generation, query translation and query evaluation efficiency, using both real and synthetic datasets.

**Keywords:** *Integration, Schema Mappings, Query Translation, Schema Transformation, Data Transformation, SPARQL endpoint, Linked Data, XML Data, Semantic Web, XML Schema to OWL, SPARQL to XQuery, SPARQL Update, SPARQL 1.1, XML Schema 1.1, OWL 2.*

---

# 1 INTRODUCTION

The *Linked-Open Data*[3], *Open-Government*[4] and *Linked Life Data*[5] initiatives have played a major role in the development of the so called *Web of Data* (*WoD*). In the WoD, a large number of organizations, institutes and companies (e.g., *DBpedia*, *GeoNames*, *PubMed*, *Data.gov*, *ACM*, *NASA*, *BBC*, *MusicBrainz*, *IEEE*, etc.) adopt the *Linked Data* practices. Utilizing the Semantic Web (*SW*) technologies [121], they publish their data and offer SPARQL endpoints (i.e., SPARQL-based search services). Nowadays, there are hundreds of large inter-linked RDF datasets from various domains which comprise the WoD. It is challenging though, to make information that is stored in non-RDF data sources (e.g., Relational databases, XML repositories, etc.) available in the WoD.

The SW infrastructure supports the management of RDF datasets [8][9][10], accessed by the SPARQL query language [12]. Since the WoD applications and services have to coexist and interoperate with the existing applications that access legacy systems, it is essential for the WoD infrastructure to provide transparent access to information stored in heterogeneous legacy data sources. Publishing legacy data that adopt the Linked Data practices and offer SPARQL endpoints over it, has become a major research and development objective for many organizations.

In the current Web infrastructure the XML/XML Schema [1][2][3] are the dominant standards for information exchange as well as for the representation of semi-structured information. As a consequence, many international standards in several domains (e.g., Digital Libraries, GIS, Multimedia, e-Learning, Government, Commercial) have been expressed in XML Schema syntax. For example, the *Dublin Core* [17] and *METS* [18] standards are used by digital libraries, the *MPEG-7* [20] and *MPEG-21* [21] standards are utilized for multimedia content and service description, the *MARC 21* [22], *MODS* [23], *TEI* [24], *EAD* [25] and *VRA Core* [26] standards are used by cultural heritage institutions (e.g., libraries, archives, museums, etc.) and the *IEEE LOM* [28] and *SCORM* [29] standards are exploited in e-learning environments. The universal adoption of XML for web data exchange and the expression of several standards using XML Schema, have resulted in a large number of XML datasets accessed using the XQuery query language [5]. For example, *Oracle* has at least *7000* customers using the XQuery feature in its products [31].

Since the SW and XML worlds have different data models, different semantics and use different query languages to access data [121], it is crucial to develop frameworks, including models and adaptable software based on them, as well as methodologies that will provide interoperability between the SW and the XML infrastructures, thus facilitating transparent XML querying in the WoD using SW technologies.

The scenario of transforming all the legacy data into SW data is clearly unrealistic due to: (a) The different data models adopted and enforced by different standardization bodies (e.g., consortiums, organizations, institutions); (b) Ownership issues; (c) The existence of systems that access the legacy data; (d) Scalability requirements (large volumes of data involved); and (e) Management requirements, e.g., support of updates. Thus, a realistic integration of the two worlds has to be established.

The W3C community has realized the need to bridge different worlds (e.g., Relational, XML, SW, etc.) under several scenarios. Tim Berners Lee introduced[6] the *Double Bus Architecture*[7], a W3C Design Issue. The Double Bus Architecture assumes that the WoD users and applications use the SPARQL query language to ask for content from the underlying XML and Relational data sources. In the context of the relational and SW worlds, the W3C *RDB2RDF* working group [100] has been established, which is attempting to bridge the relational and SW worlds [101][103]. In addition, a large number of approaches has been proposed for bridging the relational databases with the SW through SPARQL to SQL translation [104] − [115]. In the context of the SW and XML worlds, two W3C working groups (*GRDDL* [82] and *SAWSDL* [83]) focus on transforming XML data to RDF data (and vice versa). Moreover, W3C investigates the *XSPARQL*[8] approach for merging XQuery and SPARQL for transforming XML to RDF data (and vice versa).

The recent efforts in bridging the SW and XML worlds focus on data transformation (i.e., XML data to RDF data and vice versa). However, despite the significant body of related work on SPARQL to SQL translation, to the best of our knowledge, there is no work addressing the SPARQL to XQuery translation problem. Given the high importance of XML and the related standards in the Web, this is a major shortcoming in the state of the art. Finally, as far as the Linked Data context is concerned, publishing legacy data and offering SPARQL endpoints over them, has recently become a major research challenge. In spite of the fact that several systems (e.g., *D2R Server* [106], *SparqlMap* [107], *Quest* [108], *Virtuoso* [109], *TopBraid Composer*[9]) offer SPARQL endpoints[10] over relational data, to the best of our knowledge, there is no system supporting XML data.

This paper presents SPARQL2XQuery, a framework that provides transparent access over XML in the WoD. Using the SPARQL2XQuery Framework, XML datasets can be turned into SPARQL endpoints. The SPARQL2XQuery Framework provides a method for SPARQL to XQuery translation, with respect to a set of predefined mappings between ontologies[11] and XML Schemas. To this end, our Framework supports both manual and automatic mapping specifications between ontologies and XML Schemas, as well as a schema transformation mechanism.

## 1.1 Motivating Example

Here, we outline two scenarios in order to illustrate the need for bridging the SW and XML worlds in several circumstances. In our examples, three hypothetical autonomous partners are involved: (a) Digital Library X (which belongs to an institution or a company), (b) Organization A and (c) Organization Z. Each has adopted different technologies to represent and manage their data. Assume that, Digital Library X has adopted XML-related technologies (i.e., XML, XML Schema, and XQuery) and its contents are described in XML syntax, while both organizations have chosen SW technologies (i.e., RDF/S, OWL, and SPARQL).

**1st Scenario.** Consider that Digital Library X wants to publish their data in the WoD using SW technologies, a common scenario in the Linked Data era. In this case, a schema transformation and a query translation mechanism are required. Using the schema transformation mechanism, the XML Schema of Digital Library X will be transformed to an ontology. Then, the query translation mechanism will be used to translate the SPARQL queries posed over the generated ontology, to XQuery queries over the XML data.

**2nd Scenario.** Consider WoD users and/or applications that express their queries or have implemented their query APIs using the ontologies of Organization A and/or Organization Z. These users and applications should be able to have direct access to Digital Library X from the SW environment, without changing their working environment (e.g., query language, schema, API, etc.). In this scenario, a mapping model and a query translation mechanism are required. In such a case, an expert specifies the mappings between the Organization ontologies and the XML Schema of Digital Library X. These mappings are then exploited by the query translation mechanism, in order to translate the SPARQL queries posed over the Organization ontologies, to XQuery queries to be evaluated over the XML data of Digital Library X. It should be noted that in most real-world situations, an XML Schema may be mapped to more than two ontologies.

---

Note that in the first scenario, Digital Library X may want to publish its data in the WoD, using existing, well accepted vocabularies (e.g., *FOAF*, *SIOC*, *DOAP*, *SKOS*, etc.). The same may hold for the second scenario, where the queries or the APIs may be expressed over well-known vocabularies (which are manually mapped to the XML Schema of Digital Library X).

## 1.2 Framework Overview

In this paper, we present the SPARQL2XQuery Framework, which bridges the heterogeneity gap and creates an interoperable environment between the SW (*OWL/RDF/SPARQL*) and XML (*XML Schema/XML/XQuery*) worlds. An overview of the system architecture of the SPARQL2XQuery Framework is presented in Figure 1.

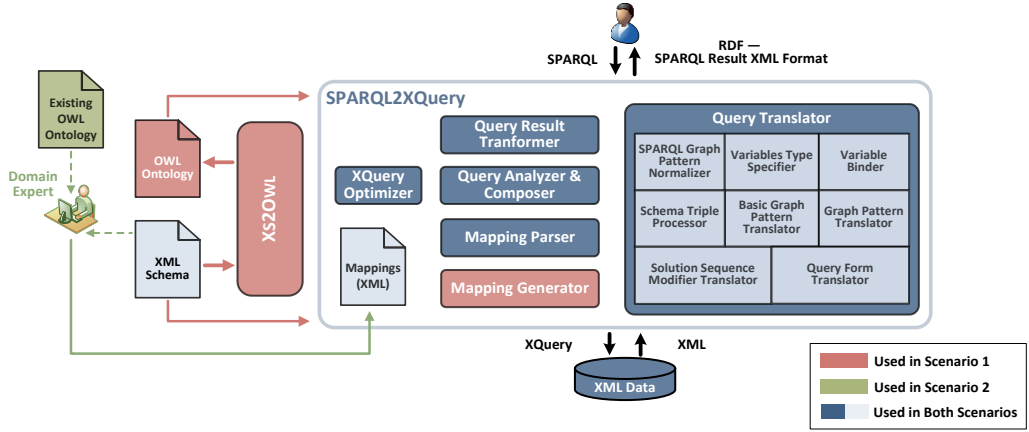

**Figure 1: SPARQL2XQuery Architectural Overview.** In the first scenario, the XS2OWL component is used to create an OWL ontology from the XML Schema. The mappings are automatically generated and stored. In the second scenario, a domain expert specifies the mapping between existing ontologies and the XML Schema. In both scenarios, SPARQL queries are processed and translated into XQuery queries for accessing the XML data. The results are transformed in the preferred format and returned to the user.

As shown in Figure 1, our working scenarios involve existing XML data that follow one or more XML Schemas. Moreover, the SPARQL2XQuery Framework supports two different scenarios:

**1st Scenario: Querying XML data based on automatically generated ontologies.** This is achieved through the XS2OWL component [61] that we have developed and integrated in the SPARQL2XQuery Framework. In particular, the XS2OWL component automatically generates OWL ontologies that capture the XML Schema semantics. Then, the SPARQL2XQuery Framework automatically detects, generates and maintains mappings between the XML Schemas and the OWL ontologies generated by XS2OWL. In this case, the following steps take place:

(a) Using the XS2OWL component, the XML Schema is expressed as an OWL ontology.

(b) The *Mapping Generator* component takes as input the XML Schema and the generated ontology, and automatically generates, maintains and stores the mappings between them in XML format.

(c) The SPARQL queries posed over the generated ontology are translated by the *Query Translator* component to XQuery expressions.

(d) The query results are transformed by the *Query Result Transformer* component into the desired format (SPARQL Query Result XML Format [13] or RDF format).

In this context, our approach can be viewed as a fundamental component of *hybrid ontology-based integration* [39] frameworks (e.g., [40][41]), where the schemas of the XML data sources are represented as OWL ontologies and these ontologies, possibly along with other ontologies, are further mapped to a global ontology.

**2ⁿᵈ Scenario: Querying XML data based on existing ontologies.** In this scenario, XML Schema(s) are manually mapped by an expert to existing ontologies, resulting in the mappings that are used in the SPARQL to XQuery translation. In this case the following steps take place:

(a) An XML Schema is manually mapped to an existing RDF/S–OWL ontology.

(b) The SPARQL queries posed over the ontology are translated to XQuery expressions.

(c) The query results are transformed in the desired format.

In both scenarios, the systems and the users that pose SPARQL queries over the ontology are not expected to know the underlying XML Schemas or even the existence of XML data. They express their queries only in standard SPARQL, in terms of the ontology that they are aware of, and they are able to retrieve XML data. Our Framework is an essential component in the WoD environment that allows setting SPARQL endpoints over the existing XML data.

The SᴘᴀʀQʟ2XQᴜᴇʀʏ Framework supports the following operations:

(a) **Schema Transformation.** Every XML Schema can be automatically transformed in an OWL ontology, using the Xs2Oᴡʟ component.

(b) **Mapping Generation.** The mappings between the XML Schemas and their OWL representations can be automatically detected and stored as XML documents.

(c) **Query Translation.** Every SPARQL query that is posed over the OWL representation of the XML Schemas (first scenario), or over the existing ontologies (second scenario), is translated in an XQuery query.

(d) **Query Result Transformation.** The query results are transformed in the preferred format.

### 1.3 Paper Contributions

The main contributions of this paper are summarized as follows:

1. We introduce the Xs2Oᴡʟ Transformation Model, which facilitates the transformation of XML Schema into OWL ontologies. As far as we know, this is the first work that fully captures the XML Schema semantics.

2. We introduce a mapping model for the expression of mappings from RDF/S–OWL ontologies to XML Schemas, in the context of SPARQL to XQuery translation.

3. We propose a method and a set of algorithms that provide a comprehensive SPARQL to XQuery translation. To the best of our knowledge, this is the first work addressing this issue.

4. We integrate the SᴘᴀʀQʟ2XQᴜᴇʀʏ Framework with the Xs2Oᴡʟ component, thus facilitating the automatic generation and maintenance of the mappings exploited in the SPARQL to XQuery translation.

5. We propose a small number of XQuery rewriting/optimization rules which are applied on the XQuery expressions produced by the translation, aiming at the generation of more efficient XQuery expressions. In addition, we experimentally study the effect of these rewriting rules on the XQuery performance.

6. We describe an extension of the SᴘᴀʀQʟ2XQᴜᴇʀʏ Framework in the context of supporting the SPARQL 1.1 update operations.

7. We conduct a thorough experimental evaluation, in terms of: (a) schema transformation time; (b) mapping generation time; (c) query translation time; and (d) query evaluation time, using both real and synthetic datasets.

### 1.4 Paper Outline

The rest of the paper is organized as follows. The related work is discussed in Section 2. The transformation of XML Schemas into OWL ontologies is detailed in Section 3. The mapping model that has been developed in the context of the SPARQL to XQuery translation is described in Section 4. An overview of the query translation procedure is presented in Section 5. The SPARQL to XQuery translation is described comprehensively in Sections 6 to 9. The XQuery rewriting/optimization rules are outlined in Section 10. Section 11 briefly discusses the support of SPARQL update operations. The experimental evaluation is presented in Section 12. The paper concludes in Section 14, where our future directions are also outlined.

## 2    RELATED WORK

A large number of *data integration* [37] and *data exchange* (also known as *data transformation/translation*) [38] systems have been proposed in the existing literature. In the context of XML, the first research efforts have attempted to provide interoperability and integration between the relational and XML worlds [44] − [51][68]. In addition, several approaches have focused on data integration and exchange over heterogeneous XML data sources [52] − [60].

In the context of interoperability support between the SW and XML worlds [121], numerous approaches for transforming XML Schemas to ontologies, and/or XML data to RDF data and vice versa have been proposed. The most recent ones combine SW and XML technologies in order to transform XML data to RDF and vice versa. Among the published results, the most relevant to our approach are those that utilize the SPARQL query language.

In the rest of this section, we present an overview of the published research that is concerned with the interoperability and integration between the SW and XML worlds (Section 2.1). The latest approaches are described in Section 2.2. Finally, a discussion about the drawbacks and the limitations of the current approaches is presented in Section 2.3.

### 2.1    Bridging the Semantic Web and XML worlds — An Overview

In this section, we summarize the literature related to interoperability and integration issues between the SW and XML worlds. We categorize these systems into *data integration systems* (Table 1) and *data exchange systems* (Table 2).

**Table 1. Overview of the Data Integration Systems in the SW and XML Worlds**

| Data Integration Systems | | | | | |
|---|---|---|---|---|---|
| **System** | **Environment Characteristics** | | | **Operations** | |
| | **Data Models** | **Schema Definition Languages** | **Query Languages** | **Query Translation** | **Schema Transformation** |
| STYX (2002) [64][65] | XML | DTD / Graph | OQL / XQuery | OQL → XQuery | No |
| ICS–FORTH SWIM (2003) [66][67][68] | Relational / XML | DTD / Relational / RDF Schema | SQL / XQuery / RQL | RQL → SQL & RQL → XQUERY | No |
| PEPSINT (2004) [69][70][71][72] | XML | XML Schema / RDF Schema | XQuery / RDQL | RDQL → XQuery | XML Schema → RDF Schema |
| Lehti & Fankhauser (2004) [73] | XML | XML Schema / OWL | XQuery / SWQL | SWQL → XQuery | XML Schema → OWL |
| SPARQL2XQuery | XML | XML Schema / OWL | XQuery / SPARQL | SPARQL → XQuery | XML Schema → OWL (Xs2OwL) |

Table 1 provides an overview of the *data integration systems* in terms of the *Environment Characteristics* and the supported *Operations*. The environment characteristics include the *Data Models* of the underlying data sources, the involved *Schema Definition Languages* and the supported *Query Languages*. The operations include the *Query Translation* and the *Schema*

*Transformation*. Regarding the schema transformation, if the method does not support schema transformation, the value is "no". Notice that the last row of each table describes our SPARQL2XQuery Framework. Note that the SPARQL2XQuery Framework does not deal with the problem of integrating data form different XML data sources; thus, it should be considered as an interoperability system or a core component of integration systems. Hence, it fits better in Table 1 than Table 2.

Table 2 provides an overview of the data exchange systems and is structured in a similar way with Table 1. If the value of the fifth column (*Use of an Existing Ontology*) is "yes", the method supports mappings between XML Schemas and existing ontologies and, as a consequence the XML data are transformed according to the mapped ontologies.

The data integration systems (Table 1) are generally older and they do not support the current standard technologies (e.g., XML Schema, OWL, RDF, SPARQL, etc.). Notice also, that, although the data exchange systems shown in Table 2 are more recent, they do not support an integration scenario neither they provide query translation methods. Instead, they focus on data and schema transformation, exploring how the RDF data can be transformed in XML syntax and/or how the XML Schemas can be expressed as ontologies and vice versa.

**Table 2. Overview of the Data Exchange Systems in the SW and XML Worlds**

| Data Exchange Systems | | | | | |
|---|---|---|---|---|---|
| **System** | **Environment Characteristics** | | **Operations** | | |
| | **Data Models** | **Schema Definition Languages** | **Schema Transformation** | **Use Existing Ontology** | **Data Transformation** |
| Klein (2002) [74] | XML / RDF | XML Schema / RDF Schema | no | no | XML → RDF |
| WEESA (2004) [75] | XML / RDF | XML Schema / OWL | no | yes | XML → RDF |
| Ferdinand et al. (2004) [76] | XML / RDF | XML Schema / OWL–DL | XML Schema → OWL-DL | no | XML → RDF |
| García & Celma (2005) [77] | XML / RDF | XML Schema / OWL–FULL | XML Schema → OWL–FULL | no | XML → RDF |
| Bohring & Auer (2005) [78] | XML / RDF | XML Schema / OWL–DL | XML Schema → OWL–DL | no | XML → RDF |
| Gloze (2006) [79] | XML / RDF | XML Schema / OWL | no | no | XML ↔ RDF |
| JXML2OWL (2006 & 2008) [80][81] | XML / RDF | XML Schema / OWL | no | Yes | XML → RDF |
| GRDDL (2007) [82] | XML / RDF | not specified | no | No | XML ↔ RDF [12] |
| SAWSDL (2007) [83] | XML / RDF | not specified | no | No | XML ↔ RDF [12] |
| Thuy et al. (2007 & 2008) [84][85] | XML / RDF | DTD / OWL–DL | DTD → OWL–DL[12] | No | XML → RDF [12] |
| Janus (2008 & 2011) [86] [87] | XML / RDF | XML Schema / OWL–DL | XML Schema → OWL–DL | No | no |
| Deursen et al. (2008) [88] | XML / RDF | XML Schema / OWL | no | Yes | XML → RDF [12] |
| XSPARQL[8] (2008) [89][90][91] | XML / RDF | not specified | no | No | XML ↔ RDF [12] |
| Droop et al. (2007 & 2008) [93][94][95] | XML / RDF | not specified | no | No | XML → RDF [12] |
| Cruz & Nicolle (2008) [96] | XML / RDF | XML Schema / OWL | no | Yes | XML → RDF |
| XSLT+SPARQL (2008) [97] | XML / RDF | not specified | no | No | RDF → XML |
| DTD2OWL (2009) [98] | XML / RDF | DTD / OWL–DL | DTD → OWL–DL | No | XML → RDF |
| Corby et al. (2009) [99] | XML / RDF / Relational | not specified | No | No | ⟨XML → RDF [12] Relational → RDF |
| TopBraid Composer (Maestro Edition) – TopQuadrant (Commercial Product) [9] | XML / RDF | not specified / OWL | XML → OWL | No | XML ↔ RDF [12] |
| XS2OWL | XML / RDF | XML Schema 1.1 / OWL 2 | XML Schema → OWL | No | XML ↔ RDF |

---

[12] The transformation is performed in a semi-automatic way that requires user intervention.

## 2.2    Recent Approaches

In this section, we present the latest approaches related to the support of interoperability and integration between the SW and XML worlds. These approaches utilize the current W3C standard technologies (e.g., XML Schema, RDF/S, OWL, XQuery, SPARQL, etc.). Most of the latest efforts (Table 2) focus on combining the XML and the SW technologies in order to provide an interoperable environment. In particular, they merge SPARQL, XQuery, XPath and XSLT features to transform XML data to RDF and vice versa.

The W3C *Semantic Annotations for WSDL* (*SAWSDL*) Working Group [83] uses XSLT to convert XML data into RDF, and uses a combination of SPARQL and XSLT for the inverse transformation. In addition, the W3C *Gleaning Resource Descriptions from Dialects of Languages* (*GRDDL*) Working Group [82] uses XSLT to extract RDF data from XML.

*XSPARQL* [89][90][91] combines SPARQL and XQuery in order to achieve the transformation of XML into RDF and back. In the XML to RDF scenario, XSPARQL uses a combination of XQuery expressions and SPARQL Construct queries. The XQuery expressions are used to access XML data, and the SPARQL Construct queries are used to convert the accessed XML data into RDF. In the RDF to XML scenario, XSPARQL uses a combination of SPARQL and XQuery clauses. The SPARQL clauses are used to access RDF data, and the XQuery clauses are used to format the results in XML syntax. Similarly, in [99] XPath, XSLT and SQL are embedded into SPARQL queries in order to transform XML and relational data to RDF. In *XSLT+SPARQL* [97] the XSLT language is extended in order to embed SPARQL SELECT and ASK queries. The SPARQL queries are evaluated over RDF data and the results are transformed to XML using XSLT expressions.

In some other approaches, SPARQL queries are embedded into XQuery and XSLT queries [92]. In [93][94][95], XPath expressions are embedded in SPARQL queries. These approaches attempt to process XML and RDF data in parallel, and benefit from the combination of the SPARQL, XQuery, XPath and XSLT language characteristics. Finally, a method that transforms XML data into RDF and translates XPath queries into SPARQL, has been proposed in [93][94][95].

## 2.3    Discussion

In this section we discuss the existing approaches, and we highlight their main drawbacks and limitations. The existing data integration systems (Table 1) do not support the current standard technologies (e.g., XML Schema, OWL, RDF, SPARQL, etc.). On the other hand, the data exchange systems (Table 2) are more recent and support the current standard technologies, but do not support integration scenarios and query translation mechanisms. Instead, they focus on data transformation and do not provide mechanisms to express XML retrieval queries using the SPARQL query language.

The recent approaches ([82][83][89][92][93][94][95][97][99]) however present severe usability problems for the end users. In particular, the users of these systems are forced to: (a) be familiar with both the SW and XML models and languages; (b) be aware of both ontologies and XML Schemas in order to express their queries; and (c) be aware of the syntax and the semantics of each of the above approaches in order to express their queries. In addition, each of these approaches has adopted its own syntax and semantics by modifying and/or merging the standard technologies. These modifications may also result in compatibility, usability, and expandability problems. It is worth noting that, as a consequence of the scenarios adopted by these approaches, they have only been evaluated over very small data sets.

Compared to the recent approaches, in the SPARQL2XQuery Framework introduced in this paper the users (a) work only on SW technologies; (b) are not expected to know the underlying XML Schema or even the existence of XML data; and (c) they express their queries only in standard (i.e., without modifications) SPARQL syntax. Finally, the SPARQL2XQuery Framework has been evaluated over large datasets.

Moreover, with the high emphasis in the Linked Data infrastructures, publishing legacy data and offering SPARQL endpoints has become a major research challenge. Although several systems (e.g., *D2R Server* [106], *SparqlMap* [107], *Quest* [108], *Virtuoso* [109], *TopBraid Composer*[9]) offer virtual SPARQL endpoints over relational data, to the best of our knowledge there is no system offering SPARQL endpoints over XML data. Finally, in contrast with the SPARQL to XQuery translation, the SPARQL to SQL translation has been extensively studied [101] − [115]. The SPARQL2XQuery Framework introduced here can offer SPARQL endpoints over XML data and it also proposes a method for SPARQL to XQuery translation.

The interoperability Framework presented in this paper includes the Xs2Owl component which offers the functionality needed for automatically transforming XML Schemas and data to SW schemas and data. As such, the Xs2Owl component is related to the data exchange systems (Table 2). The major difference between our work and existing approaches in data exchange systems that provide schema transformation mechanisms is that the latter do not support: (a) the XML Schema identity constraints (i.e., key, keyref, unique); (b) the XML Schema user-defined simple datatypes; and (c) the new constructs introduced by XML Schema 1.1 [2]. These limitations have been overcome by the Xs2Owl component, which is integrated with the other components of the SPARQL2XQuery Framework to offer comprehensive interoperability functionality. To the best of our knowledge, this is the first work that fully captures the XML Schema semantics and supports the XML Schema 1.1 constructs. Finally, this Framework is now completely integrated with the other components of the SPARQL2XQuery Framework. Some preliminary ideas regarding the SPARQL2XQuery Framework have been presented in [119].

## 3    SCHEMA TRANSFORMATION

In this section, we describe the schema transformation process (Figure 2) which is exploited in the first usage scenario, in order to automatically transform XML Schemas into OWL ontologies. Following the automatic schema transformation, mappings between the XML Schemas and the OWL ontologies are also automatically generated and maintained by the SPARQL2XQuery Framework. These mappings are later exploited by other components of the SPARQL2XQuery Framework, for automatic SPARQL to XQuery translation.

The schema transformation is accomplished using the Xs2Owl component [61][63], which implements the Xs2Owl *Transformation Model*. The Xs2Owl transformation model allows the automatic expression of the XML Schema in OWL syntax. Moreover, it allows the transformation of XML data in RDF format and vice versa. The new version of the Xs2Owl Transformation Model which is presented here, exploits the OWL 2 semantics in order to achieve a more accurate representation of the XML Schema constructs in OWL syntax. In addition, it supports the latest versions of the standards (i.e., XML Schema 1.1 and OWL 2). In particular, the XML Schema identity constraints (i.e., key, keyref, unique), can now be accurately represented in OWL 2 syntax (which was not feasible with OWL 1.0). This overcomes the most important limitation of the previous versions of the Xs2Owl *Transformation Model*.

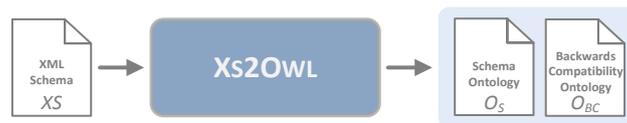

**Figure 2: The Xs2Owl Schema Transformation Process**

An overview of the Xs2Owl transformation process is provided in Figure 2. As is shown in Figure 2, the Xs2Owl component takes as input an XML Schema *XS* and generates: (a) An OWL *schema ontology $O_S$* that captures the XML Schema semantics; and (b) A *Backwards Compatibility ontology $O_{BC}$* which keeps the correspondences between the $O_S$ constructs and the *XS* constructs. $O_{BC}$ also captures systematically the semantics of the XML Schema constructs that cannot be directly captured in $O_S$ (since they cannot be represented by OWL semantics).

The *OWL Schema Ontology $O_S$*, which directly captures the XML Schema semantics, is exploited in the first scenario supported by the SPARQL2XQuery Framework. In particular, $O_S$ is utilized by the users while forming the SPARQL queries. In addition, the SPARQL2XQuery Framework processes $O_S$ and *XS* and generates a list of mappings between the constructs of $O_S$ and *XS* (details are provided in Section 4.5).

The ontological infrastructure generated by the XS2OWL component, additionally supports the transformation of XML data into RDF format and vice versa [62]. For transforming XML data to RDF, $O_S$ can be exploited to transform XML documents structured according to $XS$ into RDF descriptions structured according to $O_S$. However, for the inverse process (i.e., transforming RDF documents to XML) both $O_S$ and $O_{BC}$ should be used, since the XML Schema semantics that cannot be captured in $O_S$ are required. For example, the accurate order of the XML sequence elements should be preserved; but this information cannot be captured in $O_S$.

In the rest of this section, we outline the XS2OWL Transformation Model (Section 3.1) and we present an example that illustrates the transformation of XML Schema into OWL ontology (Section 3.2).

### 3.1 The XS2OWL Transformation Model

In this section, we outline the XS2OWL Transformation Model. A formal description of the XS2OWL Transformation Model and implementation details can be found in [120]. A listing of the correspondences between the XML Schema constructs and the OWL constructs, as they are specified in the XS2OWL Transformation Model, is presented in Table 3.

Table 3. Correspondences between the XML Schema and OWL Constructs, according to the XS2OWL Transformation Model

| XML Schema Construct | OWL Construct |
| --- | --- |
| Complex Type | Class |
| Simple Datatype | Datatype Definition |
| Element | (Datatype or Object) Property |
| Attribute | Datatype Property |
| Sequence | Unnamed Class – Intersection |
| Choice | Unnamed Class – Union |
| Annotation | Comment |
| Extension, Restriction | subClassOf axiom |
| Unique (*Identity Constraint*) | HasKey axiom [*] |
| Key (*Identity Constraint*) | HasKey axiom – ExactCardinality axiom [*] |
| Keyref (*Identity Constraint*) | In the Backwards Compatibility Ontology |
| Substitution Group | SubPropertyOf axioms |
| Alternative [+] | In the Backwards Compatibility Ontology |
| Assert [+] | In the Backwards Compatibility Ontology |
| Override, Redefine [+] | In the Backwards Compatibility Ontology |
| Error [+] | Datatype |

**Note.** The [+] indicates the new XML Schema constructs introduced by the XML Schema 1.1 specification. The [*] indicates the OWL 2 constructs.

The major difficulties that we have encountered throughout the development of the XS2OWL Transformation Model have arisen from the fact that that the XML Schema and the OWL have adopted different data models and semantics. In order to resolve some of these heterogeneity issues, we have employed the *Backwards Compatibility ontology* $O_{BC}$ which encodes XML Schema information that cannot be captured by OWL semantics. This information includes: (a) *Identification information*; (b) *Structural information*; and (c) *"Orphan" construct information*.

*Identification Information*. The OWL semantics do not allow different resources to have the same identifier (rdf:ID), while the XML Schema allows instances of different XML Schema constructs to have the same name (for example, an XML Schema element may have the same name with an XML Schema attribute, two elements of different type may also have the same name, etc.). In order to resolve this issue, the Xs2Owl component generates automatically unique identifiers for the OWL constructs in the *Schema ontology $O_S$*[13]. The correspondence between the names of the XML Schema constructs and the Schema ontology constructs is encoded in the Backwards Compatibility ontology.

*Structural Information.* The XML Schema data model describes ordered hierarchical structures, while the OWL data model allows the specification of directed unordered graph structures. As a consequence, the ordering information which is essential for some XML Schema constructs like the sequences, cannot be captured in the Schema ontology. This information is encoded in the Backwards Compatibility ontology (see [120] for details).

*"Orphan" Construct Information.* Since the XML/XML Schema and the OWL/RDF have adopted different data models and semantics, there exist "orphan" XML Schema constructs that can not be accurately represented by OWL constructs. Examples of "orphan" XML Schema constructs are the abstract and final attributes of the XML Schema type. In the context of the Xs2Owl, information about the "orphan" XML Schema constructs is encoded in the Backwards Compatibility ontology.

```
<xs:schema xmlns:xs="http://www.w3.org/2001/XMLSchema">

    <xs:complexType name="Person_Type">
        <xs:sequence>
            <xs:element ref="LastName" minOccurs="1" maxOccurs="unbounded"/>
            <xs:element name="FirstName" type="xs:string" minOccurs="1" maxOccurs="unbounded"/>
            <xs:element name="Age" type="validAgeType" minOccurs="1" maxOccurs="1" />
            <xs:element name="Email" type="xs:string" minOccurs="0" maxOccurs="unbounded"/>
        </xs:sequence>
        <xs:attribute name="SSN" type="xs:integer"/>
    </xs:complexType>

    <xs:complexType name="Student_Type">
        <xs:complexContent>
            <xs:extension base="Person_Type">
                <xs:sequence>
                    <xs:element name="Dept" type="xs:string"/>
                </xs:sequence>
            </xs:extension>
        </xs:complexContent>
    </xs:complexType>

    <xs:element name="Persons">
        <xs:complexType>
            <xs:sequence>
                <xs:element name="Person" type="Person_Type" minOccurs="0" maxOccurs="unbounded"/>
                <xs:element name="Student" type="Student_Type" minOccurs="0" maxOccurs="unbounded"/>
            </xs:sequence>
        </xs:complexType>
    </xs:element>

    <xs:element name="LastName" type="xs:string"/>

    <xs:element name="Nachname" substitutionGroup="LastName" type="xs:string"/>

    <xs:simpleType name="validAgeType" >
        <xs:restriction base="xs:float">
            <xs:minInclusive value="0.0"/>
            <xs:maxInclusive value="150.0"/>
        </xs:restriction>
    </xs:simpleType>

</xs:schema>
```

**Figure 3: An XML Schema describing Persons (Persons XML Schema)**

---

[13] This is achieved by the *identity generation rules* implemented in the Xs2Owl transformation model. The identity generation rules verify the generation of unique identifiers for all the $O_S$ OWL constructs. These rules exploit the hierarchical structure of the XML Schema, as well as the types of the XML Schema constructs to generate unique identifiers. More details can be found in [120].

## 3.2 XML Schema Transformation Example

We present here a concrete example that demonstrates the expression of an XML Schema in OWL using the Xs2Owl component.

We introduce here an XML Schema (referred in the rest of the paper as the Persons XML Schema), which will be used in the rest of this paper. The Persons XML Schema is presented in Figure 3 and describes the personal information of a sequence of persons (which may be students). The root element Persons may contain any number of Person elements of type Person_Type, and any number of Student elements of type Student_Type. The complex type Person_Type represents persons and contains the SSN attribute and several simple elements (i.e., LastName, FirstName, validAgeType and Email). The complex type Student_Type extends the complex type Person_Type and represents students. In addition to the elements and attributes defined in the context of Person_Type, the complex type Student_Type has the Dept element. The simple type validAgeType is a restriction of the float type. Finally, the top-level element Nachname is an element that may substitute the LastName element, as is specified in its substitutionGroup attribute.

The constructs of the Schema ontology $O_S$ that is automatically generated by the XS2OWL for the Persons XML Schema (referred in the rest of this paper as the Persons Ontology) are presented in Table 4 and Table 5. In particular:

–  Information about the classes is provided in Table 4. The table includes: (a) the name of the corresponding XML Schema complex type (XML Schema Complex Types column); (b) the class rdf:ID (rdf:ID column); and (c) the superclass rdf:IDs (rdfs:subClassOf column).

–  Information about the datatype properties (*DTP*) and the object properties (*OP*) is provided in Table 5. The table includes (a) the name of the corresponding XML Schema element or attribute (XML Schema Elements & Attributes column); (b) the property type, i.e., *DTP* or *OP* (Type column); (c) the property rdf:ID (rdf:ID column); (d) the rdf:IDs of the superproperties (rdfs:subPropertyOf column); (e) the property domains (rdfs:domain column); and (f) the property ranges (rdfs:range column).

**Table 4. Representation of the Persons XML Schema Complex Types in the Schema Ontology ($O_S$)**

| XML Schema Complex Types | Ontology Classes | |
|---|---|---|
| | **rdf:ID** | **rdfs:subClassOf** |
| Person_Type | Person_Type | owl:Thing |
| Student_Type | Student_Type | Person_Type |
| Persons (*unnamed complex type*) | NS_Persons_UNType | owl:Thing |

**Table 5. Representation of the Persons XML Schema Elements and Attributes in the Schema Ontology ($O_S$)**

| XML Schema Elements & Attributes | Ontology Properties | | | | |
|---|---|---|---|---|---|
| | **Type** | **rdf:ID** | **rdfs:subPropertyOf** | **rdfs:domain** | **rdfs:range** |
| LastName | DTP | LastName__xs_string | — | Person_Type | xs:string |
| FirstName | DTP | FirstName__xs_string | — | Person_Type | xs:string |
| Age | DTP | Age__validAgeType | — | Person_Type | validAgeType |
| Nachname | DTP | Nachname__xs_string | LastName__xs_string | Person_Type | xs:string |
| Email | DTP | Email__xs_string | — | Person_Type | xs:string |
| SSN | DTP | SSN__xs_integer | — | Person_Type | xs:integer |
| Dept | DTP | Dept__xs_string | — | Student_Type | xs:string |
| Person | OP | Person__Person_Type | — | NS_Persons_UNType | Person_Type |
| Student | OP | Student__Student_Type | — | NS_Persons_UNType | Student_Type |
| Persons | OP | Persons__NS_Persons_UNType | — | owl:Thing | NS_Persons_UNType |

The constructs of the Backwards Compatibility ontology generated by the XS2OWL are available in [120]. The XML Schema of Figure 3 and the Schema ontology $O_S$ generated by XS2OWL are depicted in Figure 4.

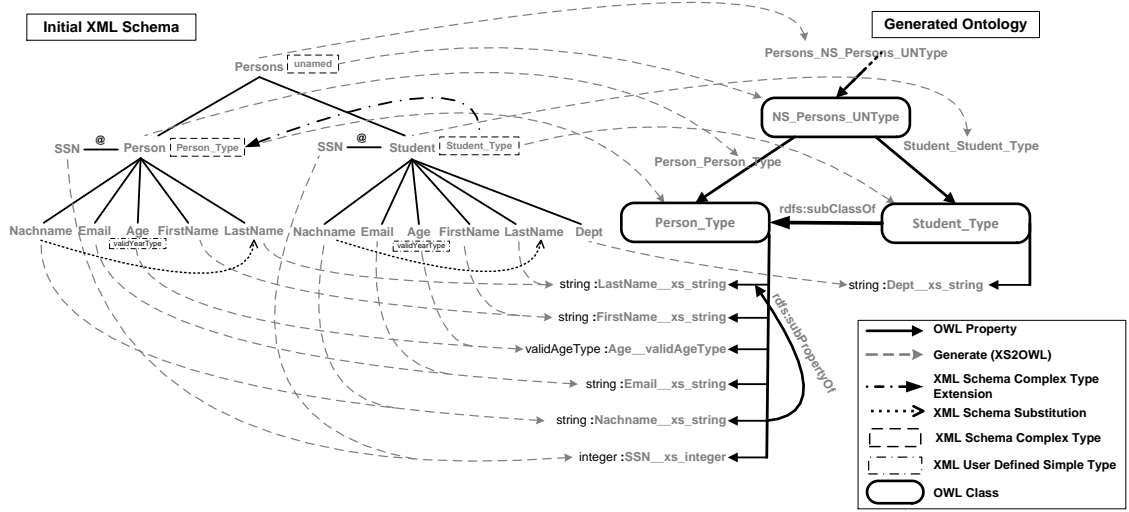

**Figure 4 : The Persons XML Schema of Figure 3 and the Persons Schema Ontology generated by XS2OWL with their correspondences drawn in dashed grey lines**

## 4 MAPPING MODEL

In the SW, the OWL–RDF/S have been adopted as schema definition languages; in the XML world, the XML Schema language is used. The proposed mapping model is defined in the context of the SPARQL to XQuery translation, for the definition of mappings between ontologies and XML Schemas. In particular, the SPARQL2XQUERY mapping model specifies: (a) the supported mappings; (b) the mapping representation; and (c) the necessary operators for formal mapping manipulation.

Mapping *conceptualization*, *definition* and *representation* have been extensively studied under several scenarios (e.g., schema integration, schema matching, data integration, data exchange, etc.). In each scenario, these concepts (i.e., conceptualization, definition, etc.) differ based on the scenario settings. For example, in the classical data integration scenario [37], the local sources are defined as views over a global schema (i.e., *local-as-view − LAV*), or the global schema is defined as a collection of views over the local schemas (i.e., *global-as-view − GAV*). In addition, several similar approaches (e.g., *global-local-as-view − GLAV*, etc.) have been extensively studied and used in data integration systems. Furthermore, in a typical data exchange setting [38], mappings that specify the relations between a source and a target schema are defined as sets of *source-to-target tuple-generating-dependencies* (*st-tgds*). The mappings are used in order to generate instances of the target schema, based on the source data. Nevertheless, our work is not concerned neither with defining views over heterogeneous XML sources nor with defining dependencies for data transformations as is the case in XML data integration and exchange systems (e.g., [52] − [60]). Our mappings can be considered as an interoperability layer between the SW and XML worlds, aiming to provide formal, flexible and precise mapping definitions, as well as generation of efficient XQuery queries. Note that in this work we do not consider the problem of integrating data from different XML data sources.

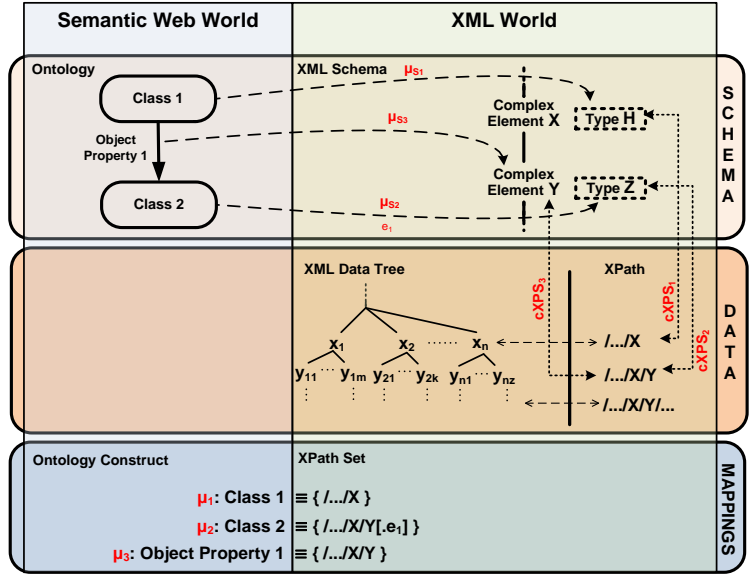

**Figure 5: Associations between the SW and XML Worlds. At the *Schema level*, associations between ontology constructs and XML Schema constructs are obtained. At the *Data level*, the XML data follows the XML Schema and every XML node can be addressed using XPath expressions. Based on the associations between the ontology and the XML Schema, the ontology constructs are associated with the corresponding XPath expressions. In the figure, $\mu_{S_i}$ represents a *schema mapping*, $cXPS_i$ a *correspondence between an XML Schema and XPath Sets*, $\mu_i$ a *mapping representation* and $e_1$ a *mapping condition*.**

We define our mapping model in the context of providing transparent XML querying in the SW world. In the proposed model, the mappings can be simply considered as pairs of ontology constructs (i.e., classes, properties, etc.) and path expressions over the XML data (i.e., XPath). The defined mappings are used for translating the SPARQL queries to XQuery expressions. The adoption of the XPath [4] notion in our mapping model, besides the wide acceptability of XPath, aims to benefit from several XPath properties (e.g., flexibility, expressivity), which are outlined below.

Using XPath expressions we can *precisely* indicate the involved XML nodes. For instance, consider a mapping that aims to indicate the persons whose age is between 20 and 30 (the person definitions follow the Persons schema of Figure 3). Using XPath, this mapping can be expressed as */Persons/Person[./age>20 and ./age<30]*. Moreover, the XPath *expressivity* enhanced with the large XPath library of built-in functions and operators [6], allows our mapping model to support flexible and expressive mapping expressions.

The expression of mappings as XPath expressions allows us to *include both schema and data information*. As schema information, we consider the hierarchical structure of data imposed by the XPath expressions. As data information, we consider conditions over data values (e.g., *age>20*, etc.). The exploitation of the data structuring allows minimizing the number of the considered mappings, resulting in the creation of non redundant or irrelevant queries.

For example, consider a mapping that maps an ontology property name to the Persons XML schema of Figure 3. Assume that the ontology property name is mapped to the XPaths: */Persons/Person/name* and */Persons/Student/name*. Consider now an ontology query aiming to return the names (i.e., the values of the name ontology property) of the persons indicated by the mapping of the previous example (i.e., persons with age between 20 and 30). By examining the property mappings, we can easily notice that the second XPath expression is not relevant to our query. Thus, in this case, the only relevant mapping is the path */Persons/Person/name*.

Finally, the adoption of XPath expressions allows the definition of mappings using other mappings (as "building blocks"). This feature can be exploited in the XQuery expressions for (a) associating different variables and/or (b) for using already evaluated results. The aforementioned can lead to the generation of efficient XQuery queries. For instance, consider an XQuery variable $v, that contains the results of the evaluation of the persons mappings over an XML dataset. Using the $v variable, we can easily "construct" the mappings for the name property as $v/name. In this way, each person can be associated with his name(s) using For XQuery clauses.

Figure 5 outlines the associations between the SW (left side) and XML (right side) worlds. In particular, it presents an ontology, an XML Schema and the associations among them, in both the schema and data levels. At the *schema level* (Ontology/XML Schema), associations between the ontology constructs (i.e., classes, properties, etc.) and the XML Schema constructs (i.e., elements, complex types, etc.) are obtained. Moreover, at the *data level*, the XML data follow the XML Schema. As a result, we can identify the occurrences of the XML Schema constructs in the XML data, and address them using a set of XPath expressions. Finally, the mappings in the context of SPARQL to XQuery translation can be simply considered as associations between ontology constructs and XPath expressions (in the bottom layer of Figure 5).

In the rest of this section, we introduce the XPath Set notion (Section 4.1), we define the schema mappings (Section 4.2), we present the association between the schema and data levels (Section 4.3), we define the mapping representation (Section 4.4), and finally we outline the automatic mapping generation process (Section 4.5).

## 4.1 Preliminaries

In our mapping model, XPath expressions are exploited in order to address XML nodes at the data level. In this section, we provide the basic notions regarding the XPath expressions (Section 4.1.1) and we introduce operators for handling sets of XPath expressions (Section 4.1.2). Finally, Section 4.1.3 specifies the basic XML Schema and Ontology constructs involved in mapping model.

### 4.1.1 Basic XPath Notions

In this section, we introduce some preliminary notions regarding the XPath and XPath Set expressions.

Let $xp \in \mathbf{XP}$ be an XPath expression, where $\mathbf{XP}$ is the set of the XPath expressions. $xp$ is expressed using a fragment of the XPath language, which involves: (a) a set of *node names* $\mathbf{N} = \{n_1, ... n_i\}$; (b) the *child operator* (/); (c) the *predicate operator* ([ ]); (d) the *wildcard operator* (*); (e) the *attribute access operator* (@); (f) the XPath *comparison and set operators* $\mathbf{XPO} = \{!=, <, <=, >, >=, |, =, \text{union}, \text{intersect}\}$; (g) the XPath *built-in functions* $\mathbf{XPF} = \{$ empty, exists, length,... $\}$; and (h) a set of *constants* $\mathbf{C}$.

The *root* node is the first node of an XPath expression. A node $a$ is a *leaf* node if it has no successors (i.e., it is the last XPath node). For example, in the XPath expression $xp = /n_1/n_2/.../n_v$, the nodes $n_1$ and $n_v$ correspond to the *root* and *leaf* nodes respectively. Moreover, $n_1$ is *parent* of $n_2$ and $n_2$ is *child* of $n_1$. The *length* of an XPath is the number of the successive nodes when traversing the path from the beginning (the length of an XPath including only the root node is one). The function *length*: $\mathbf{XP} \rightarrow \mathbb{N}^*$ assigns a length $z \in \mathbb{N}^*$ to an XPath $xp \in \mathbf{XP}$. The function *leaf*: $\mathbf{XP} \rightarrow \mathbf{N}$ assigns the name of the leaf node $n \in \mathbf{N}$ to an XPath $xp \in \mathbf{XP}$. For example, let the XPath $xp = /n_1/n_2/.../n_n$, then $length(xp) = n$ and $leaf(xp) = n_n$. For the XPath expression $xp$ with $length(xp) = n$ we define as $xp(i)$ $(1 \leq i \leq n)$ the $i^{th}$ node $xp$, with $xp(1)$ being the *root* node. In case of predicate existences in the $i^{th}$ node, $x(i)$ refers both to the $i^{th}$ node and to the predicates. As an example, let the XPath $xp = /a/b/c[./d=10]/@e$, then, $x(1) = a$, $x(2) = b$, $x(3) = c$, $x(4) = c[./d=10]$ and $x(5) = @e$.

In what follows, we introduce the notions required in order to specify the semantics of the wildcards (*) and predicates ([ ]) operators while handling the XPath expression.

> **Definition 1. (Loosely Equal Nodes).** Two XPath nodes $v$ and $w$ are defined to be *loosely equal*, denoted as $v \dotdiv w$ *if and only if*: (a) $v'$ and $w'$ result, respectively, from $v$ and $w$ if the predicates [ ] are removed; and (b) $(v' = w')$ or $(v' = * \text{ or } w' = *)$.

Intuitively, two XPath nodes are *loosely equal nodes* if they are the same when we do not consider their predicates, or at least one of them is the wildcard (*) node.

**Definition 2. (Loosely Equal XPaths).** Two XPaths $x$ and $y$ are defined to be *loosely equal*, denoted as $x \approx y$ *if and only if*: (a) they have equal lengths: $length(x)=length(y)=n$; and (b) $\forall i \in \{1,...,n\} \Rightarrow x(i) \dot{\approx} y(i)$.

Intuitively, two XPath nodes are, *loosely equal XPaths* if they have the same length and all their nodes are *loosely equal nodes*.

**Definition 3. (Prefix XPath).** An XPath $x$ is defined to be a *prefix* of an XPath $y$, denoted as $x \widetilde{\subset} y$ *if and only if*: $\exists i: i \leq l$ and $x(j) \dot{\approx} y(j), \forall j \in \{1,..., i\}$ with $l=length(x)$ where $length(x) \leq length(y)$.

Intuitively, an XPath $x$ is *prefix* of another XPath $y$, if a part of $x$ starting from the beginning of $x$ is a *loosely equal XPath* of a path starting from the beginning of $y$.

**Definition 4. (k–Prefix XPath).** An XPath $x$ is defined to be a *k–prefix* of an XPath $y$, denoted as $x \overset{k}{\widetilde{\subset}} y$ *if and only if*: $\exists k: k \leq l$ and $x(i) \dot{\approx} y(i), \forall i \in \{1,..., k\}$ with $l=length(x)$ where $length(x) \leq length(y)$.

Intuitively, an XPath $x$ is *k–prefix* of another XPath $y$, if a part of *length* $k$ of x starting from the beginning of $x$ is a *loosely equal XPath* to a part of $y$ (of $k$ *length*) starting from the beginning of $y$.

Finally, we introduce the *XPath Set* notion.

**Definition 5. (XPath Set).** The set **XPS** = { $xp_1, xp_2,…, xp_n$ }, where $xp_i \in$ **XP** is defined to be an *XPath Set*.

### 4.1.2    XPath Set Operators

In this section, we introduce and formally define a collection of *XPath Set operators* used for handling XPath Sets.

**Common Ancestors Operator.** The *Common Ancestors* operator is a binary operator written as **X ⋖ Y**, where **X** and **Y** are XPath Sets. The result of this operator is the XPath Set that contains the members (XPaths) of the left set **X**, which are *prefixes* of members (i.e., have the same ancestors) of the right set **Y**. The operator is formally defined as:

$$\mathbf{X} \lessdot \mathbf{Y} = \{ \ z: z = x_i \ / \ \exists \ y_j \in \mathbf{Y} : x_i \overset{k_i}{\widetilde{\subset}} y_j \ \}, \text{ where } x_i \in \mathbf{X} \text{ and } length(x_i) = k_i$$

**Example 1.**    Let **X** = { /a/b , /a/b/d , /e/*/f } and **Y** = { /a/b/c/d , /e/h/*/k} then **X ⋖ Y** = { /a/b , /e/*/f }.    ∎

**Descendants of Common Ancestors Operator**. The *Descendants of Common Ancestors* operator is a binary operator written as **X ⋗ Y**, where **X** and **Y** are XPath Sets. The result of this operator is the XPath Set that contains the members (XPaths) of the right set **Y**, the *prefix* XPaths of which are members of the left set **X**. The operator is formally defined as:

$$\mathbf{X} \gtrdot \mathbf{Y} = \{ \ z: z = y_j \ / \ \exists \ x_i \in \mathbf{X} : x_i \overset{k_i}{\widetilde{\subset}} y_j \ \}, \text{ where } x_i \in \mathbf{X} \text{ and } length(x_i) = k_i$$

**Example 2.**    Let **X** = { /a/b , /e/*/f } and **Y** = { /a/b/c/d , /a/p/q , /e/h/*/k } then **X ⋗ Y** = { /a/b/c/d , /e/h/*/k }.    ∎

**Suffixes of Common Ancestors Operator.** The *Suffixes of Common Ancestors* operator is a binary operator written as **X ≫ Y**, where **X** and **Y** are XPath Sets. The result of this operator is the XPath Set that contains the *suffix parts* of the members of the right set **Y**, the *prefix* XPaths of which are contained in the left set **X** (i.e., XPaths contained in **Y** with their ancestors contained in **X**). A *suffix part* of a **Y** member is formed by removing the XPath parts corresponding to the lengthiest prefix XPath included in **X**. The operator is formally defined as:

$$\mathbf{X} \gg \mathbf{Y} = \{ \ z: z = /y_j(k_i+1)/y_j(k_i+2)/... /y_j(k_j) \ / \ \exists \ x_i \in \mathbf{X} : x_i \overset{k_i}{\widetilde{\subset}} y_j \text{ and } \nexists \ x_i' \in \mathbf{X} : x_i \overset{k_i'}{\widetilde{\subset}} y_j \ , \ k_i \leq k_i' \ \}, \text{ where } x_i \in \mathbf{X}, \ y_j \in \mathbf{Y}, \text{ and}$$
$$length(x_i) = k_i, \ length(x_j) = k_j, \ k_i < k_j$$

**Example 3.**  Let **X** = *{ /a/b , /e/*/f }* and **Y** = *{ /a/b/c/d , /e/h/*/k }* then **X** ≫ **Y** = *{ /c/d , /k }*. ∎

**Example 4.**  Let **X** = *{ /a/b , /a/b/c }* and **Y** = *{ /a/b/c/d }* then **X** ≫ **Y** = *{ /d }*. ∎

**XPath Set Union Operator.** The *XPath Set Union* operator is a binary operator written as **X** $\overline{\text{U}}$ **Y**, where **X** and **Y** are XPath Sets. The result of this operator differs from the result of the classic set theory Union operator when a member of **X** and/or **Y** includes the wildcard operator (*) or predicates ([ ]). In these cases the more specific XPaths are excluded from the result set.

In order to formally define the *XPath Set Union* operator, we firstly introduce some special union operators: (a) the *Node Union* operator among XPath nodes; and (b) the *Loose XPath Union* operator among *loosely equal XPaths* (*Definition 2*). These operators are going to be exploited in the definition of the *XPath Set Union* operator among XPath Sets.

(a)  The *Node Union* operator is a binary operator written as $v \; \dot{\vee} \; w$, where $v$ and $w$ are nodes. Let $e$, $e1$ and $e2$ be XPath expressions. The operator is formally defined as:

$$v \; \dot{\vee} \; w = \begin{cases} * & \text{if } (v = *) \text{ or } (w = *) \\ k & \text{if } (v = k) \text{ or } (w = k) \\ k[e] & \text{if } (v = k[e] \text{ or } w \neq *) \text{ or } (w = k[e] \text{ or } v \neq *) \\ k[e1 \mid e2] & \text{if } (v = k[e1]) \text{ or } (w = k[e2]) \end{cases}$$

(b)  The *Loose XPath Union* operator is a binary operator written as $x \; \tilde{\vee} \; y$ and is applied to $x, y \in \mathbf{XP}$ when $x$ and $y$ are *loosely equal* i.e., $x \approx y$. The operator is formally defined as:

$$x \; \tilde{\vee} \; y = \{ \; z \colon z = \text{'/' } x(1) \; \dot{\vee} \; y(1) \text{ '/' } x(2) \; \dot{\vee} \; y(2) \text{ '/' } \dots \text{ '/' } x(n) \; \dot{\vee} \; y(n), \text{ where } n = length(x) = length(y) \; \}$$

Finally, the *XPath Set Union* operator is formally defined as:

$$\mathbf{X} \; \overline{\text{U}} \; \mathbf{Y} = \{ z \colon z = x \; \tilde{\vee} \; y \text{ if } x \approx y \} \cup \{ x \in \mathbf{X} \text{ if not } x \approx y, \forall y \in \mathbf{Y} \} \cup \{ y \in \mathbf{Y} \text{ if not } y \approx x \; \forall x \in \mathbf{X} \}$$

**Example 5.**  Let **X** = *{ /a , /a/b , /d/* , /e/*/f }* and **Y** = *{ /d/g , /a/b/c , /e/h/* }* then **X** $\overline{\text{U}}$ **Y** = *{ /a , /a/b , /d/* , /a/b/c , /e/*/* }*. ∎

**Example 6.**  Let **X** = *{ /a/* , /c/*/d[./e>10] , /g/h[./m>10]}* and **Y** = *{ /a/b , /c/f/d , /g/h[./n>20]}* then **X** $\overline{\text{U}}$ **Y** = *{ /a/* , /c/*/d , /g/h[./m>10 | ./n>20] }*. ∎

**XPath Set Intersection Operator.** The *XPath Set Intersection* operator is a binary operator written as **X** $\overline{\cap}$ **Y**, where **X** and **Y** are XPath Sets. The result of this operators differs from the result of the classic set theory Intersection operator when a member of **X** and/or **Y** includes the wildcard operator (*) or predicates ([ ]). In these cases the more general XPaths are excluded from the result set.

In order to formally define the *XPath Set Intersection* operator, we firstly introduce some special intersection operators: (a) the *Node Intersection* operator among XPath nodes; and (b) the *Loose XPath Intersection* operator among *loosely equal XPaths* (*Definition 2*). These operators are going to be exploited in the definition of the *XPath Set Intersection* operator among XPath Sets.

(a)  The *Node Intersection* operator is a binary operator written as $v \; \dot{\wedge} \; w$, where $v$ and $w$ are nodes. Let $e$, $e1$ and $e2$ be XPath expressions. Formally the operator is defined as:

$$v \mathbin{\dot{\wedge}} w = \begin{cases} * & \textit{if } (v = *) \textit{ or } (w = *) \\ k & \textit{if } (v = k \textit{ or } w \neq k[e]) \textit{ or } (w = k \textit{ or } v \neq k[e]) \\ k[e1] & \textit{if } (v = k[e1] \textit{ or } w \neq k[e2]) \textit{ or } (w = k[e1] \textit{ or } v \neq k[e2]) \\ k[e1][e2] & \textit{if } (v = k[e1]) \textit{ or } (w = k[e2]) \end{cases}$$

(b) The *Loose XPath Intersection* operator is a binary operator written as $x \mathbin{\widetilde{\wedge}} y$, is applied to $x, y \in \mathbf{XP}$ when $x$, and $y$ are *loosely equal* i.e., $x \approx y$. The operator is formally defined as:

$$x \mathbin{\widetilde{\wedge}} y = \{ z \colon z = \text{'}/\text{'} \, x(1) \mathbin{\dot{\wedge}} y(1) \text{'}/\text{'} \, x(2) \mathbin{\dot{\wedge}} y(2) \text{'}/\text{'} \dots \text{'}/\text{'} \, x(n) \mathbin{\dot{\wedge}} y(n), \text{ where } n = length\ (x) = length(y) \}$$

Finally, the *XPath Set Intersection* operator is formally defined as:

$$\mathbf{X} \mathbin{\overline{\cap}} \mathbf{Y} = \begin{cases} x \mathbin{\widetilde{\wedge}} y & \text{if } x \approx y \\ \varnothing & \text{elsewhere} \end{cases}$$

**Example 7.**     Let $\mathbf{X} = \{ /a, /a/b, /d/*, /e/*/f \}$ and $\mathbf{Y} = \{ /d/g, /a/b/c, /e/h/* \}$ then $\mathbf{X} \mathbin{\overline{\cap}} \mathbf{Y} = \{ /d/g, /e/h/f \}$.     ∎

**Example 8.**     Let $\mathbf{X} = \{ /a/*, /c/*/d[./e>10], /g/h[./m>10] \}$ and $\mathbf{Y} = \{ /a/b, /c/f/d, /g/h[./n>20] \}$ then
$\mathbf{X} \mathbin{\overline{\cap}} \mathbf{Y} = \{ /a/b, /c/f/d[./e>10], /g/h[./m>10] [./n>20] \}$.     ∎

**XPath Set Concatenation Operator.** The *XPath Set Concatenation* operator is a binary operator written as $\mathbf{X} \oplus \mathbf{Y}$, where $\mathbf{X}$ and $\mathbf{Y}$ are XPath Sets. The result of this operator is the set that contains the XPaths formed by appending a member of $\mathbf{Y}$ on every member of $\mathbf{X}$. The operator is formally defined as:

$$\mathbf{X} \oplus \mathbf{Y} = \{ z \colon z = x \textit{ concatenate } y, \forall x \in \mathbf{X}, \forall y \in \mathbf{Y} \} \text{ [14]}$$

**Example 9.**     Let $\mathbf{X} = \{ /a, /a/b \}$ and $\mathbf{Y} = \{ /c/d, /e/f \}$ then $\mathbf{X} \oplus \mathbf{Y} = \{ /a/c/d, /a/e/f, /a/b/c/d, /a/b/e/f \}$.     ∎

### 4.1.3    Basic XML Schema & Ontology Constructs

Here, we specify the basic XML Schema and ontology constructs involved in the proposed mapping model.

Let an XML Schema $XS$; (a) $\mathbf{XT}$ is the set of the (complex and simple) *Types* defined in $XS$. Let $\mathbf{XST}$ be the set of the *Simple Types* of $XS$ and $\mathbf{XCT}$ be the set of the *Complex Types* of $XS$. Then, $\mathbf{XT} = \mathbf{XST} \cup \mathbf{XCT}$; (b) $\mathbf{XE}$ is the set of the *Elements* defined in $XS$; and (c) $\mathbf{XAttr}$ is the set of the *Attributes* defined in $XS$. As *XML Schema Constructs* we defined the set $\mathbf{XC} = \mathbf{XT} \cup \mathbf{XE} \cup \mathbf{XAttr}$. Let $xc_1$, $xc_2 \in XC$ be XML constructs. We denote by $xc_1.xc_2$ that the definition of $xc_2$ is nested in the definition of $xc_1$.

Let also an OWL Ontology $OL$; (a) $\mathbf{C}$ is the set of the *OL Classes*; (b) $\mathbf{DT}$ is the set of the *OL Datatypes*; and (c) $\mathbf{Pr}$ is the set of the (datatype and object) *OL Properties*. Let $\mathbf{OP}$ be the set of the *OL Object Properties* and $\mathbf{DTP}$ be the set of the *OL Datatype Properties*. Then, $\mathbf{Pr} = \mathbf{DTP} \cup \mathbf{OP}$. As *Ontology Constructs* we define the set $\mathbf{OC} = \mathbf{C} \cup \mathbf{DT} \cup \mathbf{Pr}$.

In addition, we define a function *Domain:* $\mathbf{Pr} \rightarrow \mathbb{P}(\mathbf{C})$, which assigns the *powerset* (i.e., $\mathbb{P}$) of $\mathbf{C}$ as domain to a property $pr \in \mathbf{Pr}$. We also define a function *Range:* $\mathbf{Pr} \rightarrow \mathbf{A}$, which assigns the range $\mathbf{r} \subseteq \mathbf{A}$ to an ontology property $pr \in \mathbf{P}$ where $\mathbf{A} = \mathbf{DT}$ if $pr \in \mathbf{DTP}$

---

and $\mathbf{A} = \mathcal{P}(\mathbf{C})$ if $pr \in \mathbf{OP}$. The image (i.e., range) of the functions *Range* and *Domain* of a Property $pr$ (i.e., *Domain(pr)* and *Range(pr)*), are denoted, for the sake of simplicity, as *pr.domain* and *pr.range* in the rest of the paper.

## 4.2   Schema Mappings

In this section we define the *Schema Mappings*, which are used to define associations between disparate schema structures over ontologies and XML Schemas in the context of SPARQL to XQuery translation. In our mapping model, the schema mappings may be also enriched with data level information (e.g., conditions over data values), resulting into precise and flexible mappings. Note that since in our context SPARQL queries expressed over ontologies are translated to XQuery queries expressed over XML Schemas, the schema mappings are defined in a directional way from ontologies to XML Schemas.

Given an ontology *OL* and an XML Schema *XS*, let **oc** be a set of *OL* constructs and **xc** a set of *XS* constructs. A *Schema Mapping* ($\mu_S$) between *OL* and *XS* is an expression of the form:

$$\mu_S: OE \overset{E}{\longmapsto} XE,$$

where *OE* is an expression containing **oc** constructs, conjunctions ($\wedge$) or/and disjunctions ($\vee$), *XE* is an expression containing **xc** constructs, conjunctions or/and disjunctions and **E** is a set of conditions applied over the **xc** members.

A schema mapping represents a association among **oc** and **xc** under the conditions specified in **E**. We can simply say that the **oc** members are mapped to the **xc** members under the conditions specified in **E**. The **E** conditions can be simply considered as tree expressions applied over the **xc** constructs.

In more detail, a mapping condition $e \in \mathbf{E}$ is a *tree expression* referring to *XS* constructs and/or XML data that follow *XS*. In particular, a mapping condition $e$ is *applied* on a set of XML Schema constructs $xca \subseteq \mathbf{xc}$ and it may also *refer* (i.e., include) to several constructs independent on $xca$. In addition, a condition $e$ may contain (a) tree paths, (b) operators and functions (e.g., *intersection*, *union*, $<$, $>$, $=$, $\neq$, *ends-with*, *concat*, etc.), as well as (c) constants (e.g., 25, 3.4, "John", etc.). It is remarkable that every XML Schema construct can be referred in a condition expression. Moreover, a mapping condition $e$ may be applied to specific constructs or may be applied to the whole *XE* expression. To sum up, a schema mapping condition $e$ could be any condition which can be expressed in XPath syntax [4]; this way, the high expressiveness of the XPath expressions (including the built-in functions [6]) may be exploited in a mapping condition, and, together with the flexibility of applying independent conditions over different XML constructs, it leads to rich, flexible and expressive schema mappings.

For example, let $c$ be an ontology class and $w$, $z$ be XML Schema complex types. In addition, let the conditions $e_1$ and $e_2$ be applied, respectively, over $w$ and $z$ (denoted as $w \langle\!\langle e_1 \rangle\!\rangle$, $z \langle\!\langle e_2 \rangle\!\rangle$) and a condition $e_3$ applied over the whole *XE* expression (not over a specific construct). A schema mapping $\mu_S$ of the class $c$ to the disjunction of the complex types $w$ and $z$ under the conditions $e_1$ and $e_2$, respectively on $w$ and $z$, and both under the condition $e_3$ is denoted as: $\mu_S: c \overset{\{\ e_3,\ w \langle\!\langle e_1 \rangle\!\rangle,\ z \langle\!\langle e_2 \rangle\!\rangle\}}{\longmapsto} w \vee z$, where, according to the schema mapping definition, $c$ is the *OE* expression, $w \vee z$ is the *XE* expression and $\{e_3, w \langle\!\langle e_1 \rangle\!\rangle, z \langle\!\langle e_2 \rangle\!\rangle\}$ is the condition set **E**. Since the condition $e_3$ is not applied over a specific construct (i.e., it is applied over all the constructs included in *XE*), it holds that $\mathbf{E} = \{w \langle\!\langle e_1 \wedge e_3 \rangle\!\rangle, z \langle\!\langle e_2 \wedge e_3 \rangle\!\rangle\}$.

Regarding the ontology properties, let $pr$ be an ontology property and $q$ be an XML Schema element or attribute. The schema mapping $\mu_S: pr \longmapsto q$ corresponds to $pr.domain \longmapsto d$ and $pr.range \longmapsto q$, where $d$ is the (complex) XML element in which $q$ is defined. In addition, the domain and range of an ontology property $pr$, might be individually mapped to different XML Schema elements/attributes. For instance, let $q$, $v$ be XML Schema elements/attributes, then $\mu_{S_1}: pr.domain \longmapsto q$ and $\mu_{S_2}: pr.range \longmapsto v$.

We can also observe from Figure 5 that the following three schema mappings are obtained: $\mu_{S_1}$: *Class1* $\longmapsto$ *Type H*, $\mu_{S_2}$: *Class2* $\overset{\{Z_{\ell\{\varphi_1\}}\}}{\longmapsto}$ *Type Z* and $\mu_{S_3}$: *Object Property* $\longmapsto$ *Complex Element Y*.

### 4.2.1    *Schema Mapping Specification*

In the first SPARQL2XQuery scenario, where the XS2OWL component is exploited, the schema mappings between the constructs of the XML Schemas and the generated ontologies are automatically specified through the XS2OWL transformation process (Section 3.1). Note that in this case, none of the schema mappings is conditional (i.e., the condition set **E** is equal to the empty set).

We have presented in Figure 4 (with dashed grey lines) the automatically specified schema mappings of the schema transformation example of Section 3.2. Note that the arrows represent the schema transformation process and the schema mappings follow the inverse direction. In this example (Figure 4), we can observe several schema mappings, for instance: $\mu_{S_1}$: *Person_Type* $\longmapsto$ *Person*, $\mu_{S_2}$: *Dept__xs_string* $\longmapsto$ *Student.Dept*, $\mu_{S_3}$: *SSN__xs_integer* $\longmapsto$ *Person.SSN* $\vee$ *Student.SSN*, etc. For each of these schema mappings, the condition set **E** is equal to the empty set and is omitted.

In the second SPARQL2XQuery scenario, an existing ontology is manually mapped to an XML Schema by a domain expert. The mapping process is guided by the language level correspondences (summarized in Table 3), which have also been adopted by the XS2OWL transformation model. For example, ontology classes can associated with XML Schema complex types, ontology object properties with XML elements of complex type, etc. Then, at the schema level, schema mappings between the ontology and XML Schema constructs have to be manually specified (e.g., the person class is mapped to the person_type complex type), following the language level correspondences.

**Example 10. Schema Mapping Specification**

In Figure 6, we present an example of the manual mapping specification scenario, where two existing ontologies that describe the data of two organizations (Organization A and Organization Z) have been manually mapped to an XML Schema. The mappings are presented with dashed grey lines.

In this example, the XML Schema is an extension of the previously presented Persons XML Schema (Figure 3). Here, the Persons schema has been extended by adding the complex element Courses of type Couses_Type as a sub-element of the Student element. The Courses element contains two simple sub-elements, ID and Grade, of type xs:integer and xs:float respectively. These extensions were made in order to be able to define more complex manual mappings in our examples.

Regarding the involved ontologies, the ontology of Organization A has the AGUFIL Group class, where AGUFIL stands for "Adult Gmail Users with the First name Identical to Last name". Moreover, the ontology of Organization Z has the MIT CS Student class which describes the Computer Science Students of the MIT institute. Each of these ontologies has several (self-explained) properties.

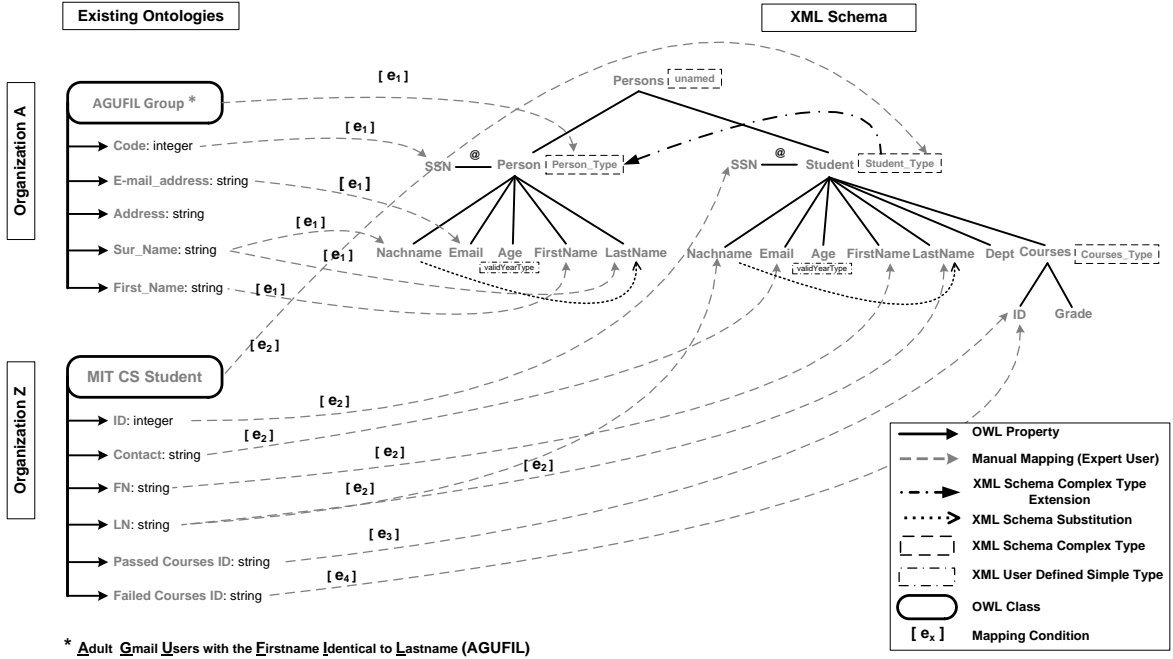

**Figure 6: Manually Specified Schema Mappings between existing Ontologies and an XML Schema (extension of the Persons XML Schema). A domain expert has manually mapped the ontologies to the XML Schema. The mappings are drawn with dashed grey lines.**

We can observe from Figure 6 that several schema mappings can be obtained. For instance, the class AGUFIL Group from Organization A can be mapped to the Person_Type XML complex type (see μs₁ above), under the $e_1$ condition (see above) that restricts the persons to those who are older than 18 years old (i.e., are adults), their first name is the same with their last name and their email account is on the Gmail domain (i.e., ends with *gmail.com*).

$$\mu s_1\text{: } AGUFIL\ Group \overset{\{e_1\}}{\longmapsto} Person\_Type,$$

where $e_1 \equiv Age > 18 \bigwedge FirstName = Lastname \bigwedge email.ends\text{-}with("gmail.com")$

In a similarly way, the class MIT CS Student from Organization Z can be mapped to the Student_Type XML complex type (μs₂), under the $e_2$ condition (see above) that restricts the students to those who have the *CS* as department and their email account is on the MIT domain (i.e., ends with *mit.edu*).

$$\mu s_2\text{: } MIT\ CS\ Student \overset{\{e_2\}}{\longmapsto} Student\_Type,$$

where $e_2 \equiv Dept = "CS" \bigwedge email.ends\text{-}with("mit.edu")$

In Organization A, the ontology property Code can be mapped to the SSN attribute of the Person class under the $e_1$ condition (μs₃). Similarly, the property ID can be mapped to the SSN attribute of the Student class under the $e_2$ condition (μs₄). The Sur_Name property can be mapped to the union of the LastName and Nachname sub-elements of the Person element under the $e_1$ condition (μs₅). Finally, the same holds for the LN property and the LastName and Nachname subelements of the Person element (μs₆).

$$\mu s_3\text{: } Code \overset{\{e_1\}}{\longmapsto} Person.SSN$$

$$\mu s_4\text{: } ID \overset{\{e_2\}}{\longmapsto} Student.SSN$$

$$\mu s_5\text{: } Sur\_Name \overset{\{e_1\}}{\longmapsto} Person.LastName \bigvee Person.Nachname$$

$$\mu_{S6}: LN \overset{\{e_2\}}{\longmapsto} Student.LastName \bigvee Student.Nachname$$

Similarly, in Organization Z, the property Passed_Courses_ID can be mapped to the Course ID sub-element of the Student element ($\mu_{S7}$), under the condition $e_3$ that restricts the Course IDs to those belonging to students from the CS department whose email account is on the MIT domain (i.e., ends with *mit.edu*) and also refer to courses having a passing grade (i.e., equal or greater than 5.0). A similar mapping ($\mu_{S8}$) has been defined for the Failed_Courses property and the condition $e_4$.

$$\mu_{S7}: Passed\_Courses\_ID \overset{\{e_3\}}{\longmapsto} Student.Course.ID,$$
$$\text{where } e_3 \equiv Dept = \text{"CS"} \bigwedge email.ends\text{-}with(\text{"mit.edu"}) \bigwedge Grade \geq 5.0$$

$$\mu_{S8}: Failed\_Courses\_ ID \overset{\{e_4\}}{\longmapsto} Student.Course.ID,$$
$$\text{where } e_4 \equiv Dept = \text{"CS"} \bigwedge email.ends\text{-}with(\text{"mit.edu"}) \bigwedge Grade < 5.0 \qquad \blacksquare$$

## 4.3 Correspondences between XML Schema Constructs and XPath Sets — Associating Schema and Data

We have already defined the schema mappings between ontology constructs and XML Schema constructs (Section 4.2). Since we want to translate SPARQL queries into XQuery expressions that are evaluated over XML data, we should identify the correspondences between the ontology constructs (referred in the SPARQL queries) and the XML data, with respect to the predefined schema mappings. In this section we attempt to express the associations that hold between the XML Schema constructs and the XML data nodes using XPath Set expressions.

At the data level, the XML data is valid with respect to the XML Schema(s) it follows. As a result, for each XML Schema construct we can identify its corresponding XML data nodes and address them using XPath expressions. In this way, we can define the associations between XML schema constructs and XML data.

Given a SPARQL query, for all the ontology constructs referred in the query: (a) we identify the XML Schema constructs based on the predefined schema mappings; and (b) we determine the corresponding XPath Sets for the identified XML Schema constructs. As a result, the ontology constructs referred in the SPARQL query are directly associated with XML data through XPaths.

Formally, let $D$ be an XML dataset, valid with respect to an XML Schema $XS$. A *correspondence of an XML Schema construct xc to an XPath Set xps* is a function $cXPS: \mathbf{XC} \rightarrow \mathbf{XPS}$ that assigns the XPath Set $xps \in \mathbf{XPS}$ to the XML construct $xc \in \mathbf{XC}$, where $xps$ addresses all the corresponding XML nodes of $xc$ in $D$.

For example, we can observe from Figure 5 that we have the following three *XML Schema Constructs to XPath Set Correspondences*: $cXPS$ *(Type H)* = { /.../X}, $cXPS$ *(Type Z)* = { /.../X/Y} and $cXPS$ *(Complex Element Y)* = { /.../X/Y}.

**Table 6. Correspondences between schema mapping expressions ("XML part") and XPath Sets**

| Schema Mapping Expression | XPath Set Correspondences |
|---|---|
| *w* | *cXPS (w)* |
| *e* $\bigvee$ *p* | *xe* $\overline{\bigcup}$ *xp* |
| *e* $\bigwedge$ *p* | *xe* $\overline{\bigcap}$ *xp* |
| *w⟨⟨ce⟩⟩* | *cXPS (w)* [ *xce* ] |
| *w* $\bigvee$ *z* | *cXPS (w)* $\overline{\bigcup}$ *cXPS (z)* |
| *w* $\bigwedge$ *z* | *cXPS (w)* $\overline{\bigcap}$ *cXPS (z)* |
| ( *w* $\bigvee$ *z* )⟨⟨ce⟩⟩ | *cXPS (w)* [*xce*] $\overline{\bigcup}$ *cXPS (z)* [*xce*] |
| ( *w* $\bigwedge$ *z* )⟨⟨ce⟩⟩ | *cXPS (w)* [*xce*] $\overline{\bigcap}$ *cXPS (z)* [*xce*] |

Table 6 summarizes the correspondences between the "XML part" (i.e., referring to the XML Schema constructs) of the schema mapping expressions and XPath Sets. In the left column of the Table 6, *w* and *z* are XML Schema constructs, *e* and *p* XML conditions and *ce* an expression comprised of XML conditions $e_i$, with $i \geq 1$ and possibly conjunctions and disjunctions. In the right column, *xe*, *xp* and *xce* are the XPath expressions corresponding to the *e*, *p* and *ce* conditions respectively. Notice that the conditions are applied over the XPaths using the XPath predicates "[ ]".

### Example 11. Correspondences between XML Schema Constructs and XPath Sets

Consider the schema mapping $\mu_{S5}$: *Sur_Name* $\overset{\{e_1\}}{\longmapsto}$ *Person.LastName* $\lor$ *Person.Nachname* where $e_1 \equiv$ *Age >18* $\land$ *FirstName = Lastname* $\land$ *email.ends-with("gmail.com")* that has been defined in Example 10. This schema mapping maps the Sur_Name property to the union of the LastName and Nachname sub-elements of the Person element under the condition $e_1$. The condition $e_1$ restricts the persons to those who are older than 18 years old, their first name is the same with their last name and their email account is on the Gmail domain.

From the schema mapping $\mu_{S5}$ we have the ("XML part"): *(Person.LastName $\lor$ Person.Nachname)*《$e_1$》, thus, based on Table 6 we come out with the XPath Set correspondence: *cXPS (Person.LastName)*[xe$_1$] $\overline{\cup}$ *cXPS (Person.Nachname)*[xe$_1$], where *xe$_1$* is the XPath expression corresponding to condition *e$_1$*:

$$xe_1 \equiv ./Age >18 \text{ and } ./firstName = ./Lastname \text{ and ends-with}(./email, \text{ "gmail.com"})$$

As a consequence, for the schema mapping $\mu_{S5}$ the corresponding XPath Set is:

*[/Persons/Person[./Age >18 and ./firstName = ./Lastname and ends-with(./email, "gmail.com")]/LastName, /Persons/Person[./Age >18 and ./firstName = ./Lastname and ends-with(./email, "gmail.com")]/ Nachname }.*

In a similar way, consider the schema mapping $\mu_{S7}$ of Example 10: *Passed_Courses_ID* $\overset{\{e_3\}}{\longmapsto}$ *Student.Course.ID* where $e_3 \equiv$ *Dept = "CS"* $\land$ *email.ends-with("mit.edu")* $\land$ *Grade $\geq$ 5.0*. This schema mapping maps the Passed_Courses_ID to the Course ID sub-element of the Student element, under the condition *e$_3$*. The *e$_3$* condition, restricts the Course IDs to those belonging to students from the CS department whose email account is on the MIT domain and also refer to the courses having a passing grade.

From the schema mapping $\mu_{S7}$ we have: *Student.Course.ID*《$e_3$》, thus, based on Table 6 we come out with the XPath Set correspondence: *cXPS (Student.Course.ID)*[xe$_3$], where *xe$_3$* is the XPath expression corresponding to condition *e$_3$*:

$$xe_3 \equiv ./Dept = \text{"CS" and ends-with}(./email, \text{"mit.edu"}) \text{ and } ./Courses/Grade \geq 5.0.$$

Hence, for the schema mapping $\mu_{S7}$ the corresponding XPath Set is:

*{ /Persons/Student[./Dept = "CS" and ends-with(./email , "mit.edu") and ./Courses/Grade $\geq$ 5.0 ]/Courses/ID }.* ∎

### 4.4 Schema Mapping Representation

In the previous sections we have defined the associations at the schema level (*Schema Mappings* – Section 4.2), as well as the associations between the XML Schema and the XML data (*Correspondences between XML Schema Constructs and XPath Sets* – Section 4.3). Here we exploit these associations in order to define and represent the schema mappings in the context of SPARQL to XQuery translation.

In particular, we specify the association of the ontology constructs with XPath Sets through the exploitation of (a) the predefined *schema mappings* between the ontology and XML Schema constructs; and (b) the determined XPath Set for the mapped XML constructs (i.e., *correspondences between XML Schema constructs and XPath Sets*).

To sum up, a mapping in the context of SPARQL to XQuery translation (or simply a *mapping* in the rest of the paper) is represented as the association of an ontology construct with XPath Sets. Thus, this *mapping* representation forms a "*direct association*" between the ontology constructs and the XML data using XPath expressions.

Formally, given an ontology *OL* and an XML Schema *XS*, let **oc** be a set of *OL* constructs, **xc** a set of *XS* constructs and μs a schema mapping between **oc** and **xc**. A *Mapping* (μ) between *OL* and *XS* in the context of the SPARQL to XQuery translation is an expression of the form:

μ: *oc* ≡ *sxps*, where *sxps* is an XPath Set corresponding to **xc** constructs under the schema mapping μs.

For example, as we can observe from Figure 5 that the following three *mappings* can be obtained: μ1: *Class1*≡*{ /.../X}*, μ2: *Class2*≡*{ /.../X/Y[.E]}* and μ3: *Object Property*≡*{ /.../X/Y}*.

In the rest of the paper, for every ontology class *c*, the associated XPath Set is denoted as $\mathbf{X}_c$ (*Class XPath Set*). In addition, for every ontology property *pr*, the associated XPath Set is denoted as $\mathbf{X}_{pr}$ (*Property XPath Set*). Furthermore, for the *pr* domains and ranges, the associated XPath Sets are denoted as $\mathbf{X}_{prD}$ (*Property Domains XPath Set*) and $\mathbf{X}_{prR}$ (*Property Ranges XPath Set*) respectively.

### Example 12. Representing Schema Mappings in the Context of SPARQL to XQuery Translation

Consider the schema mappings μS2, μS3, μS5, μS6, μS7 and μS8 of Example 10 between the ontology and the XML Schema presented in Figure 6. The representations of these schema mappings in the context of SPARQL to XQuery translation are listed below:

μ2: MIT CS Student ≡ $\mathbf{X}_{\text{MIT CS Student}}$ = *{/Persons/Student[./Dept = "CS" and ends-with(./email , "mit.edu")]}*

μ3: Code ≡ $\mathbf{X}_{\text{Code}}$ = *{/Persons/Person[./Age >18 and ./firstName = ./Lastname and ends-with(./email, "gmail.com")]/@SSN}*

μ5: Sur_Name ≡ $\mathbf{X}_{\text{Sur\_Name}}$ = *{/Persons/Person[./Age >18 and ./firstName = ./Lastname and ends-with(./email, "gmail.com")]/LastName}*

μ6: LN ≡ $\mathbf{X}_{\text{LN}}$ = *{/Persons/Student[./Dept = "CS" and ends-with(./email , "mit.edu")]/LastName}*

μ7: Passed_Courses_ID ≡ $\mathbf{X}_{\text{Passed\_Courses\_ID}}$ = *{/Persons/Student[{ ./Dept = "CS" and ends-with(./email , "mit.edu") and ./Courses/Grade ≥ 5.0]/Courses/ID}*

μ8: Failed_Courses_ID ≡ $\mathbf{X}_{\text{Failed\_Courses\_ID}}$ = *{/Persons/Student[./Dept = "CS" and ends-with(./email, "mit.edu") and ./Courses/Grade < 5.0]/Courses/ID}.* ∎

### 4.5 Automatic Mapping Generation

In the first SPARQL2XQuery scenario, the mappings are automatically generated. In particular, the generation of the mappings is carried out by the *Mapping Generator* component, which takes as input an XML Schema and the OWL ontology generated by XS2OWL for this XML Schema. In the first phase, the Mapping Generator component parses the input files and obtains the schema mappings between the XML Schema and the generated ontology by exploiting the XS2OWL Transformation Model. Then, using the XML Schema, the Mapping Generator component determines the *XML Schema construct to XPath Set Correspondences* for all the XML constructs. Finally, the component generates an XML document that contains the associations of all the ontology constructs with the XPath Sets. In particular, it generates the sets $\mathbf{X}_c$, $\mathbf{X}_{pr}$, $\mathbf{X}_{prD}$ and $\mathbf{X}_{prR}$ for all the ontology classes and properties.

### Example 13. Automatic Mapping Generation

Consider the XML Schema of Figure 3 and the corresponding ontology generated by XS2OWL (Table 4 and Table 5). Based on the automatically specified schema mappings (Figure 4), the Mapping Generator component generates the mapping representations listed below. It should be mentioned that in this case the mappings are trivial, since the ontology is an OWL representation of the XML Schema.

| Generated Mappings between the XML Schema and the Ontology of **Figure 4** |
| --- |

*Classes:*

Person_Type = $X_{Person\_Type}$ = { /Persons/Person }

Student_Type = $X_{Student\_Type}$ = { /Persons/Student }

NS_Persons_UNType = $X_{NS\_Persons\_UNType}$ = { /Persons }

*Object Properties:*

Persons__NS_Persons_UNType = $X_{Persons\_\_NS\_Persons\_UNType}$ = { /Persons }

  Persons__NS_Persons_UNType.domain = $X_{Persons\_\_NS\_Persons\_UNTypeD}$ = { /Persons }

  Persons__NS_Persons_UNType.range = $X_{Persons\_\_NS\_Persons\_UNTypeR}$ = { /Persons }

Person__Person_Type = $X_{Person\_\_Person\_Type}$ = { /Persons/Person }

  Person__Person_Type.domain = $X_{Person\_\_Person\_TypeD}$ = { /Persons/Person }

  Person__Person_Type.range = $X_{Person\_\_Person\_TypeR}$ = { /Persons/Person }

Student__Student_Type = $X_{Student\_\_Student\_Type}$ = { /Persons/Student }

  Student__Student_Type.domain = $X_{Student\_\_Student\_Type\ D}$ = { /Persons }

  Student__Student_Type.range = $X_{Student\_\_Student\_Type\ R}$ = { /Persons/Student }

*Datatype Properties:*

FirstName__xs_string = $X_{FirstName\_\_xs\_string}$ = { /Persons/Person/FirstName, /Persons/Student/FirstName }

  FirstName__xs_string.domain = $X_{FirstName\_\_xs\_stringD}$ = { /Persons/Person, /Persons/Student }

  FirstName__xs_string.range = $X_{FirstName\_\_xs\_stringR}$ = { /Persons/Person/FirstName, /Persons/Student/FirstName }

LastName__xs_string = $X_{LastName\_\_xs\_string}$ = { /Persons/Person/LastName, /Persons/Student/LastName }

  LastName__xs_string.domain = $X_{LastName\_\_xs\_stringD}$ = { /Persons/Person, /Persons/Student }

  LastName__xs_string.range = $X_{LastName\_\_xs\_stringR}$ = { /Persons/Person/LastName, /Persons/Student/LastName }

Age__xs_integer = $X_{Age\_\_xs\_integer}$ = { /Persons/Person/Age, /Persons/Student/Age }

  Age__xs_integer.domain = $X_{Age\_\_xs\_integerD}$ = { /Persons/Person, /Persons/Student }

  Age__xs_integer.range = $X_{Age\_\_xs\_integerR}$ = { /Persons/Person/Age, /Persons/Student/Age }

Email__xs_string = $X_{Email\_\_xs\_string}$ = { /Persons/Person/Email, /Persons/Student/Email }

  Email__xs_string.domain = $X_{Email\_\_xs\_stringD}$ = { /Persons/Person, /Persons/Student }

  Email__xs_string.range = $X_{Email\_\_xs\_stringR}$ = { /Persons/Person/Email, /Persons/Student/Email }

Nachname__xs_string = $X_{Nachname\_\_xs\_string}$ = { /Persons/Person/Nachname, /Persons/Student/Nachname }

  Nachname__xs_string.domain = $X_{Nachname\_\_xs\_stringD}$ = { /Persons/Person, /Persons/Student }

  Nachname__xs_string.range = $X_{Nachname\_\_xs\_stringR}$ = { /Persons/Person/Nachname, /Persons/Student/Nachname }

SSN__xs_integer = $X_{SSN\_\_xs\_integer}$ = { /Persons/Person/@SSN, /Persons/Student/@SNN }

  SSN__xs_integer.domain = $X_{SSN\_\_xs\_integerD}$ = { /Persons/Person, /Persons/Student }

  SSN__xs_integer.range = $X_{SSN\_\_xs\_integerR}$ = { /Persons/Person/@SSN, /Persons/Student/@SNN}

Dept__xs_string = $X_{Dept\_\_xs\_string}$ = { /Persons/Student/Dept }

  Dept__xs_string.domain = $X_{Dept\_\_xs\_stringD}$ = { /Persons/Student }

  Dept__xs_string.range = $X_{Dept\_\_xs\_stringR}$ = { /Persons/Student/Dept }

∎

## 5 INTRODUCING THE QUERY TRANSLATION PROCESS

In this section, we give an overview of the SPARQL query language (Section 5.1), we introduce several basic notions (Section 5.2), and finally we summarize the query translation process (Section 5.3).

### 5.1 SPARQL Query Language Overview

SPARQL [35] is a W3C recommendation and it is today the standard query language for RDF data. The evaluation of a SPARQL query is based on graph pattern matching. The SPARQL Where clause consists of a *Graph Pattern*. The Graph Pattern is defined recursively and contains *Triple patterns* and SPARQL operators. The operators of the SPARQL algebra that can be applied on Graph Patterns are: AND, UNION, OPTIONAL and FILTER. *Triple patterns* are just like *RDF triples* but each of the *subject*, *predicate* and *object* parts may be a variable.

SPARQL allows four query forms: Select, Ask, Construct and Describe. In addition, SPARQL provides various solution sequence modifiers that can be applied on the initial solution sequence in order to create another, user desired, sequence. The supported SPARQL solution sequence modifiers are: Distinct, Order By, Reduced, Limit, and Offset. Finally, the SPARQL query results may be *RDF Graphs*, *SPARQL solution sequences* and *Boolean values*.

#### 5.1.1 RDF and SPARQL Syntax

In this section, we provide a set of formal definitions of the syntax of RDF and SPARQL (based on [33] and [12]).

Let **I** be the set of the IRIs (*Internationalized Resource Identifiers*)[15], **L** the set of the RDF Literals, and **B** be the set of the Blank nodes. In addition, assume the existence of an infinite set **V** of variables disjoint from the previous sets (**I**, **B**, **L**).

---

[15] *http://www.ietf.org/rfc/rfc3987.txt*

**Definition 6. (RDF Triple).** A triple $\langle s, p, o \rangle \in (\mathbf{I} \cup \mathbf{B}) \times \mathbf{I} \times (\mathbf{I} \cup \mathbf{B} \cup \mathbf{L})$ is called *RDF triple*[16]. $s$, $p$ and $o$ represent, respectively, the subject, predicate and object of an RDF triple. The subject $s$ can either be an IRI or a Blank node. The predicate $p$ must be an IRI. The object $o$ can be an IRI, a Blank node or an RDF Literal.

**Definition 7. (RDF Dataset).** An *RDF Dataset* (or RDF Graph) is a set of RDF triples.

**Definition 8. (Triple Pattern).** A triple $\langle s, p, o \rangle \in (\mathbf{I} \cup \mathbf{B} \cup \mathbf{V}) \times (\mathbf{I} \cup \mathbf{V}) \times (\mathbf{I} \cup \mathbf{B} \cup \mathbf{L} \cup \mathbf{V})$ is called *Triple pattern*.

In the rest of the paper, when we refer to variables, we also refer to Blank nodes, since they are semantically equivalent.

**Definition 9. (Graph Pattern).** A *Graph Pattern* (*GP*) is a SPARQL graph pattern expression defined recursively as follows: (a) A *Triple pattern* is a graph pattern. (b) If $P_1$ and $P_2$ are graph patterns, then the expressions ($P_1$ AND $P_2$), ($P_1$ OPT $P_2$), and ($P_1$ UNION $P_2$) are graph patterns. (c) If $P$ is a graph pattern and $R$ is a SPARQL *built-in condition*, then the expression ($P$ FILTER $R$) is a graph pattern.

Note that a SPARQL *built-in condition* (or else *Filter expression*) is constructed using IRIs, RDF literals, variables and constants, as well as operators (e.g., &&, ||, !, =, !=, >, <, ≤, ≥, +, -, *, /, bound, lang, regex, etc.) (Refer to [12] for a complete list). With $var(gp)$ we denote the set of variables occurring in a graph pattern $gp$.

**Definition 10. (Union–Free Graph Pattern).** A SPARQL graph pattern (*GP*) that does not contain UNION operators is a *Union–Free Graph Pattern* (*UF–GP*).

**Definition 11. (Basic Graph Pattern).** A finite sequence of conjunctive *triple patterns* and possible *Filters* is called *Basic Graph Pattern* (*BGP*).

## 5.2 Query Translation Preliminaries

Here, we introduce some essential query translation notions. Let $\mathbf{I}_{RDF}$ be the set containing the IRIs of the RDF vocabulary (e.g., rdf:type, rdf:Property, etc.), $\mathbf{I}_{RDFS}$ the set containing the IRIs of the RDF Schema vocabulary (e.g., rdfs:subClassOf, rdfs:domain, etc.) and $\mathbf{I}_{OWL}$ the set containing the IRIs of the OWL vocabulary (e.g., owl:equivalentClass, owl:FunctionalProperty, etc.). Moreover, let $\mathbf{I}_{CL}$ be the set containing the IRIs of the classes of an ontology and $\mathbf{I}_{PR}$ the set containing the IRIs of the properties of an ontology.

From the above sets, we define the set $\mathbf{I}_{VC}$, containing all the IRIs of the *RDF/S* and *OWL* vocabularies $\mathbf{I}_{VC} = \mathbf{I}_{RDF} \cup \mathbf{I}_{RDFS} \cup \mathbf{I}_{OWL}$. Moreover, we define the set $\mathbf{I}_{OL}$, containing the *IRIs* that refer to ontology classes and properties $\mathbf{I}_{OL} = \mathbf{I}_{CL} \cup \mathbf{I}_{PR}$.

**Definition 12. (Schema Triple Pattern).** A *Schema Triple Pattern* is a triple pattern which refers to the ontology structure and/or semantics. In particular, a *Schema Triple Pattern* is a triple pattern that contains concepts and properties of the *RDF/S* and *OWL* vocabularies, or, a triple pattern having *IRIs* that refer to ontology classes or properties. Formally, a *Schema Triple Pattern* is defined as follows:

A triple $\langle s, p, o \rangle \in (\mathbf{I}_{VC} \cup \mathbf{I}_{OL} \cup \mathbf{B} \cup \mathbf{V}) \times (\mathbf{I}_{VC} \cup \mathbf{I}_{OL}) \times (\mathbf{I}_{VC} \cup \mathbf{I}_{OL} \cup \mathbf{B} \cup \mathbf{L} \cup \mathbf{V})$ is called *Schema Triple Pattern* (or simply *Schema Triple*). We use $schemaTr(gp)$ to denote the set of *Schema Triples* occurring in a graph pattern $gp$.

**Example 14. Schema Triple Patterns**

For example, the triple patterns: (a) *?x rdfs:subClassOf ?y* and (b) *?z rdf:type owl:FunctionalProperty* are Schema Triple Patterns, which have the rdfs:subClassOf, the rdf:type and the owl:FunctionalProperty terms of the RDF/S and OWL vocabularies as predicates and objects. Similarly, the triple patterns (c) *?x rdfs:subClassOf ns:Person*, (d) *ns:hasName*

---

*rdfs:domain ?x* and (e) *ns:Person owl:equivalentClass ?x* which have the ontology class ns:Person and the ontology property ns:hasName as subject and predicate respectively. ∎

Below we introduce the notion of *semantically corresponding* queries.

Let $SL_1$ and $SL_2$ be schema definition languages, and, $s_1$ and $s_2$ be schemas expressed in $SL_1$ and $SL_2$, respectively. Let **M** be a set of mappings between $s_1$ and $s_2$. Let $D_1$ be a set of instances (i.e., dataset) over $s_1$. A *Data Transformation* (*DTr*) from a set of instances $D_1$ to a set of instances $D_2$ w.r.t. **M**, is the transformation of $D_1$ into instances of $s_2$ w.r.t **M**; resulting $D_2$. Thus, the *Data Transformation* can be considered a function $DTr$: $\{D_1\} \times \mathbf{M} \longrightarrow D_2$, where $D_1$ and $D_2$ are sets of instances over the schemas $s_1$ and $s_2$, and, $s_1$ and $s_2$ are expressed in $SL_1$ and $SL_2$ schema definition languages.

Let $SL_1$ and $SL_2$ be schema definition languages, and, $s_1$ and $s_2$ be schemas expressed in $SL_1$ and $SL_2$, respectively. Let **M** be a set of mappings between $s_1$ and $s_2$. Let $D_1$ be a set of instances over $s_1$. Let $D_2 = DTr(D_1, \mathbf{M})$ the *Data Transformation* of the set of instances $D_1$ w.r.t. **M**, where $D_2$ is a set of instances over $s_2$. Let $QL_1$ and $QL_2$ be query languages, and, $Q_1$ and $Q_2$ be queries expressed in $QL_1$ and $QL_2$, respectively. We say that $Q_1$ is *semantically correspondent* to $Q_2$ w.r.t. **M** if and only if the solutions returned from the evaluation of $Q_1$ over $D_1$ are the same as the evaluation of $Q_2$ over $D_2$.

In our problem, $SL_1$ and $SL_2$ are, respectively, the XML Schema and OWL schema definition languages. Moreover, $QL_1$ and $QL_2$ are, respectively, the XQuery and SPARQL query languages.

### 5.3    Query Translation Overview

In this section we present an overview of the SPARQL to XQuery query translation process, which is performed by the *Query Translator* component. The Query Translator takes as input a SPARQL query and the mappings between an ontology and an XML Schema and translates the SPARQL query to semantically corresponding XQuery expressions w.r.t. the mappings.

The query translation process is based on a generic method and a set of algorithms for translating SPARQL queries to XQuery expressions following strictly the SPARQL semantics. The translation covers all the syntax variations of the SPARQL grammar [12]; as a result, it can handle every SPARQL query. In addition, the translation process is generic and scenario independent, since the mappings are represented in an abstract formal form as XPath Sets. The mappings may be automatically generated or manually specified.

The objectives for the development of the query translation process have been the following: (a) Development of a generic method for the SPARQL to XQuery translation; (b) Capability of translating every query compliant to the SPARQL grammar; (c) Obeying strictly the SPARQL semantics; (d) Independence from query engines and storage environments; (e) Production of as simple XQuery expressions as possible; (f) Construction of XQuery expressions so that their correspondence with SPARQL can be easily understood; and (g) Construction of XQuery expressions that produce results that do not need any further processing.

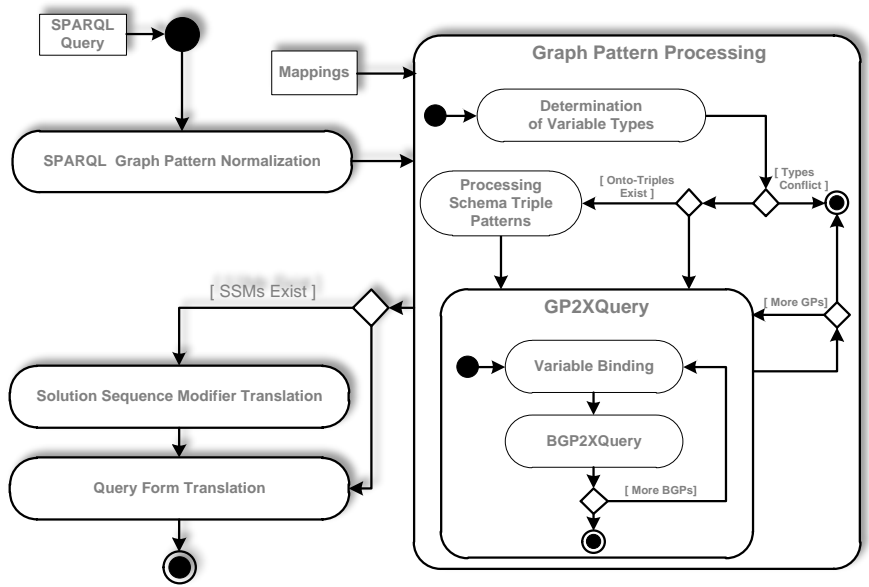

**Figure 7: UML activity diagram describing the SPARQL to XQuery translation process. The activity takes as input a SPARQL query and the mappings between an OWL ontology and an XML Schema and generates the semantically corresponding XQuery expressions**

Figure 7 presents, in the form of a UML diagram, the entire translation process. As is shown in Figure 7, the translation process takes as input a SPARQL query and the mappings between an ontology and an XML Schema. The *SPARQL Graph Pattern Normalization* activity (Section 6.1) rewrites the Graph Pattern (*GP*) of the SPARQL query in an equivalent normal form, resulting into a simpler and more efficient translation process. The *Graph Pattern Processing* follows, in order to translate the Graph Pattern of the SPARQL query (the query **Where** clause) to XQuery expressions. Afterwards, the solution sequence modifiers (*SSMs*) that may be contained in the query are translated (Section 9.1). Finally, based on the SPARQL query form, the generated XQuery is enhanced with appropriate expressions in order to achieve the desired structure of the results (e.g., either construct an RDF graph or return a SPARQL result sequence) (Section 9.2).

The Graph Pattern Processing is a composite activity with various sub-activities (Figure 7). Initially, an activity identifies the types of the SPARQL variables, in order to determine the form of the results, as well as to perform consistency checking in variable usage (Section 6.2). Afterwards, an activity (Section 6.3) processes the Schema Triples (*Definition 12*) that may exist in the pattern and determines the variable bindings for them (i.e., assigns the appropriate XPaths to variables). These bindings are going to be used in the next steps as initial variable bindings. Finally, a translation algorithm (*GP2XQuery*) that translates *GPs* to XQuery expressions is exploited (Section 8.2). Throughout the *GP2XQuery* translation, for each Basic Graph Pattern (*BGP*) contained in the *GP*, a *Variable Binding* phase (Section 7) and a BGP to XQuery translation (Section 8.3) are preformed.

# 6    QUERY NORMALIZATION, VARIABLE TYPES & SCHEMA TRIPLES

## 6.1    SPARQL Graph Pattern Normalization

In this section, we describe the *SPARQL Graph Pattern normalization* phase, which rewrites the Graph Pattern (*GP*, *Definition 9*) of a SPARQL query, and transforms it to an equivalent *normal form* (see below). The SPARQL graph pattern normalization is based on the *GP* expression equivalences proved in [33] and on query rewriting techniques.

> **Definition 13.    (Well Designed Graph Pattern)** [33]**.** A Union–Free Graph Pattern (*Definition 10*) $P$ is *well designed* if for every sub-pattern $P' = (P_1$ OPT $P_2)$ of $P$ and for every variable *?X* occurring in $P$, the following condition holds: if *?X* occurs both inside $P_2$ and outside $P'$ then it also occurs in $P_1$.

The graph pattern equivalences differ for the *well designed GPs* (*Definition 13*) and the *non–well designed GPs*[17]. Thus, in case of OPT existence, it is essential for this phase to identify if the *GP* is *well designed* or not (if OPT does not exist, *GP* is always *well designed*). This clarification is performed by validating the *well design* condition over the *GP*. Finally, every *GP* is transformed to a *normal form* formally described as follows:

$$P_1 \text{ UNION } P_2 \text{ UNION } P_3 \text{ UNION } \cdots \text{ UNION } P_n, \text{ where } P_i\ (1{\leq}i{\leq}n) \text{ is a Union–Free Graph Pattern.} \quad \textbf{(1)}$$

The new *GP* normal form allows an easier and more efficient translation process, as well as the creation of more efficient XQuery queries since: (a) The normal form contains a sequence of Union–Free Graph Patterns, each of which can be processed independently. (b) The *normal form* contains larger Basic Graph Patterns. The larger Basic Graph Patterns result in a more efficient translation process, since they reduce the number of the *variable bindings*, as well as the *BGP* to XQuery translation processes that are required (more details can be found in Section 8.2). (c) The larger Basic Graph Patterns result in more sequential conjunctions (i.e., ANDs) intrinsically handled by XQuery expressions, thus more efficient XQuery queries (more details in can be found Section 8.2).

Note that in almost all cases, the "real-world" (i.e., user defined) SPARQL graph patterns are initially expressed in normal form [34], thus this phase is often avoided.

## 6.2    Variable Type Determination

In this section we describe the *variable type determination* phase. This phase identifies the type of every SPARQL variable referenced in a Union–Free Graph Pattern (*UF–GP*, *Definition 10*). The determined variable types are used to specify the form of the results and, consequently, the syntax of the Return XQuery clause. Moreover, the variable types are exploited for generating more efficient XQuery expressions. In particular, the variable types are exploited by the *processing Schema Triple patterns* and the *variable binding* phases, in order to reduce the possible bindings by pruning the redundant bindings. Finally, through the variable type determination, a consistency check is performed in variable usage, in order to detect possible conflicts (i.e., the same variable may be determined with different types in the same *UF–GP*). In such a case, the *UF–GP* can not be matched against any RDF dataset, thus, this *UF–GP* is pruned and is not translated, resulting into more efficient XQuery expressions that speed up the translation process (see *Example 16*). In Table 7 we define the variable types that may occur in triple patterns.

---

[17] A graph pattern that is not compatible with Definition 13 is called a *non-well designed* graph pattern.

**Table 7. Variable Types**

| Notation | Name | Description |
|----------|------|-------------|
| *CIVT* | *Class Instance Variable Type* | Represents class instance variables |
| *LVT* | *Literal Variable Type* | Represents literal value variables |
| *UVT* | *Unknown Variable Type* | Represents unknown type variables |
| *DTPVT* | *Data Type Predicate Variable Type* | Represents data type predicate variables |
| *OPVT* | *Object Predicate Variable Type* | Represents object predicate variables |
| *UPVT* | *Unknown Predicate Variable Type* | Represents unknown predicate variables |

*6.2.1    Variable Type Determination Rules*

Here we describe the rules that are used for the determination of the variable types. Let *OL* be an ontology, *UF–GP* be a Union–Free Graph Pattern expressed over *OL*, **mDTP** (*Mapped Data Type Properties Set*) be the set of the mapped datatype properties of *OL*, **mOP** (*Mapped Object Properties Set*) be the set of the mapped object properties of *OL*, $\mathbf{V}_{\text{UFGP}}$ (*UF–GP Variables Set*) be the set of the variables that are defined in the *UF–GP*[18] and $\mathbf{L}_{\text{UFGP}}$ (*UF–GP Literal Set*) be the set of the literals referenced in the *UF–GP*.

The *variable type determination* is a function *VarType*: $\mathbf{V}_{\text{UFGP}} \longrightarrow \mathbf{VT}$ that assigns a variable type *vt* ∈ **VT** to every variable *v* ∈ $\mathbf{V}_{\text{UFGP}}$, where **VT** = {*CIVT, LVT, UVT, DTPVT, OPVT, UPVT*} includes all the variable types. The relation between the domain and range of the function *VarType* is defined by the determination rules presented below.

Here, we enumerate the determination rules that are applied iteratively for each triple in the given *UF–GP*. The final result of the rules is not affected by the order in which the rules are applied neither by the order in which the triple patterns are parsed. As $T_x$ is denoted the type of a variable *x*.

Given a (non-Schema) triple pattern *t* ∈ ⟨*s, p, o*⟩, where *s* is the subject part, *p* the predicate part and *o* the object part, we define the following rules:

**Rule 1:** If *s* ∈ $\mathbf{V}_{\text{UFGP}}$ ⟹ $T_s$ = *CIVT*. If the subject is a variable, then the variable type is *Class Instance Variable Type* (*CIVT*).

**Rule 2:** If *p* ∈ **mDTP**, and *o* ∈ $\mathbf{V}_{\text{UFGP}}$ ⟹ $T_o$ = *LVT*. If the predicate is a datatype property and the object is a variable, then the type of the object variable is *Literal Variable Type* (*LVT*).

**Rule 3:** If *p* ∈ **mOP**, and *o* ∈ $\mathbf{V}_{\text{UFGP}}$ ⟹ $T_o$ = *CIVT*. If the predicate is an object property and the object is a variable, then the type of the object variable is *Class Instance Variable Type* (*CIVT*).

**Rule 4:** $T_p$ = *DTPVT* ⟺ $T_o$ = *LVT* | *p*, *o* ∈ $\mathbf{V}_{\text{UFGP}}$. If the predicate variable type is *Data Type Predicate Variable Type* (*DTPVT*), then the type of the object variable is *Literal Variable Type* (*LVT*). The inverse also holds.

**Rule 5:** $T_p$ = *OPVT* ⟺ $T_o$ = *CIVT* | *p*, *o* ∈ $\mathbf{V}_{\text{UFGP}}$. If the predicate variable type is *Object Predicate Variable Type* (*OPVT*), then the type of the object variable is *Class Instance Variable Type* (*CIVT*). The inverse also holds.

**Rule 6:** If *o* ∈ $\mathbf{L}_{\text{UFGP}}$, and *p* ∈ $\mathbf{V}_{\text{UFGP}}$ ⟹ $T_p$ = *DTPVT*. If the object is a literal value, then the type of the predicate variable is *Data Type Predicate Variable Type* (*DTPVT*).

The unknown variable types *UTV* and *UPTV* do not result in conflicts in case that a variable has been also defined to have another type since they can be just ignored. All the variable types are initialized to the *Unknown Predicate Variable Type* (*UPVT*) if they appear in the predicate part of a triple; otherwise, they are initialized to the *Unknown Variable Type* (*UVT*).

---

[18] The $\mathbf{V}_{\text{UFGP}}$ set does not include the variables that occur only in *Schema triple patterns* (*Definition 12*), since the Schema triple patterns are omitted from the *variables type determination* phase.

As a result of the variable initialization, the following rule holds: If $s$, $p$, $o \in \mathbf{V}_{\text{UFGP}}$, and $T_p = UPVT$, and $T_o = UVT \Rightarrow T_p = UPVT$ and $T_o = UVT$. If a triple has subject, predicate and object variables, the predicate variable type is *Unknown Predicate Variable Type* (*UPVT*) and the object variable type is *Unknown Variable Type* (*UVT*), no change is needed since they cannot be specified.

The variable type determination phase, including the variable initialization, the determination rules and the conflict check is also presented as an algorithm in [120].

Below we provide two examples in order to demonstrate the variable type determination phase. The examples use sequences of triple patterns expressed over the `Persons` ontology. The second example (*Example 16*) presents a case of variable type conflict.

### Example 15. Determination of the Variable Types

Consider the following sequence of triple patterns:

> *?e rdfs:subClassOf Person_Type.*
> *?y ?p ?k. ?x rdf:type ?e.*
> *?x Dept__xs_string ?dept.*
> *?p rdfs:domain ?e.*
> *?y ?p "Johnson".*

Since the *Schema Triple patterns* are pruned in the determination of the variable types, the result comprises of the following three triple patterns:

> $t_1 = \langle$ ?y ?p ?k $\rangle$
> $t_2 = \langle$ ?x Dept__xs_string ?dept $\rangle$
> $t_3 = \langle$ ?x ?p "Johnson" $\rangle$.

Initially, it holds that:

> $T_y = UVT$, $T_p = UPVT$, $T_k = UVT$, $T_x = UVT$ and $T_{\text{dept}} = UVT$.

Using the determination rules presented above, the following hold:

> For $t_1 = \langle$ ?y ?p ?k $\rangle$ hold:
> > $T_y = CIVT$ (*Rule 1*), $T_p = UPVT$ (*no change*) and $T_k = UVT$ (*no change*).
> For $t_2 = \langle$ ?x Dept__xs_string ?dept $\rangle$ hold:
> > $T_x = CIVT$ (*Rule 1*) and $T_{\text{dept}} = LVT$ (*Rule 4*).
> For $t_3 = \langle$ ?x ?p "Johnson" $\rangle$ hold:
> > $T_x = CIVT$ (*no change*) and $T_p = DTPVT$ (*Rule 6*).

Finally it holds that:

> $T_y = CIVT$, $T_p = DTPVT$, $T_k = UVT$, $T_x = CIVT$ and $T_{\text{dept}} = LVT$.      ■

### Example 16. Variable Type Usage Conflicts

Assume the following sequence of triple patterns:

> *?n ?p ?k.*
> *?y FirstName__xs_string ?n.*

Initially, it holds that:

> $T_n = UVT$, $T_p = UPVT$, $T_k = UVT$ and $T_y = UVT$.

Using the rules presented above, the following hold:

For $t_1 = \langle$ ?n  ?p  ?k $\rangle$ hold:

$T_n = CIVT$ (*Rule 1*), $T_p = UPVT$ (*no change*) and $T_k = UVT$ (*no change*).

For $t_2 = \langle$ ?y FirstName__xs_string ?n $\rangle$ hold:

$T_y = CIVT$ (*Rule 1*) and $T_n = LVT$ (*Rule 4*).

Finally it holds that:

$T_y = CIVT$, $T_p = UPVT$, $T_k = UVT$ and $T_n = ?$ *Conflict* (from $t_1$: $T_n = CIVT$, from $t_2$: $T_n = LVT$). ∎

### 6.2.2    Variable Result Form

For the formation of the result set we follow the Linked Data principles for publishing data. The resources are identified using *Uniform Resource Identifiers* (*URI*) in order to have a unique and universal name for every resource. The form of the results depends on the variable types. The following result forms are adopted for each variable type: (a) For *CIVT* variables, every result item is a combination of the URI of the XML Document that contains the node assigned to the variable with the XPath of the node itself (including the node context position). In XML, every element and/or attribute can be uniquely identified using XPath expressions and document–specific context positions. For example: *http://www.music.tuc.gr/xmlDoc.xml#/Persons/Student[3]*. (b) For *DTPVT*, *OPVT* and *UPVT* variables, every result item consists of the XPath of the node itself (without the position of the node context). For example: */Persons/Student/FirstName*. (c) For *LVT* variables, every result item is the text representation of the node content. (d) For *UVT* variables, two cases are distinguished: (i) If the assigned node corresponds to a simple element, then the result form is the same with that of the *LVT* variables; and (ii) If the assigned node corresponds to a complex element, the result form is the same with that of the *CIVT* variables.

For the construction of the proper result form, XQuery functions (e.g., func:CIVT( )) formed using standard XQuery expressions, are used in the Return XQuery clauses.

### 6.3    Schema Triple Pattern Processing

In this section we present the *schema triple pattern processing*. This phase is performed in order to support schema-based queries. As schema-based queries are considered queries which contain triple patterns that refer to the ontology structure and/or semantics (i.e., *Schema Triple Patterns, Definition 12*). In the schema triple pattern processing context, the *Schema Triple Patterns* contained in the query are processed against the ontology so that the schema information can be used throughout the translation.

At first, ontology constructs are bound to the variables contained in the Schema Triples. Then, using the predefined mappings, the ontology constructs are replaced with the corresponding XPath Sets. As a result of this processing, XPaths are bound to the variables contained in the Schema Triples. These bindings will be used as initial bindings by the *variable binding* phase (Section 7). Note that as specified in *Definition 12*, triple patterns having a variable on their predicate part are not defined as schema triples, since they can deal either with data or with schema info. Hence, these triples are considered as *non*-schema triple patterns.

The schema triple patterns can be analyzed over the ontology, using a query or an inference engine. It should be noted that, in our approach we do not consider the semantics (e.g., entailment, open/close world assumptions, etc.) adopted in the evaluation of schema triples over the ontology. Since, the schema triple processing uses the results (i.e., ontology constructs) of the schema triple evaluation. Here, we have adopted simple entailment semantics (like the current SPARQL specification [12]). However, inferred results adhering to the RDFS or OWL entailments can be used if the SPARQL engine performs a query expansion step before evaluating the schema triples query, or an RDFS/OWL reasoner has been used. Currently, W3C works on defining the entailment regimes in the forthcoming SPARQL 1.1 [15], which specify exactly what answers we get for several common entailment relations such as *RDFS entailment* or *OWL Direct Semantics entailment*. Finally, note that the SW is based on the *Open World Assumption* (*OWA*), while the XML world is based on the *Closed World Assumption* (*CWA*). This means that in the SW whatever is not explicitly stated is considered to be *unknown*, while in the XML world whatever is not explicitly stated in considered to be *false*.

In order to demonstrate the Schema triple processing phase, we provide an example that outlines the procedure, which uses a sequence of triple patterns expressed over the `Persons` ontology.

**Example 17. Processing Schema Triple Patterns**

Assume the following sequence of triple patterns (the same with that of *Example 15*):

> *?e rdfs:subClassOf Person_Type.*
>
> *?y ?p ?k. ?x rdf:type ?e.*
>
> *?x Dept__xs_string ?dept.*
>
> *?p rdfs:domain ?e. ?y ?p "Johnson".*

From the previous triple pattern sequence, we isolate and process the following *Schema Triples*:

$st_1 = \langle$ ?e rdfs:subClassOf Person_Type $\rangle$, $st_2 = \langle$ ?x rdf:type ?e $\rangle$ and $st_3 = \langle$ ?p rdfs:domain ?e $\rangle$.

In the first step of processing, the variables of the *Schema Triples* are bound to ontology constructs that result in:

?e = $\langle$ Student_Type $\rangle$ (from the processing of *Schema Triple* $st_1$)

?x = $\langle$ Student_Type $\rangle$ (from the processing of *Schema Triples* $st_1$ & $st_2$)

?p = $\langle$ LastName__xs_string, FirstName__xs_string, Age__validAgeType, Email__xs_string, SSN__xs_integer, Nachname__xs_string, Dept__xs_string $\rangle$ (from the processing of *Schema Triple* $st_3$).

Note that the *?p* variable should be bound only to datatype properties, since from the *variable type determination* (*Example 15*) we have $T_p = DTPVT$. Thus, in case that *?p* has been bound to object properties, these will be pruned.

In the next processing step, we determine the XPath Sets of the variables (based on the mappings specified in *Example 13*), that result in:

$\mathbf{X}_e$ = { /Persons/Student }

$\mathbf{X}_x$ = { /Persons/Student }

$\mathbf{X}_p$ = { /Persons/Person/FirstName, /Persons/Student/FirstName, /Persons/Person/LastName, /Persons/Student/Last-Name, /Persons/Person/Age, /Persons/Student/Age, /Persons/Person/Email, /Persons/Student/Email, /Persons/Person/Nachname, /Persons/Student/Nachname, /Persons/Person/@SSN, /Persons/Student/@SSN, /Persons/Student/Dept }.

These bindings are going to be the initial bindings for the *variable binding* phase. ∎

# 7 VARIABLE BINDING

In this section, we describe the *variable binding* phase. In our context the term *"variable bindings"* is used to describe the assignment of XPaths to the variables referenced in a given Basic Graph Pattern (*BGP*, *Definition 11*), thus enabling the translation of the *BGPs* in XQuery expressions.

Intuitively, this phase considers the graph structure(s) constructed by the triples patterns defined in the Basic Graph Pattern, as well as the mappings, in order to determine the appropriate set of bindings. This set of bindings is going to be used in the construction of the XQuery expressions. It should be noted that, due to the form of the mappings (i.e., XPaths Sets) the (hierarchical) structure of XML data is also considered by the variable binding phase.

Additional schema information and/or semantics possibly expressed in the SPARQL query are exploited in the variable binding phase by using the bindings determined in the Schema Triple processing phase (Section 6.3). For this reason, the Schema Triples are omitted (i.e., pruned) from this phase and the determined Schema Triple bindings are used as initial bindings.

The variable binding algorithm is presented in Section 7.1, the variable binding rules are described in Section 7.2 and the XPath Set relations for triple patterns are discussed in Section 7.3.

## 7.1 Variable Binding Algorithm

### 7.1.1 Preliminaries

An RDF triple $\langle s, p, o \rangle$ is a sub graph of the directed RDF graph, where $s, o$ are graph nodes and $p$ is a directed graph edge, directed from $s$ to $o$. As $\mathbf{X}_s$, $\mathbf{X}_p$ and $\mathbf{X}_o$ we denote the XPath Set correspond to subject, predicate and object XPath Sets respectively. Moreover, let $\mathbf{X}_{pD}$ and $\mathbf{X}_{pR}$ be the XPath Sets corresponding, respectively, to the predicate domains and ranges.

Considering the hierarchical structure of XML data and the structure of the directed RDF graph, the following relations must hold for the XPath Sets of the triple pattern parts:

(a) $\exists x_s \in \mathbf{X}_s$ and $\exists x_{pD} \in \mathbf{X}_{pD} : x_s \widetilde{\sqsubset} x_{pD}$. The subject XPath Set ($\mathbf{X}_s$) contains XPaths that prefix the XPaths contained in the predicate domains XPath Set ($\mathbf{X}_{pD}$).

(b) $\exists x_{pD} \in \mathbf{X}_{pD}$ and $\exists x_{pR} \in \mathbf{X}_{pR} : x_{pD} \widetilde{\sqsubset} x_{pR}$. The predicate domains XPath Set ($\mathbf{X}_{pD}$) contains XPaths that prefix the XPaths contained in the predicate ranges XPath Set ($\mathbf{X}_{pR}$).

(c) $\exists x_{pR} \in \mathbf{X}_{pR}$ and $\exists x_o \in \mathbf{X}_o : x_{pR} \widetilde{\sqsubset} x_o$. The predicate ranges XPath Set ($\mathbf{X}_{pR}$) contains XPaths that prefix the XPaths contained in the object XPath Set ($\mathbf{X}_o$).

Thus, from (a), (b) and (c), we conclude to the *Subject–Predicate–Object Relation*, formally defined in (2):

$$\exists x_s \in \mathbf{X}_s, \ \exists x_{pD} \in \mathbf{X}_{pD}, \ \exists x_{pR} \in \mathbf{X}_{pR}, \ \exists x_o \in \mathbf{X}_o : x_s \widetilde{\sqsubset} x_{pD} \widetilde{\sqsubset} x_{pR} \widetilde{\sqsubset} x_o \tag{2}$$

The *Subject–Predicate–Object Relation* must holds for every single triple pattern. Thus, the variable binding algorithm uses this relation in order to determine the appropriate bindings for the entire set of the conjunctive triple patterns (i.e., *BGP*), starting from the bindings of any single triple pattern part (*subject, predicate, object*).

> **Definition 14. (Shared Variable).** A variable contained in a Union–Free Graph Pattern is called a *Shared Variable* when it is referenced in more than one triple patterns of the same Union–Free Graph Pattern regardless of its position in those triple patterns.

In case of shared variables, the algorithm uses the XPath Set Operators (i.e., $\prec$, $\succ$, $\widetilde{\sqcap}$, Section 4.1.2), in order to determine the maximum set of bindings that satisfy the *Subject–Predicate–Object Relation* for the entire set of triple patterns (i.e., the entire *BGP*). As a result, all the XML nodes that satisfy the *BGP* are identified.

The variable binding algorithm does not determine the XPaths for *Literal Variable Type* (*LVT*) shared variables, since the literal equality (e.g., string equality, integer equality, etc.) is independent of the XML structure (i.e., XPath expressions). For example, consider that we want to identify the students with the same First Name and Last Name values. In this case, let the XPaths be */Persons/Student/FirstName* and */Persons/Student/LastName*. Thus, the bindings for variables of *LVT* type cannot be determined at this step. Instead, they will be handled by the *BGP2XQuery* algorithm (Section 8.3), which exploits a combination of the mappings and the determined variable bindings.

For this phase we introduce the "special" XPath set value "$\ominus$". The value "$\ominus$", can be considered as the not initialized value, similar to the *null* value, however, different than the empty set $\varnothing$. Regarding the "$\ominus$" value, (a) the *Intersection* ($\widetilde{\sqcap}$), (b) the *Descendants of Common Ancestors* ($\succ$) and (c) the *Common Ancestors* ($\prec$) operators have the following semantics. Let the XPath Set $\mathbf{A}$, where $\ominus \notin \mathbf{A}$ and $\mathbf{A} \neq \varnothing$ and the XPath Set $\mathbf{e} = \{\ominus\}$. We have: **(a)** *Intersection* ($\widetilde{\sqcap}$): (i) $\mathbf{A}\widetilde{\sqcap}\mathbf{e}=\mathbf{e}\widetilde{\sqcap}=\mathbf{A}$

(ii) $e \cap e=e$; **(b)** *Descendants of Common Ancestors* ($\succ$): (i) $A \succ e = e$ (ii) $e \succ A=A$ (iii) $e \succ e=e$; **(c)** *Common Ancestors* ($\prec$): (i) $A \prec e=A$, (ii) $e \prec A=e$ (iii) $e \prec e=e$.

### 7.1.2   Algorithm Overview

Here we outline the *Variable Binding* algorithm (*Algorithm 1*), which takes as input (a) a Basic Graph Pattern (*BGP*); (b) a set of initial bindings ($\mathbf{X}^{Sch}$); (c) the types of variables that are present in the *BGP* (*varTypes*); and (d) the mappings of the *BGP* ontology constructs (**M**). The variable types are determined by the determining of variable types phase and the initial bindings are those resulting from the Schema Triple processing.

---

**Algorithm 1: Variable Binding Algorithm**

**Input:** Basic Graph Pattern *BGP*, Initial Bindings $\mathbf{X}^{Sch}$,
     Variable Types *varTypes*, Mappings **M**

**Output:** Variable Bindings $\mathbf{X}_v$

1.  **for each** variable $v$ **in** *BGP*   *//initialize the bindings*
2.     **if** $v \in var(schemaTr(BGP))$   *//if the variable v are included at schema triples*
3.         $\mathbf{X}_v^0 = \mathbf{X}_v^{Sch}$
          *//initialize the bindings from the bindings determined the from schema triple processing*
4.     **else**
5.         $\mathbf{X}_v^0 = \{\ominus\}$   *//initialize with the "special" value "$\ominus$"*
6.     **end if**
7.  **end for**
8.  it = 0   *//iteration counter initialization*
9.  **repeat**
11.     **for each** triple *t* **in** *BGP*   *//loop over all the BGP triples*
12.         **if** $s \in \mathbf{V}$   *//if the subject is a variable*
13.            $\mathbf{X}_s^{i+1} = \mathrm{B}_s(\ t, \mathbf{X}_s^i, \mathbf{X}_{pD}^i, \mathbf{X}_o^i, \mathbf{M}\ )$
            *//determine the subject bindings of the current iteration (i.e., t+1)*
14.         **end if**
15.         **if** $p \in \mathbf{V}$   *//if the predicate is a variable*
16.            $\mathbf{X}_p^{i+1} = \mathrm{B}_p(\ t, \mathbf{X}_s^i, \mathbf{X}_p^i, \mathbf{M}, varTypes\ )$
            *//determine the predicate bindings of the current iteration (i.e., i+1)*
17.         **end if**
18.         **if** $o \in \mathbf{V}$   *//if the object is a variable*
19.            $\mathbf{X}_o^{i+1} = \mathrm{B}_o(\ t, \mathbf{X}_s^i, \mathbf{X}_p^i, \mathbf{X}_o^i, \mathbf{M}, varTypes\ )$
            *//determine the object bindings of the current iteration (i.e., i+1)*
20.         **end if**
21.     **end for**
22.     i = i + 1   *//increase the counter*
23.  **until** ($\forall\ v \in var(BGP) \Rightarrow \mathbf{X}_v^i = \mathbf{X}_v^{i-1}$)
     *//loop until the bindings of the previous iteration are equal with the bindings of this iteration*
24.  **return** $\mathbf{X}_v$, $\forall\ v \in var(BGP)$ *//return all the variable bindings for this basic graph pattern*

---

In the beginning of the algorithm (*Lines 1~7*), the variables that are not included in any Schema Triple, thus, no binding has been previously determined (from the Schema Triple Processing phase), are initialized here with the "special" value "$\ominus$" (*Line 5*). The rest of the variables (included in a Schema Triple) are initialized to the initial bindings (*Line 3*). Then, the algorithm performs an iterative process (*Lines 11~21*) where it determines, at each step, the bindings of the entire *BGP* (triple by triple). The determination of the bindings of a single triple is performed using binding rules (*Lines 13, 16 & 19*). Each part of the triple (subject–predicate–object) uses a binding rule (Section 7.2). This iterative process continues until the bindings for all the variables found in the successive iterations are equal (*Line 23*). This implies that no further modifications in the variable bindings are to be made and that the current bindings are the final ones. Thus, the variable binding algorithm ensures that all the variables have been bound to the largest XPath sets with respect to: (a) the structure of the RDF data; (b) the structure of the XML data; and (c) the mappings between them. Note that $\mathbf{X}_w^i$ denotes the determined XPath Set at the $i^{th}$ iteration of the algorithm for the $w$ triple part.

## 7.2   Variable Binding Rules

In this section we present the *Variable Binding Rules* applied by the variable binding algorithm (*Lines 13, 16 & 19*) in order to determine the bindings for all the parts (i.e., subject, predicate and object) of a single triple pattern.

Initially, in order to define the binding rules we distinguish the *Triple Pattern Types*. According to the specified types we have defined the variable binding rules presented above.

Let the sets **V**, **L**, **I**, **B** (as defined in Section 5.1.1). We define four different types of triple patterns: (a) a triple pattern $\langle s, p, o \rangle \in (\mathbf{V} \cup \mathbf{B}) \times \mathbf{I} \times \mathbf{L}$, where the subject part $s$ is a variable or a blank node, the predicate part $p$ is an IRI and the object part $o$ is a literal, is defined to be of Type 1; (b) a triple pattern $\langle s, p, o \rangle \in (\mathbf{V} \cup \mathbf{B}) \times \mathbf{I} \times (\mathbf{V} \cup \mathbf{B})$, where the subject and object parts $s$, $o$ are variables or blank nodes and the predicate part $p$ is an IRI, is defined to be of Type 2; (c) a triple pattern $\langle s, p, o \rangle \in (\mathbf{V} \cup \mathbf{B}) \times (\mathbf{V} \cup \mathbf{B}) \times \mathbf{L}$, where the subject and predicate parts $s$, $p$ are variables or blank nodes and the object part $o$ is a literal, is defined to be of Type 3; (d) a triple pattern $\langle s, p, o \rangle \in (\mathbf{V} \cup \mathbf{B}) \times (\mathbf{V} \cup \mathbf{B}) \times (\mathbf{V} \cup \mathbf{B})$, where all the parts $s$, $p$, $o$ are variables or blank nodes, is defined to be of Type 4.

Given a triple pattern $t$ $\langle s, p, o \rangle$, where $s$ is the subject part, $p$ the predicate part and $o$ the object part, we define the following binding rules depending on the triple pattern types. The following rules are applied to the triple pattern parts that are variables or blank nodes. Firstly, the subject binding rule is applied, then the predicate binding rule, and finally the object binding rule, in order to determine the new XPath Set for every variable or blank node part.

In what follows, $\mathbf{X}_w{}^i$ denotes the determined XPath Set at the $i^{th}$ iteration of the binding algorithm for the $w$ triple part. In particular, $\mathbf{X}_s{}^i$ denotes the XPath Set corresponding to the subject part $s$, $\mathbf{X}_p{}^i$ denotes the XPath Set corresponding to the predicate part $p$, $\mathbf{X}_{pD}{}^i$ denotes the XPath Set corresponding to the domains of the predicate part $p$ and $\mathbf{X}_o{}^i$ denotes the XPath Set corresponding to the object part $o$.

### 7.2.1    Subject Binding Rule

Here we present the binding rule $B_S$ (3), which is applied in order to determine the XPath Set of the subject part ($\mathbf{X}_s{}^{i+1}$). The subject binding rule takes as input (a) the triple for which the determination is performed ($t$); (b) the previously determined bindings for the subject part ($\mathbf{X}_s{}^i$); (c) the previously determined bindings for the domains of the predicate part ($\mathbf{X}_{pD}{}^i$); (d) the previously determined bindings for the object part ($\mathbf{X}_o{}^i$); and (e) the mappings (**M**). Note that with the term *previously determined bindings* we refer to the bindings determined in the previous algorithm iteration.

$$B_S\ (t, \mathbf{X}_s{}^i, \mathbf{X}_{pD}{}^i, \mathbf{X}_o{}^i, \mathbf{M}) =$$

$$= \begin{cases} \mathbf{X}_s{}^i \ \overline{\cap}\ \mathbf{X}_{pD}{}^i & \textbf{if Type 1} \\[2mm] \mathbf{X}_s{}^i \ \overline{\cap}\ \mathbf{X}_{pD}{}^i \lessdot \mathbf{X}_o{}^i & \textbf{if Type 2} \\[2mm] \begin{cases} \mathbf{X}_{S1} & \textbf{if } \mathbf{X}_s{}^i = \ominus \textbf{ and } \mathbf{X}_{pD}{}^i = \ominus \quad \{1\} \\ \mathbf{X}_s{}^i \ \overline{\cap}\ \mathbf{X}_{pD}{}^i & \textbf{else} \end{cases} & \textbf{if Type 3} \\[4mm] \begin{cases} \mathbf{X}_{S1} & \textbf{if } \mathbf{X}_s{}^i = \ominus \textbf{ and } \mathbf{X}_{pD}{}^i = \ominus \textbf{ and } \mathbf{X}_o{}^i = \ominus \quad \{2\} \\ \mathbf{X}_{pD}{}^i \ \overline{\cap}\ \mathbf{X}_s{}^i \lessdot \mathbf{X}_o{}^i & \textbf{else} \end{cases} & \textbf{if Type 4} \end{cases} \qquad \textbf{(3)}$$

$\{1\}$ holds if the type of the triple pattern is Type 3, in case that the subject XPath Set ($\mathbf{X}_s{}^i$) and the predicate XPath Set ($\mathbf{X}_p{}^i$) have not been determined (are equal to $\ominus$). Moreover, $\{2\}$ holds for triple patterns of Type 4, in case that $\mathbf{X}_s{}^i$, $\mathbf{X}_p{}^i$ and the object XPath Set ($\mathbf{X}_o{}^i$) have not been determined.

In the above cases, we assign to the subject XPath Set ($\mathbf{X}_s{}^i$) the XPath Set union of the sets of all the mapped classes (exploiting **M**). Thus, $\mathbf{X}_{S1} = \mathbf{X}_{c_1} \ \overline{\cup}\ \mathbf{X}_{c_2} \ \overline{\cup}\ \cdots \ \overline{\cup}\ \mathbf{X}_{c_n}$, for each mapped class $c_i$, where $\mathbf{X}_{c_i}$ is the XPath Set corresponding to $c_i$, $\forall i \in \{1, ..., n\}$.

Similarly are defined the predicate ($B_P$) and the object ($B_O$) binding rules, which are available in [120].

Here, we provide an example that illustrates the variable binding phase.

**Example 18. Variable Binding**

Assume the following sequence of triple patterns (the same with that of *Example 15*):

> *?e rdfs:subClassOf Person_Type.*
>
> *?y ?p ?k. ?x rdf:type ?e.*
>
> *?x Dept__xs_string ?dept.*
>
> *?p rdfs:domain ?e. ?y ?p "Johnson".*

The Schema Triple patterns are not taken into account, resulting in the following triple patterns:

$t_1 = \langle$ ?y ?p ?k $\rangle$

$t_2 = \langle$ ?x Dept__xs_string ?dept $\rangle$

$t_3 = \langle$ ?x ?p "Johnson" $\rangle$.

From the Schema Triple processing (*Example 17*), the following initial bindings have been determined (the operands of the bindings rules are omitted for the sake of simplicity):

$\mathbf{X}_x^{\text{Sch}} = \mathbf{X}_x^0 = \{$ /Persons/Student $\}$

$\mathbf{X}_p^{\text{Sch}} = \mathbf{X}_p^0 = \{$ /Persons/Person/FirstName, /Persons/Student/FirstName, /Persons/Person/LastName, /Persons/Student/Last-Name, /Persons/Person/Age, /Persons/Student/Age, /Persons/Person/Email, /Persons/Student/Email, /Persons/Person/Nachname, /Persons/Student/Nachname, /Persons/Person/@SSN, /Persons/Student/@SSN, /Persons/Student/Dept $\}$

$\mathbf{X}_y^0 = \ominus$

$\mathbf{X}_k^0 = \ominus$

$\mathbf{X}_{\text{dept}}^0 = \ominus$

From the Determination of the Variable Types the following variable types (*Example 15*) have determined:

$T_y = CIVT$, $T_p = DTPVT$, $T_k = UVT$, $T_x = CIVT$ and $T_{\text{dept}} = LVT$.

From the Variable Binding Algorithm (Section 7.1) the following hold:

***1st Iteration***

For $t_1 = \langle$ ?y ?p ?k $\rangle$ (Type 4) hold:

$\mathbf{X}_y^1 = \mathbf{X}_y^0 \; \overline{\sqcap} \; \mathbf{X}_{pD}^0 \lessdot \mathbf{X}_k^0 = \ominus \; \overline{\sqcap} \; \{$ /Persons/Person, /Persons/Student $\} \lessdot \ominus = \{$ /Persons/Person, /Persons/Student$\}$

$\mathbf{X}_p^1 = \mathbf{X}_y^0 > \mathbf{X}_p^0 = \{$ /Persons/Person, /Persons/Student$\} > \{$ /Persons/Person/FirstName, /Persons/Student/FirstName, /Persons/Person/LastName, /Persons/Student/LastName, /Persons/Person/Age, /Persons/Student/Age, /Persons/Person/Email, /Persons/Student/Email, /Persons/Person/Nachname, /Persons/Student/Nachname, /Persons/Person/@SSN, /Persons/Student/@SSN, /Persons/Student/Dept$\} = \{$ /Persons/Person/FirstName, /Persons/Student/FirstName, /Persons/Person/LastName, /Persons/Student/LastName, /Persons/Person/Age, /Persons/Student/Age, /Persons/Person/Email, /Persons/Student/Email, /Persons/Student/Nachname, /Persons/Person/@SSN, /Persons/Student/@SSN, /Persons/Student/Dept$\}$

$\mathbf{X}_k^1 = \text{\textit{Non Determinable}}$

For $t_2 = \langle$ ?x Dept__xs_string ?dept $\rangle$ (Type 2) hold:

$\mathbf{X}_x^1 = \mathbf{X}_x^0 \; \overline{\sqcap} \; \mathbf{X}_{\text{Dept\_\_xs\_stringD}} \lessdot \mathbf{X}_{\text{dept}}^0 = \{$ /Persons/Student$\} \; \overline{\sqcap} \; \{$/Persons/Student$\} \lessdot \ominus = \{$ /Persons/Student $\}$

$\mathbf{X}_{\text{dept}}^1 = \text{\textit{Non Determinable}}$

For $t_3 = \langle$ ?x  ?p "Johnson" $\rangle$ (Type 3) hold:

$\mathbf{X}_x{}^1 = \mathbf{X}_x{}^0 \cap \mathbf{X}_{pD}{}^0 = \{$ /Persons/Student$\} \cap \{$ /Persons/Person, /Persons/Student $\} = \{$ /Persons/Student$\}$

$\mathbf{X}_p{}^1 = \mathbf{X}_x{}^0 > \mathbf{X}_p{}^0 = \{$ /Persons/Student$\} > \{$ /Persons/Person/FirstName, /Persons/Student/FirstName, /Persons/Person/Last-Name, /Persons/Student/LastName, /Persons/Person/Age, /Persons/Student/Age, /Persons/Person/Email, /Persons/Student/Email, /Persons/Student/Nachname, /Persons/Person/@SSN, /Persons/Student/@SSN, /Persons/Person/Dept$\} = \{$ /Persons/Student/FirstName, /Persons/Student/LastName, /Persons/Student/Age, /Persons/Student/Email, /Persons/Student/Nachname, /Persons/Student/@SSN, /Persons/Student/Dept$\}$

## 2nd Iteration

For $t_1 = \langle$ ?y  ?p  ?k $\rangle$ (Type 4) hold:

$\mathbf{X}_y{}^2 = \mathbf{X}_y{}^1 \cap \mathbf{X}_{pD}{}^1 < \mathbf{X}_k{}^1 = \{$ /Persons/Person, /Persons/Student$\} \cap \{$ /Persons/Student$\} < \ominus = \{$ /Persons/Student$\}$

$\mathbf{X}_p{}^2 = \mathbf{X}_y{}^1 > \mathbf{X}_p{}^1 = \{$ /Persons/Student $\} > \{$ /Persons/Student/FirstName, /Persons/Student/LastName, /Persons/Student/Age, /Persons/Student/Email, /Persons/Student/Nachname, /Persons/Student/@SSN, /Persons/Student/Dept$\} = \{$ /Persons/Student/FirstName, /Persons/Student/LastName, /Persons/Student/Age, /Persons/Student/Email, /Persons/Student/Nachname, /Persons/Student/@SSN, /Persons/Student/Dept$\}$

$\mathbf{X}_k{}^2 = $ *Non Determinable*

For $t_2 = \langle$ ?x  Dept__xs_string  ?dept $\rangle$ (Type 2) hold:

$\mathbf{X}_x{}^2 = \mathbf{X}_x{}^1 \cap \mathbf{X}_{Dept\_\_xs\_stringD} < \mathbf{X}_{dept}{}^1 = \{$ /Persons/Student$\} \cap \{$/Persons/Student$\} < \ominus = \{$ /Persons/Student$\}$

$\mathbf{X}_{dept}{}^2 = $ *Non Determinable*

For $t_3 = \langle$ ?x  ?p "Johnson" $\rangle$ (Type 3) hold:

$\mathbf{X}_x{}^2 = \mathbf{X}_x{}^1 \cap \mathbf{X}_{pD}{}^1 = \{$ /Persons/Student$\} \cap \{$ /Persons/Student$\} = \{$ /Persons/Student$\}$

$\mathbf{X}_p{}^2 = \mathbf{X}_x{}^1 > \mathbf{X}_p{}^1 = \{$ /Persons/Student$\} > \{$ /Persons/Student/FirstName, /Persons/Student/LastName, /Persons/Student/Age, /Persons/Student/Email, /Persons/Student/Nachname, /Persons/Student/@SSN, /Persons/Student/Dept$\} = \{$/Persons/Student/FirstName, /Persons/Student/LastName, /Persons/Student/Age, /Persons/Student/Email, /Persons/Student/Nachname, /Persons/Student/@SSN, /Persons/Student/Dept$\}$

## 3rd Iteration

Nothing changes from the second iteration, so the Variable Binding Algorithm terminates.

## Finally

The following final bindings have been determined:

$\mathbf{X}_x = \{$ /Persons/Student$\}$

$\mathbf{X}_p = \{$ /Persons/Student/FirstName, /Persons/Student/LastName, /Persons/Student/Age, /Persons/Student/Email, /Persons/Student/Nachname, /Persons/Student/@SSN, /Persons/Student/Dept$\}$

$\mathbf{X}_y = \{$ /Persons/Student $\}$

$\mathbf{X}_k = $ *Non Determinable*

$\mathbf{X}_{dept} = $ *Non Determinable*. ∎

### 7.3    XPath Set Relations for Triple Patterns

In several cases, XPath Sets that correspond to different SPARQL variables must be associated. For example, let the triple pattern *?x FirstName__xs_string ?y*; the variable *x* corresponds to Persons and Students and the variable *y* to their first name(s). The variable binding phase will result in two XPath Sets: one for all the Persons and Students corresponding to variable *x* (i.e., $\mathbf{X}_x$ = {/Persons/Person, /Persons/Student}) and one for all the First Names corresponding to variable *y* (i.e., $\mathbf{X}_y$ = {/Persons/Person/FirstName, /Persons/Student/FirstName}). However, the association of persons and their names still has to be done. We introduce the *extension relation* which can hold among different XPath Sets and can be exploited in order to associate them.

> **Definition 15.   (Extension Relation).** An XPath Set **D** is said to be an *extension* of an XPath Set **A** if all the XPaths in **D** are descendants of the XPaths of **A**. This relation can be achieved if the *XPath Set Concatenation* ( ⊕ ) operator (Section 4.1.2) is applied to the XPath Set **A** having as right operand an XPath Set **C**, and as result the XPath Set **D**, which will be an *extension* of **A** (i.e., **A**⊕**C**=**D**, **D** is an *extension* of **A**).

In the above example, the XPath Set $\mathbf{X}_y$ is an *extension of* the XPath Set $\mathbf{X}_y$. The XPath Set $\mathbf{X}_y$ may result from the XPath Set $\mathbf{X}_x$ since: $\mathbf{X}_x$ ⊕ {/FirstName} = {/Persons/Person, /Persons/Student} ⊕ {/FirstName} = {/Persons/Person/FirstName, /Persons/Student/FirstName} which is equal to the XPath Set $\mathbf{X}_y$.

Based on the *Subject–Predicate–Object Relation* defined in (2), the *extension relation* holds for the XPath Sets and results from the *Variable Binding Algorithm*. It implies that the XPath Set bound to the object part corresponds to an *extension* of the XPath Set bound to one of the predicate and subject parts. Moreover, the XPaths bound to the predicate part correspond to an *extension* of the XPath Sets bound to the subject part. Thus, the *extension relation* is exploited by the translation process, using also the For and Let XQuery clauses, in order to associate different XQuery variables.

Note that the notion of *extension* is also used to describe relations between XQuery variables. If the *extension* relation holds for the XPaths used in the For/Let clauses that assign values to the variables, then the *extension* relation also holds between these variables. In particular, consider the following For or Let (For/Let) XQuery clauses: For/Let *$v₁* in/:= *expr₁* and For/Let *$v₂* in/:= *expr₂*; if the XPath expressions occurring in *expr₂* are *extensions* of the XPath expressions occurring in *expr₁*, then the XQuery variable *$v₂* is also said to be an *extension* of the *$v₁* XQuery variable. For example, consider the XQuery expressions: Let *$x:=/Persons/Person* union */Persons/Student* and For *$y in $x/FirstName*. In these XQuery expressions, the XQuery variable $y is said to be an *extension* of the XQuery variable $x.

## 8    GRAPH PATTERN TRANSLATION

In this section, we describe the *Graph Pattern translation* phase, which translates a Graph Pattern (*GP*, *Definition 9*) into semantically corresponding (Section 5.2) XQuery expressions.

The XQuery and SPARQL basic notions are introduced in Section 8.1, an overview of the graph pattern translation is presented in Section 8.2, the basic graph pattern translation is described in Section 8.3 and we close with a discussion on the major challenges that we faced during the graph pattern translation in Section 8.4.

### 8.1    Preliminaries

In this section we provide an overview of the semantics of the SPARQL graph patterns (most of them defined in [33]), as well as some preliminary notions regarding the XQuery syntax representation.

**Definition 16.** **(SPARQL Graph Pattern Solution).** A *SPARQL Graph Pattern solution* $\omega$: $\mathbf{V} \rightarrow (\mathbf{I} \cup \mathbf{B} \cup \mathbf{L})$ is a partial function that assigns RDF terms of an RDF dataset to variables of a SPARQL graph pattern. The domain of $\omega$, $dom(\omega)$, is the subset of $\mathbf{V}$ where $\omega$ is defined. The *empty graph pattern solution* $\omega_{\varnothing}$ is a graph pattern solution with an empty domain. The SPARQL graph pattern evaluation result is a set $\mathbf{\Omega}$ of graph pattern solutions.

Two Graph Pattern solutions $\omega_1$ and $\omega_2$ are *compatible* when for all $x \in dom(\omega_1) \cap dom(\omega_2)$ it holds that $\omega_1(x) = \omega_2(x)$. Furthermore, two graph pattern solutions with disjoint domains are always *compatible*, and the empty graph pattern solution $\omega_{\varnothing}$ is *compatible* with any other graph pattern solution.

Let $\mathbf{\Omega}_1$ and $\mathbf{\Omega}_2$ be sets of Graph Pattern solutions. The *Join*, *Union*, *Difference* and *Left Outer Join* operations between $\mathbf{\Omega}_1$ and $\mathbf{\Omega}_2$ are defined as follows: (a) $\mathbf{\Omega}_1 \bowtie \mathbf{\Omega}_2 = \{\omega_1 \cup \omega_2 \mid \omega_1 \in \mathbf{\Omega}_1, \omega_2 \in \mathbf{\Omega}_2$ are *compatible graph pattern solutions*$\}$, (b) $\mathbf{\Omega}_1 \cup \mathbf{\Omega}_2 = \{\omega \mid \omega_1 \in \mathbf{\Omega}_1$ or $\omega_2 \in \mathbf{\Omega}_2\}$, (c) $\mathbf{\Omega}_1 \setminus \mathbf{\Omega}_2 = \{\omega \in \mathbf{\Omega}_1 \mid$ for all $\omega' \in \mathbf{\Omega}_2, \omega$ and $\omega'$ are not *compatible*$\}$, (d) $\mathbf{\Omega}_1 \bowtie \mathbf{\Omega}_2 = (\mathbf{\Omega}_1 \bowtie \mathbf{\Omega}_2) \cup (\mathbf{\Omega}_1 \setminus \mathbf{\Omega}_2)$.

The semantics of the SPARQL graph pattern expressions are defined as a function $[[.]]_D$, which takes a graph pattern expression and an RDF dataset $D$ and returns a set of graph pattern solutions.

**Definition 17.** **(SPARQL Graph Pattern Evaluation).** Let $D$ be an RDF dataset over $(\mathbf{I} \cup \mathbf{B} \cup \mathbf{L})$, $t$ a triple pattern, $P$, $P_1$, $P_2$ graph patterns and $R$ a built-in condition. Given a graph pattern solution $\omega$, we denote as $\omega \vDash R$ that $\omega$ satisfies $R$ (the Filter operator semantics are described in detail in [116]). The evaluation of a graph pattern over $D$, denoted by $[[.]]_D$, is defined recursively as follows: (a) $[[t]]_D = \{\omega \mid dom(\omega) = var(t)$ and $\omega(t) \in D\}$, (b) $[[(P_1 \text{ AND } P_2)]]_D = [[P_1]]_D \bowtie [[P_2]]_D$, (c) $[[(P_1 \text{ OPT } P_2)]]_D = [[P_1]]_D \bowtie [[P_2]]_D$, (d) $[[(P_1 \text{ UNION } P_2)]]_D = [[P_1]]_D \cup [[P_2]]_D$ and (e) $[[(P \text{ FILTER } R)]]_D = \{\omega \in [[P]]_D \mid \omega \vDash R\}$

Finally, we introduce the SPARQL *Return Variable* notion, which is exploited throughout the SPARQL to XQuery translation, as well as some basic notions regarding the XQuery syntax.

**Definition 18.** **(SPARQL Return Variable).** A *SPARQL return variable* is a variable for which the SPARQL query would return some information. The *Return Variables* ($RV$) of a SPARQL query constitute the *Return Variables* set $\mathbf{RV} \subseteq \mathbf{V}$. In particular: (a) Select and Describe SPARQL queries, the $RV$ consists of the variables referred after the query form clause; in case of wildcard (*) use, $\mathbf{RV} = \mathbf{V}$; (b) for Ask SPARQL queries, $\mathbf{RV} = \varnothing$; (c) for Construct SPARQL queries, $\mathbf{RV}$ consists of the variables referred in the query graph template (i.e., the variables that belong to the graph template variable set $\mathbf{GTV}$), thus, $\mathbf{RV} = \mathbf{GTV}$.

Due to the fact that the term "*predicate*" is used in the SPARQL and XPath languages, in the rest of this paper we will refer to the XPath predicate as *XPredicate*. Moreover, the XQuery variable $\$doc$ is defined to be initialized by the clauses: let $\$doc := \text{fn:doc ("URI")}$, or let $\$doc := \text{fn:collection ("URI")}$; where URI is the address of the XML document or document collection that contains the XML data over which the produced XQuery will be evaluated.

Finally, we define the abstract syntax representation of the XQuery For and Let clauses $xC$ as follows: (a) *for $\$var$ in expr*; and (b) *let $\$var := expr$*, where $\$var$ is an XQuery variable named *var* and *expr* is a sequence of XPath expressions. As *xC.var* we denote the name of the XQuery variable defined in $xC$, as *xC.expr* we denote the XPath expressions of $xC$ and as *xC.type* we denote the type (For or Let) of the XQuery clause $xC$. Finally, as *xE* we denote a sequence of XQuery expressions.

## 8.2 Graph Pattern Translation Overview

The graph pattern ($GP$) concept is defined recursively. The *Basic Graph Pattern translation* phase (Section 8.3) translates the basic components of a $GP$ (i.e., *BGPs*) into XQuery expressions, which in several cases have to be associated in the context of a $GP$. That is, to apply the SPARQL operators (i.e., UNION, AND, OPT and FILTER) that may occur outside the *BGPs*. The *GP2XQuery* algorithm traverses the SPARQL evaluation tree resulting from the $GP$, so as to identify and handle the SPARQL operators.

Particularly, the SPARQL `UNION` operator corresponds to the union operation applied to the graph pattern solutions of its operand graphs (*Definition 17*). The implementation of the `UNION` operator is straightforward in XQuery. The `FILTER` operator restricts the query solutions to the ones for which the filter expression is true. The translation of the `FILTER` operator in the context of *BGPs* is presented in Section 0. The same holds for the translation of the filters occurring outside the *BGPs*. The SPARQL `AND` and `OPT` operators correspond to the *Join* and *Left Outer Join* operators respectively, applied to the graph pattern solutions of their operand graphs (*Definition 17*). The semantics of the *Join* and *Left Outer Join* operators in SPARQL differ slightly from the relational algebra join semantics, in the case of *unbound*[19] variables[20]. In particular, the existence of an *unbound* variable in a SPARQL join operation does not produce an *unbound* result. In other words, the join in the SPARQL semantics, is defined as a non null-rejecting join. The semantics of the *compatible mappings* in the case of un-bound variables have been discussed in [115][116][33].

Note however that SPARQL does not provide the *minus* operator at syntax level. The minus operator can be expressed as a combination of optional patterns and filter conditions which include the *bound* operator (like the *Negation as Failure* (*NAS*) in logic programming[21]). The semantics of the SPARQL minus operator have been extensively studied in [35].

The unbound variable semantics in conjunction with the `OPT` operator result in a "special" type of *GPs*. This type is well known as *non-well designed GPs* (*Definition 13*) with some of its properties being different from the rest of the *GPs* (i.e., the *well designed* ones)[22]. In particular, in the context of translating the `AND` and `OPT` operators, the possible evaluation strategies differ for the well designed and the non-well designed graph patterns (for more details see [33]). As a result, in order to provide an efficient translation for the `AND` and `OPT` operators, we must not handle all graph patterns in a uniform way. Below we outline the translation for both well-designed and non-well designed graph patterns in XQuery expressions.

### 8.2.1    Well Designed Graph Patterns

Every well-designed Union–Free Graph Pattern $P_i$ contained in the normal form (1) is transformed in the form of (4) after the graph pattern normalization phase (Section 6.1):

$$( \cdots (t_1 \text{ AND } \cdots \text{ AND } t_k) \text{ OPT } O_1) \text{ OPT } O_2) \cdots ) \text{ OPT } O_n), \tag{4}$$

where each $t_i$ is a triple pattern, $n{\geq}0$ and each $O_j$ has the same form (4) [33].

We can observe from (4) that the `AND` operators are occurring only between triple patterns (expressed with "." in the SPARQL syntax) in the context of Basic Graph Patterns (*BGPs*). As a consequence, in the case of well designed *GPs*, the `AND` operators are exclusively handled by the *BGP2XQuery* algorithm, as described in Section 8.3. In particular, the *BGP2XQuery* algorithm uses associated For/Let XQuery clauses to resemble nested loop joins. In addition, throughout the For/Let XQuery clauses creation, the *BGP2XQuery* algorithm exploits the *extension relation* (*Definition 15*) in order to use the already evaluated XQuery values, providing a more efficient join implementation.

Considering the well designed *GP* definition (*Definition 13*), as well as the form (4), we conclude that the following holds for the operands of an `OPT` operator: For the expressions of the form $P = (P_1 \text{ OPT } P_2)$ occurring in (4), every variable occurring both inside $P$ and outside $P$, it occurs for sure in $P_1$. As a result, the variables occurring outside $P$ have always bound values, imposed from the $P_1$ evaluation. Note that the above property holds only for well designed *GPs* and not for the non-well designed ones. Exploiting this property, we can provide an efficient implementation of the `OPT` operators, which are going to use the already evaluated results (produced from the left operand evaluation) in the evaluation of the

---

right operand. Consider for example the well designed graph pattern $P = (t_1$ OPT $(t_2$ OPT $t_3)$, where $t_1$, $t_2$ and $t_3$ triple patterns. The evaluation of $P$ over a dataset $D$ will be $[[t_1]]_D \bowtie (([[t_1]]_D \bowtie [[t_2]]_D)) \bowtie (([[t_1]]_D \bowtie [[t_2]]_D \bowtie [[t_3]]_D))$.

The *GP2XQuery* algorithm traverses the SPARQL execution tree in a depth-first manner, the *BGP2XQuery* algorithm translates the *BGPs* occurring in *GP*. In case of OPT operators, the XQuery expressions resulting from the translation of the right operands use the XQuery values already evaluated from the translation of the left operand, reducing the required computations.

### 8.2.2    Non-Well Designed Graph Patterns

The evaluation strategy outlined above can not be applied in the case of the *non-well designed GPs*. The unbound variables semantics and the "confused" use of variables in the OPT operators of the non-well designed *GPs* do not allow the use of the intermediate results during the graph pattern evaluation.

For example, consider the following non-well designed graph pattern $P = ((?x \; p_1 \; ?y)$ OPT $(?x \; p_2 \; ?z))$ OPT $(?w \; p_3 \; ?z)$. The evaluation of the expression $((?x \; p_1 \; ?y)$ OPT $(?x \; p_2 \; ?z))$ will possibly return results with unbound values for the variable $?z$. In the evaluation strategy adopted for the well designed *GPs*, the results from the evaluation of $((?x \; p_1 \; ?y)$ OPT $(?x \; p_2 \; ?z))$ expression (intermediate results) and in particular the results from the variable $?z$, will be used to evaluate the OPT $(?w \; p_3 \; ?z)$ expression. The unbound values that possibly occur for variable $?z$, will reject the evaluation of the OPT$(?w \; p_3 \; ?z)$ expression. However, this rejection is not consistent with the unbound variable semantics. Due to that, an unbound $?z$ value resulting from the evaluation of expression $((?x \; p_1 \; ?y)$ OPT $(?x \; p_2 \; ?z))$, will not reject a bound value $?z$ resulting from the evaluation of expression OPT$(?w \; p_3 \; ?z)$.

As a result, for the non-well designed *GPs*, we are forced to independently evaluate the *BGPs*, so that the AND and OPT operators will be applied over the results produced from the *BGP* evaluation. In the context of SPARQL to XQuery translation, the *GP2XQuery* algorithm traverses the SPARQL execution tree in a button-up fashion and the *BGP* are independently translated by the *BGP2XQuery* algorithm. Finally, the AND and OPT operators are applied using XQuery clauses among the XQuery expressions resulting from the *BGP2XQuery* translation, taking also into consideration the semantics of the *compatible mappings* for unbound variables.

## 8.3    Basic Graph Pattern Translation

This section describes the translation of Basic Graph Pattern (*BGP*, *Definition 11*) into XQuery expressions.

### 8.3.1    BGP2XQuery Algorithm Overview

We outline here the *BGP2XQuery* algorithm, which translates *BGPs* into XQuery expressions. The algorithm is not executed triple by triple. Instead, it processes the subjects, predicates, and objects of all the triples separately. For each SPARQL variable included in the *BGP*, the algorithm creates For or Let XQuery clauses, using the variable bindings, the input mappings, and the *extension relation* (*Definition 15*). In every case, the name of an XQuery variable is the same with that of the corresponding SPARQL variable, so the correspondences between the SPARQL and XQuery queries can be easily captured. Regarding the literals included in the *BGP*, the algorithm translates them as XPath conditions using *XPredicates*. The translation of SPARQL Filters depends on the Filter expression. Most of the Filters are translated as XPath conditions expressed using *XPredicates*, however some "special" Filter expressions are translated as conditions expressed in XQuery Where clauses. Finally, the algorithm creates an XQuery Return clause that includes the *Return Variables* that were defined in the *BGP*. The translation of *BGPs* is described in detail in the following sections.

### 8.3.2 For or Let Clause?

A crucial issue in the XQuery expression construction is the enforcement of the appropriate solution sequence based on the SPARQL semantics. To achieve this, for a SPARQL variable *v*, we create a For or a Let clause according to the algorithm presented below (*Algorithm 2*). Intuitively, the algorithm chooses between the construction of For and Let clauses in order to produce the desired solution sequence. For example, consider a SPARQL query which returns persons and their first names. For a person *A*, that has two first name *n₁* and *n₂*, the returned solution sequence will consist of two results *A n₁* and *A n₂*.

---

**Algorithm 2: For or Let XQuery Clause Selection** (*QF*, **RV**, *v* )

**Input:** SPARQL query form *QF*, *Return Variables* **RV**, SPARQL variable *v*

**Output:** XQuery Clause Type

1.    **if** *QF ≠ Ask*
2.      **if** (*v ∈* **RV**) **or** ( ∃*K ∈* **RV** | *K is extension of v* )
3.        **return** Create a **For** XQuery Clause
4.      **else**
5.        **return** Create a **Let** XQuery Clause
6.      **end if**
7.    **else**
8.      **return** Create a **Let** XQuery Clause
9.    **end if**

---

For the Select, Construct and Describe query forms (*Lines 1∼6*) the algorithm will create for the variable *v* a For XQuery clause if *v* is included in the **RV** or if any return variable is an *extension* (*Definition 15*) of the variable *v* (*Line 3*), otherwise it will create a Let XQuery clause (*Line 5*).

For Ask queries (*Lines 7∼9*) that do not return a solution sequence, and in order to make the generated XQueries more efficient, the algorithm will create only Let XQuery clauses (*Line 8*), in order to check if a *BGP* can be matched over XML data.

### 8.3.3 Subject Translation

The *Subject Translation* algorithm (*Algorithm 3*) translates the subject part of all the triple patterns of a given *BGP* to XQuery expressions. It should be noted that, for the rest of the paper the symbol *Nx* denotes the name of SPARQL variable *X* and the triple patterns are represented as *s p o*, where *s* is the subject, *p* the predicate and *o* the object part of the triple pattern.

---

**Algorithm 3: Subject Translation** ( *BGP*, *QF*, **RV**, *bindings* )

**Input:** Basic Graph Pattern *BGP*, SPARQL query form *QF*,
        SPARQL Return Variables **RV**, Variable Bindings *bindings*

**Output:** For or Let XQuery Clause *xC*

1. **for each** *triple* in *BGP*
2.    **if** *s ∈* **V**    *// If subject is a variable*
3.      *xC.type ←* **For or Let XQuery Clause Selection** ( *QF*, **RV**, *s* )
        *//Create a For or Let XQuery Clause*
4.      *xC.var ← Nₛ*    *// Define an XQuery Variable with the name of SPARQL Variable s*
5.      *xC.expr ←* $doc/x₁ union $doc/x₂ union … union $doc/xₙ , ∀ xᵢ ∈ **Xₛ**
        *// Set expr equal to the XPath Set of the Subject prefixed with the $doc variable*
        *//**Xₛ** is the binding XPath Set for the variable s*
6.    **end if**
7. **end for**
8. **return** *xC*

---

For each subject *s* that is a variable (*Line 2*), the algorithm creates a For or Let XQuery clause *xC*, using the *For or Let XQuery Clause Selection Algorithm* (*Line 3*) to determine the type (i.e., For or Let) of the clause. The XQuery variable *xC.var* defined in the XQuery clause being created has the same value with the name of the subject $N_s$ (i.e., the SPARQL and the XQuery variables have the same name) (*Line 4*). The XQuery expression *xC.expr* is defined using the variable bindings of the subject $X_s$ and the *$doc* variable (*Line 5*). Finally, the algorithm returns the generated For or Let XQuery clause (*Line 8*).

### 8.3.4    Predicate Translation

---

**Algorithm 4: Predicate Translation ( *BGP*, *QF*, **RV**, *bindings* )**

---

**Input:** Basic Graph Pattern *BGP*, SPARQL query form *QF*,
      SPARQL Return Variables **RV**, Variable Bindings *bindings*

**Output:** For or Let XQuery Clause *xC*

1.  **for each** *triple* **in** *BGP*
2.     **if** $p \in$ **V**    *// If predicate is a variable*
3.         *xC.type* ← **For or Let XQuery Clause Selection** ( *QF*, *RV*, *p* )
           *//Create a For or Let XQuery Clause*
4.         *xC.var* ← $N_p$    *// Define an XQuery Variable with the same name with the SPARQL Variable p*
5.         *xC.expr* ← \$ N$_s$ /x$_1$ union \$ N$_s$ /x$_2$ union … union \$ N$_s$ /x$_n$ , ∀ x$_i$ ∈ **X**$_s$⋙**X**$_p$
           *// Set expr equal to the variable corresponding to the triple subject variable suffixed with XPaths that have*
           *resulted from the **X**$_s$⋙**X**$_p$ operation. The XPath Set. **X**$_p$ is the binding XPath Set for the variable p and **X**$_s$ is*
           *the binding XPath Set for the subject s*
6.     **end if**
7.  **end for**
8.  **return** *xC*

---

The *Predicate Translation* algorithm (*Algorithm 4*) translates the predicate part of all the triple patterns of a given *BGP* to XQuery expressions. For each predicate *p* that is a variable (*Line 2*), the algorithm creates a For or Let XQuery clause *xC*, using the *For or Let XQuery Clause Selection Algorithm* (*Line 3*) to determine the type (i.e., For or Let) of the clause. The XQuery variable *xC.var* defined in the XQuery clause being created has the same value with the name of the predicate $N_p$ (*Line 4*). The XQuery expression *xC.expr* is defined using: (a) the variable bindings of the predicate **X**$_p$; (b) the variable bindings of the subject **X**$_s$; (c) the XQuery variable $\$N_s$ that represents the subject of the triple; and (d) the *extension relation* (*Definition 15*). *xC.expr* associates the subject and predicate bindings (*Line 5*). Finally, the algorithm returns the generated For or Let XQuery clause (*Line 8*).

### 8.3.5    Object Translation

The *Object Translation* algorithm (*Algorithm 5*) translates the object part of all the triple patterns of a given *BGP* to XQuery expressions. For the objects *o* that are literals (*Lines 2~12*), the algorithm creates *XPredicates* in order to translate them. If the predicate *p* of the triple is a variable, the *XPredicate* restriction is applied to the (For or Let) XQuery clause created during the translation of the predicate (*Lines 3~5*). If the predicate is not a variable, the appropriate restrictions using *XPredicates* are applied to the (For or Let) XQuery clause, created during the translation of the subject *s* of the triple (*Lines 9~12*).

For the objects *o* that are variables (*Lines 13~29*), if the predicate *p* is also a variable (*Lines 14~21*) the algorithm creates a Let XQuery clause (*Lines 15~17*), in order to assign the predicate XQuery variable to the XQuery variable of the object. If the predicate is an *IRI* (*Lines 21~28*), the algorithm creates a For or Let XQuery clause *xC* using the For or Let XQuery Clause Selection Algorithm (*Line 22*) to determine the type (i.e., For or Let) of the clause. In this case**,** the algorithm uses the variable bindings of the subject **X**$_s$, the mapping $\mu_p$ of the property defined in the predicate part and the *extension relation* (*Definition 15*) for triple patterns, in order to associate the subject, the predicate and the object bindings (*Line 24*).

**Binding Assurance Condition.** According to the SPARQL semantics, all the variables used in a *BGP* must be bound for all the solutions in the solution sequence. That is, RDF terms must be assigned to all the variables in all the solutions. In the *BGP2XQuery* translation, this is not always guaranteed when Let XQuery clauses are used to translate SPARQL variables. In these cases, we must check if a value has been bound to each variable (*Lines 7, 19, 26*). In order to perform this check, we exploit the XQuery function exists( ), which allows checking the assignment of some value to a variable. A *Binding Assurance Condition* for a variable *w* corresponds to a definition of the form "exists($w) = true" in the XQuery Where clause.

**Algorithm 5: Object Translation ( *BGP*, *QF*, **RV**, *bindings*, *mappings* )**

**Input:** Basic Graph Pattern *BGP*, SPARQL query form *QF*, SPARQL Return Variables **RV**, Variable Bindings *bindings*,
mappings between the ontology and the XML schema *mappings*

**Output:** For or Let XQuery Clause *xC*

1.  **for each** *triple* **in** *BGP*
2.     **if** $o \in \mathbf{I}$     *// If the object is a literal*
3.        **if** $p \in \mathbf{V}$     *// If the predicate is a variable*
4.           Create *XPredicate* over the *xC.expr* where *xC* is the For/Let clause created for the predicate *p*
5.           *XPredicate* ← [.= "*o*"]
6.           **if** Let XQuery Clause created for *p*
7.              Create *"Bindings Assurance Condition"* for *p*     *//see "Biding Assurance Condition" Section*
8.           **end if**
9.        **else**     *// The predicate is not a variable – it is an IRI*
10.          Create *XPredicate* $\forall x_i \in \mathbf{X}_s$ in *xC.expr*, where *xC* is the For/Let clause created for the subject *s*
11.          *XPredicate* ← [./$y_1$ = "*o*" or ./$y_2$ = "*o*" or … or ./$y_n$ = "*o*"] $\forall y_i \in \{x_i\} \gg \mu_p$
            *// $\mathbf{X}_s$ is the bindings XPath Set for the subject $S$ and $\mu_p$ is the mappings XPath Set for the property p*
12.       **end if**
13.    **else if** $o \in \mathbf{V}$   *// If the object is a variable*
14.       **if** $p \in \mathbf{V}$     *// If the predicate is a variable*
15.          *xC.type* ← Create a Let XQuery Clause
16.          *xC.var* ← $N_o$     *// Define an XQuery Variable with the name of the SPARQL Variable o*
17.          *xC.expr* ← \$ $N_p$     *// Set expr equal to the predicate Variable*
18.          **if** Let XQuery Clause created for *p*
19.             Create *"Bindings Assurance Condition"* for *o*     *//see "Biding Assurance Condition" Section*
20.          **end if**
21.       **else**     *// The predicate is not a variable – it is an IRI*
22.          *xC.type* ← **For or Let XQuery Clause Selection** ( *QF*, **RV**, *o* )     *//Create a For or Let XQuery Clause*
23.          *xC.var* ← $N_o$     *// Define an XQuery Variable with the name of the SPARQL Variable o*
24.          *xC.expr* ← \$ $N_s$ / $x_1$ union \$ $N_s$ /$x_2$ union … union \$ $N_s$ / $x_n$ $\forall x_i \in \mathbf{X}_s \gg \mu_p$
            *// Set expr equal to the variable corresponding to the triple subject suffixed with some of the XPath of the Predicate XPath Set*
            *// $\mathbf{X}_s$ is the bindings XPath Set for the subject s and $\mu_p$ is the mappings XPath Set for the property p.*
25.          **if** Let XQuery Clause created for *o*
26.             Create *"Bindings Assurance Condition"* for *o*     *//see "Biding Assurance Condition" Section*
27.          **end if**
28.       **end if**
29.    **end if**
30.    **end for**
31. **return** *xC*

---

### 8.3.6    *Filter Translation*

The *Filter Translation* algorithm (*Algorithm 6*) translates the SPARQL FILTERs that may be contained in a given *BGP* into XQuery expressions. A straightforward approach for handing SPARQL Filters would be to translate Filter expressions as conditions expressed in XQuery Where clauses. However, this approach would result in inefficient XQuery expressions, since the Filter conditions are evaluated at the final stage of the query processing.

Therefore, we attempt to provide an efficient Filter translation algorithm by applying the Filter restrictions earlier, when this is possible. The earlier the Filter conditions are applied the more efficient XQuery expressions are constructed. The conditions reduce the size of the evaluated data which are going to be used in the later stages of the query processing, similarly to the "*predicate pushdown*" technique which is used in the query optimization context.

| **Algorithm 6: Filter Translation ( *BGP* )** |
| :--- |
| **Input:** Basic Graph Pattern *BGP* |
| **Output:** Where XQuery Clause *xC* or Create *XPredicates* over XQuery clauses |
| 1.  **for each** *Filter* **in** *BGP* |
| 2.      Translate the SPARQL Operators of the *Filter* expression |
| 3.      **if** (*Filter* **is** *safe* ) |
| 4.          Create *XPredicates* for the Filter expressions |
| 5.      **else** |
| 6.          *xC* ← Create an XQuery Where Clause Condition |
| 7.      **end if** |
| 8.  **end for** |
| 9.  **return** *xC* |

The early evaluation of the Filter expressions in XQuery can be achieved using XPredicates. This way, the Filter conditions are applied when the XPath expressions are evaluated. However, not all the Filter expressions can be expressed as XPredicates conditions. There exist "special" cases, where the Filter expressions can not be evaluated in an earlier stage, because of the SPARQL variables that occur inside the Filter expression. In these cases the Filters are translated as conditions expressed in XQuery Where clauses. These "special" cases are known as *not safe* Filter expressions (*Definition 19*) and are discussed below.

**Safe Filter.** There are cases, where the evaluation of Filter expressions is not valid under the evaluation function semantics (*Definition 17*). These "special" cases are identified by the usage of the variables inside the Filter expression and the graph patterns.

> **Definition 19.  (Safe Filter Expressions).** A Filter expression $R$ is *safe* if for Filter expressions of the form $P$ FILTER $R$, it holds that, all the variables that occur in $R$ also occur in $P$ (i.e., $var(R) \subseteq var(P)$ ) [33].

In case of existence of Filter expressions which are *not safe* (*Definition 19*), the evaluation function of the SPARQL queries has to be modified in comparison to the standard SPARQL evaluation semantics. As an example, consider the following pattern: $(?x_{p1}\ ?y)$ OPT$((?x_{p2}\ ?z)$ FILTER$(?y=?z))$. Based on the evaluation function, the expression $(?x_{p2}\ ?z)$ FILTER $(?y=?z)$, is evaluated first. However, the variable *?y* inside the Filter expression does not exist in left side pattern (i.e., $?x_{p2}\ ?z$), thus, this evaluation will produce an ambiguous result. Our translation method overcomes this issue by evaluating the Filter expression as an XQuery Where clause conditions, applied after the graph pattern (translation and) evaluation.

**Filter Expressions Operators.** The SPARQL query language provides several unary and binary operations which can be used inside the Filter expressions. Some of these operators (e.g., &&, ||, !, =, !=, ≤, ≥, +, -, *, /, regex, bound, etc.) can be mapped directly to XQuery built-in functions and operators, whereas for other operators (e.g., sameTerm, lang, etc.) XQuery functions have to be implemented in order to simulate them. Finally, a few SPARQL operators can not be supported in the XQuery language. In particular, the isBlank SPARQL operator can not be implemented for the XML data model, since the *blank node* notion is not defined in XML. In addition, it is very complex to evaluate the isIRI, isLiteral and datatype SPARQL operators over XML data. The result of these operators is difficult and inefficient to be determined on-the-fly through the evaluation of the XQuery expressions over XML data. This can only be achieved via a complex and a large sequence of XQuery if – if else conditions. The if – if else conditions will exploit the mappings in order to evaluate the above operators, resulting in inefficient XQuery expressions. However, the results of these operators can be determined after the XQuery evaluation, by processing the return results and the mappings.

**Filter Evaluation.** The SPARQL query language supports three-valued logic (i.e., True, False and Error) for Filter expression evaluation. Instead, the XQuery query language supports two-valued logic or Boolean logic (i.e., True and False). In order for our method to bridge this difference, based on the semantics presented at [12] and [116], the SPARQL Error value is mapped to the XQuery False value, while, the SPARQL Error value could be easily supported by our translation by exploiting XQuery if – if else conditions throughout the Filter expression translation. These conditions would check for errors that have occurred during the evaluation of the XQuery Where clause conditions and would return the Error value. A common SPARQL error example occurs when unbound variables exist inside the Filter expression.

### 8.3.7 Return Clause Construction

The *Construct Return Clause* algorithm (*Algorithm 7*) builds the XQuery *Return* Clause. For Ask SPARQL queries (*Lines 1~2*), the algorithm creates an XQuery Return clause which, for efficiency reasons, includes only the literal *"yes"* (*Line 2*). For the other query forms (i.e., Select, Construct, Describe) (*Lines 3~6*), the algorithm creates an XQuery Return clause *xC*

---

**Algorithm 7: Construct Return Clause ( *BGP*, *QF*, **RV**, *varTypes* )**

**Input:** Basic Graph Pattern *BGP*, SPARQL query form *QF*,
      SPARQL Return Variables **RV**, Variable Types *varTypes*

**Output:** Return XQuery Clause *xC*

1.   **if** *QF* = Ask
2.      *xC* ← return("yes")              *//Create an XQuery Return clause*
3.   **else**   *//The query form is not Ask*
             *//Create an XQuery Return clause*
4.      *xC* ← return(<Result>
                <var$_1$>...</var$_1$> , <var$_2$>...<var$_2$>,...,<var$_n$>...</var$_n$></Result>)
                         ∀ *var$_i$* ∈ **RV** ∩ *var*(*BGP*)
        *// Each Return Variable included in the given BGP is inserted in the XQuery return clause*
5.      ∀ *var$_i$* ∈ **RV** ∩ *var*(*BGP*) use the *varTypes* to
         determine the result form of *var$_i$*
6.   **end if**
7.   **return** *xC*

---

that includes all the *Return Variables* (**RV**) used in the *BGP* (*Line 4*). The syntax of the return clause allows (using markup tags) the distinction of each solution in the solution sequence, as well as the distinction of the corresponding values for each variable. The structure of the return results allows the SPARQL operators AND and OPT to be applied over the results returned by different XQuery Return clauses. Finally the algorithm, for each variable included in the return clause, and based on the variable types (*varTypes*), uses the appropriate function to format the result form of the variable (Section 6.2.2) and returns the generated return XQuery clause (*Line 7*).

## 8.4 Discussion

The Graph Pattern Translation is the most complex phase of the SPARQL to XQuery translation process. The noteworthy issues that have arisen throughout this phase are outlined and discussed here.

**Creating XQuery Clauses.** Throughout the XQuery clause creation we had to overcome several difficulties, involving the *accurate solution sequence cardinality*, the *association of different XQuery variables* and the *binding assurance*.

*Associating Different XQuery Variables.* Throughout the creation of the For/Let XQuery clauses, the *BGP2XQuery* algorithm (Section 8.3) exploits the *extension relation* (*Definition 15*) in order to achieve the association of different XQuery variables. For example, consider the case where the XQuery variable *$per* that refers to *Persons* (corresponding to the XPath */Persons/Person*) should be associated with the XQuery variable *$fn*, which refers to the *First Names* of the *Persons* (corresponding to the XPath */Persons/Person/FirstName*). This can be accomplished using For XQuery clauses and defining the XQuery variable *$fn* as an *extension* of the XQuery variable *$per*, i.e., *for $per in /Persons/Person for $fn in $per/FirstName*.

*Accurate Solution Sequence Cardinality.* An interesting issue in the graph pattern translation is to ensure the generation of the appropriate solution sequence based on the SPARQL semantics. In our translation, this has been accomplished by the *For or Let XQuery Clause Selection* algorithm (Section 8.3.2) which determines the creation of the appropriate For or Let XQuery clauses.

*Binding Assurance.* In order to guarantee that all the variables defined in a *Basic Graph Pattern* are bound in all the solutions, we have developed a binding condition assurance mechanism. The binding assurance mechanism exploits the XQuery function exists( ) when it is required to guarantee the assignment of a value to the XQuery variables (Section 8.3.5).

**Implementing SPARQL Operators.** Several issues have arisen throughout the implementation of the SPARQL algebra operations (i.e., UNION, AND, OPT and FILTER) using XQuery expressions.

*Well Designed Graph Patterns vs. Non-Well Designed Graph Patterns.* The existence of *non-well designed graph patterns*, as well as the SPARQL operator semantics, which define the *Join* (i.e., AND) and *Left Outer Join* (i.e., OPT) operators as non null-rejecting forced us to handle the well designed in a different way from the non-well designed graph patterns. This way we have provided an efficient implementation for the former. In this implementation the intermediate results are used in the XQuery expression in order to reduce the computation cost (Section 8.2).

*Efficient Join & Condition Implementation.* The efficient implementation of some basic SPARQL query features using XQuery expressions is an interesting part of the translation. Consider as an example the translation of the joins occurring between the triple patterns (expressed with "**.**" in the SPARQL syntax) in the Basic Graph patterns. In the context of *BGPs*, the joins are implemented efficiently by the *BGP2XQuery* algorithm (Section 8.3) that associates For/Let XQuery clauses that resemble a nested loop join. In addition, throughout the For/Let XQuery clause creation, the *BGP2XQuery* algorithm exploits the *extension relation* (*Definition 15*) in order to use the already evaluated XQuery values providing a more efficient join implementation. Another issue, is the translation of the literal parts of the triple patterns (Section 8.3.5), which are translated as conditions over the For/Let XQuery clauses using *XPredicates*. In this way, the conditions imposed by the existence of the literals are applied early in the XQuery evaluation plan, resulting in a more efficient XQuery evaluation.

*Solution Structure.* Remarkable issues are the need of the distinction of the different solutions in the solution sequence, as well as the distinction of the corresponding values for each variable in each solution. In this way, the SPARQL solution sequence modifiers and algebra operators can be applied on the results produced by the XQuery expressions. The above issues have been resolved by exploiting "special" markup tags (e.g., *<Result>*, etc.) throughout the creation of the XQuery Return clause (Section 8.3.7).

**Handling SPARQL Filters.** Several interesting issues resulted from the translation of the Filter expressions, including handling the *safe* and *non-safe Filter expressions*, mapping the SPARQL three-valued logic to the XQuery two-valued logic, translating the SPARQL operators used in the Filter expressions, etc.

*Safe vs. non-Safe Filter Expressions.* In order to provide efficient Filter translation, we try to evaluate the Filter Expressions in an early stage of the XQuery evaluation. This is achieved using XPredicates that apply the Filter conditions over the For/Let XQuery clauses. However, there are "special" cases, where the Filter Expressions can not be evaluated in an earlier stage, due to the SPARQL variables that occur in the Filter Expression. These cases are known as *not safe* Filter expressions (*Definition 19*). They occur because of the flexibility of the SPARQL syntax in expressing queries. In order to overcome this issue, our translation method evaluates the conditions defined in these Filter expressions at the end (using Where clauses), in order to guarantee that the variables occurring in the Filter expression have already been evaluated (Section 0).

*Implementing Filter Expression Operators.* Regarding the SPARQL operators included in Filter expressions, most of them can be directly mapped to XQuery built-in functions and operators (e.g., regex, &&, ||, !, =, !=, ≤, ≥, +, -, *, /, etc.). However, for some "more complex" SPARQL operators (e.g., sameTerm, lang, etc.) we have developed native XQuery functions that simulate them. Finally, a few SPARQL operators (e.g., isBlank ) can not be implemented in the XQuery language, since they are not supported by the XML data model.

*Three-valued Logic vs. Two-valued Logic.* The evaluation of Filter expressions in the SPARQL query language is based on three-valued logic (i.e., True, False and Error), while the XQuery query language supports Boolean logic (i.e., True and False). An issue is to handle and relate the three-valued logic using the XQuery Boolean logic. In our translation, for efficiency reasons, the SPARQL Error value has been mapped to the False XQuery value. However, it is possible, but inefficient, to support the Error value in the generated XQuery expressions (Section 0).

# 9 SOLUTION SEQUENCE MODIFIERS & QUERY FORMS

## 9.1 Translating Solution Sequence Modifiers

This section describes the *Solution Sequence Modifier Translation* phase, which translates the SPARQL *Solution Sequence Modifiers* (*SSMs*) into XQuery expressions. The *SSMs* that may be contained in a SPARQL query are translated using XQuery clauses and built-in functions. The *SSMs* supported by the current SPARQL specification are the Distinct, Reduced, Order By, Limit, and Offset solution sequence modifiers.

Table 8 summarizes the XQuery expressions and built-in functions that are used for the translation of the solution sequence modifiers. Let $xE_{GP}$ be the XQuery expressions produced from the graph pattern (*GP*) translation. In Table 8, *$Results* is an XQuery variable, which is bound to the solution sequence produced by the XQuery expressions $xE_{GP}$ (i.e., let *$Results :=* *($xE_{GP}$)), *n*, *m* are positive integers and *?x*, *?y* are SPARQL variables.

**Solution Sequence Modifier Priorities.** If more than one solution modifiers are declared in the given SPARQL query, the order in which they are applied during the translation phase is the following (the order is compatible with the SPARQL query language semantics): (a) Order By, (b) Distinct / Reduced and (c) Offset / Limit.

**Table 8. Translation of the SPARQL Solutions Sequence Modifiers in XQuery Expressions**

| Solution Sequence Modifier | XQuery Expressions |
|---|---|
| **LIMIT** *n* | **return**( $Results[**position**( ) < = *n* ] ) |
| **OFFSET** *n* | **return**( $Results[**position**( ) > *n* ] ) |
| **LIMIT** *n* && **OFFSET** *m* | **return**( $Results[ **position**( ) > *m*  **and position**( ) < = *n+m* ] ) |
| **ORDER BY DESC**(?x) **ASC**(?y) | **for** $res **in** $Results<br>**order by** $res/x **descending empty least**, $res/y **empty least**<br>**return** $res |

## 9.2 Translating Query Forms

The *Query Form Translation* is the final phase of the SPARQL to XQuery translation. The current specification of the SPARQL query language supports four query forms: Select, Ask, Construct and Describe. According to the query form, the type of the returned results is different. In particular, after the translation of any solution sequence modifier, the generated XQuery is enhanced with the appropriate, for this query form, XQuery expressions in order to form the appropriate type of the results (e.g., an RDF graph, a result sequence, or a Boolean value).

**Select Queries.** The Select SPARQL queries return (all or a subset of) the variables bound in a query pattern match. To simulate this query form in XQuery, the results are returned as sequences of XML elements created by the XQuery Return clauses (see the *Build Return Clause* Algorithm, Section 8.3.7). This sequence should be contained in a root element in order to be a valid XML document. Thus, we create a root element *"Results"* containing the result sets produced by the XQuery return clause.

**Ask Queries.** The Ask SPARQL queries return a Boolean value (yes or no), indicating whether a query pattern is matched in a dataset or not. The cardinality of the solution for Ask queries is one (i.e., the value yes/no should be returned once). Thus, we check for the existence of any result and we return "yes" if one or more results exist and "no" otherwise.

**Construct Queries.** The Construct SPARQL queries return an RDF graph structured according to the *graph template* of the query. The result is an RDF graph formed by taking each query solution in the solution sequence, substituting the variables in the graph template, and combining the triples into a single RDF graph using the union operation.

In order to implement the semantics for the unbound variables, for each triple pattern of the graph template that contains variables we check if any one of these variables is unbound. In that case, the triple is not returned. Moreover, in order to enforce the semantics of the Blank node naming conventions in the RDF graph, we exploited an XQuery *positional variable* (defined using "at" term in XQuery syntax).

**Describe Queries.** The Describe SPARQL queries return an RDF graph which provides a "description" of the matching resources. The "description" semantics are not determined by the current SPARQL specification, however they are determined by the SPARQL query engines (note that several SPARQL engines do not support Describe queries). As a result, we provide an "approximate" support for this query form, by evaluating the Describe SPARQL query against the source ontology and then translating it as a Select query. The overall result is a combination of the RDF graph returned by the SPARQL Describe query and the result sequence returned by the translated XQuery.

The translation of the query forms described above is outlined in (5), where **Qx** is the set of the XQuery expressions resulted after the translation of SPARQL query form. Let the SPARQL query $Q_S = \langle QF, GP, SSM \rangle$, where *QF* is the query form, *GP* is the query graph pattern and *SSM* the solution sequence modifiers. Let $xE_Q$ be the XQuery expressions produced from the graph pattern (*GP*) and solution sequence modifier (*SSM*) translation. For the Construct query form (last case in (5)), we consider the graph template: " _:a iri:property ?x .  _:a ?p ?y . ", which consists of two triple patterns and containing the blank node "_:a".

$$
\mathbf{Qx} = \begin{cases}
\begin{array}{ll}
\textbf{let } \$\text{Results} := (xE_Q \text{ )} & \textbf{if } QF = Select \\
\textbf{return ( } \text{<Results> } \$\text{Results } \text{</Results> )} & \\
& \\
& \\
\textbf{let } \$\text{Results} := (xE_Q) & \textbf{if } QF = Ask \\
\textbf{return ( if ( empty (} \$\text{Results) ) then } \text{"no" } \textbf{else } \text{"yes" )} & \\
& \\
& \\
\textbf{let } \$\text{Results} := (xE_Q) & \textbf{if } QF = Construct \\
\textbf{for } \$\text{res } \textbf{at } \$\text{iter } \textbf{in } \$\text{Results} & \\
\textbf{return ( if ( exists( } \$\text{res/x ) ) then} & \\
\qquad\qquad \textbf{concat ( concat ( } \text{"\_:a" , } \$\text{iter ) , } \text{" iri:property " , } \textbf{string(} \$\text{res/x ) , "." )} & \\
\qquad \textbf{else ( )} & \\
\qquad \textbf{if ( exists( } \$\text{res/p ) } \textbf{and exists( } \$\text{res/y ) ) then} & \\
\qquad\qquad \textbf{concat ( concat( } \text{"\_:a" , } \$\text{iter ), } \textbf{string(} \$\text{res/p ) , } \textbf{string(} \$\text{res/y ) , "." )} & \\
\qquad \textbf{else ( ) )} & \\
\end{array}
\end{cases}
\quad \textbf{(5)}
$$

## 10    XQUERY REWRITING / OPTIMIZATION

It was pointed out in Section 5.3 that among the objectives of the proposed SPARQL to XQuery translation method was the generation of simple XQuery expressions, so that their correspondence with the SPARQL queries could be easily understood. This has led to the generation of some inefficient XQuery expressions, but it was expected that the XQuery optimizer would optimize those queries to achieve better execution performance. However, we have attempted to use the query optimizer of two XQuery engines with no improvement to be achieved for any of the queries. This observation led us to develop some XQuery rewriting rules and integrate them in our Framework. The performance evaluation studies included in the next section (Section 13.3) show that they are useful in improving the XQuery performance. Since the XQuery performance is beyond the scope of this paper, we present here a limited number of simple rewriting rules.

### 10.1    Rewriting Rules [23]

The proposed rewriting rules aim at providing more efficient XQuery expressions that benefit from: (a) The knowledge of how the XQuery expressions are generated from the SPARQL to XQuery translation; and (b) The XML Schema semantics. The rules exploit the aforementioned, in order to remove redundant XQuery clauses and variables, unnest nested For XQuery clauses and minimize the loops executed by the For XQuery clause. The rewriting rules, which are listed below, are applied sequentially on the generated XQuery queries. Firstly, *Rule 1* is applied, followed by *Rule 2* that is applied on the resulting XQuery query, and finally *Rule 3* is applied on the output of *Rule 2*.

**Rewriting Rule 1 (Changing For Clauses to Let).** The *Changing* For *Clauses to* Let rule is applied on the For XQuery clauses, from the top to the bottom (or the inverse). Intuitively, this rule exploits the schema information in order to convert the For clauses in Let ones in cases where multiple values cannot exist. The objective of this rule is to avoid the unnecessary checks for possible multiple values performed by the For clauses, in cases where Let clauses can also be used. The use of this rule results in more Let clauses, which may be removed later, when *Rule 2* is applied.

**Example 19.** *Applying the Changing For Clauses to Let Rule*

| *Initial XQuery Expressions* | | *Rewritten XQuery Expressions* |
|---|---|---|
| … | | … |
| **for** $stud **in** $doc/Persons/Student | $\Rightarrow$ | **for** $stud **in** $doc/Persons/Student |
| **for** $age **in** $stud/Age [24] | | **let** $age **:=** $stud/Age |
| **where** ( **exists**($age) ) | | **where** ( **exists**($age) ) |
| … | | … |

■

**Rewriting Rule 2 (Reducing Let Clauses).** The *Reducing* Let *Clauses* rule is applied iteratively to the Let XQuery clauses, from the bottom to the top. Intuitively, this rule removes the unnecessary Let clauses that have been produced from triple pattern translation and can be pruned. The objective of the rule is to eliminate the unnecessary XQuery clauses and variables. In addition, in the case of *Biding Assurance Condition* existence, a *predicate pushdown* is performed. In particular, the **exists** condition placed in the Where XQuery clause is evaluated in an earlier query processing stage since it is applied on the XPaths using XPath predicates.

---

[23] The formal definitions of the rewriting rules are available in [120].

[24] From the Persons XML Schema (Figure 3) we have the following cardinality constraints for the Age element: minOccurs="1" and maxOccurs="1".

**Example 20.** *Applying the Reducing Let Clauses Rule*

a)

*Initial XQuery Expressions*

*Rewritten XQuery Expressions*

… …

**for** $stud    **in** $doc/Persons/Student        ⟹        **for** $stud   **in** $doc/Persons/Student
**let** $course **:=** $stud/Course                                    **for** $grade **in** $stud/Course/Grade
**for** $grade  **in** $course/Grade                                 …

…

b)

*Initial XQuery Expressions (rewritten from Example 19)*    *Rewritten XQuery Expressions*

… …

**for** $stud   **in**   $doc/Persons/Student        ⟹        **for** $stud **in** $doc/Persons/Student[./age]
**let** $age   **:=**   $stud/age                                       …
**where** ( **exists**($age) )

…                                                                                                                              ∎

**Rewriting Rule 3 (Unnesting For Clauses).** The *Unnesting* For *Clauses* rule is applied iteratively on the For XQuery clauses, from the top to the bottom. Intuitively, this rule unnests nested For clauses that can be expressed as a single For clause. The objective of the rule is to reduce the nested For clauses. In this way some XQuery clauses and variables are removed.

**Example 21.** *Applying the Unnesting For Clauses Rule*

*Initial XQuery Expressions*

*Rewritten XQuery Expressions*

… …

**for** $stud   **in** $doc/Persons/Student        ⟹        **for** $name **in** $doc/Persons/Student/name
**for** $name **in** $stud/name                                    …

…                                                                                                                              ∎

## 11    TOWARDS SUPPORTING SPARQL UPDATE OPERATIONS

In this section, we briefly describe an extension of the SPARQL2XQuery Framework in the context of supporting the SPARQL 1.1 update operations.

In order to support SPARQL update operations, we have studied the extension of the SPARQL to XQuery translation using the recently introduced XQuery Update Facility [7]. SPARQL 1.1 supports two main categories of update operations: a) *Graph update*, which includes operations regarding graph additions and removals; and b) *Graph management*, which contains "storage-level" operations, e.g., CREATE, DROP, MOVE, COPY, etc. Our work focuses on the *graph update operations,* since the storage-level operations are out of scope of the SPARQL2XQuery Framework working scenario (i.e., interoperability/integration scenario). We are working on the support of the SPARQL update operations that are presented in Table 9. In most cases, similar methods and concepts previously defined in the SPARQL2XQuery Framework are going to be used for the update operations translation. For instance, the translation of graph patterns, triple patterns and RDF triples is also included in the update operations translation.

For each update operation (Delete Data, Insert Data, and Delete/Insert), a simplified SPARQL syntax template is presented, as well as the corresponding XQuery expressions in Table 9. In SPARQL context, we assume the following sets, let *tr* be an RDF triples set, *tp* a triple patterns set, *trp* a set of triples and/or triple patterns, and *gp* a graph pattern. In addition, in XQuery, let *xEw*, *xEI* and *xED* denote sets of XQuery expressions (i.e., FLOWR expressions) resulted from the translation of the graph patterns included in Where, Insert and Delete clauses, respectively. As $xE_c$($v_1$, $v_2$,... $v_n$) we denote that the XQuery expressions $xE_c$ are using the values assigned to XQuery variables $v_1$, $v_2$,... $v_n$. Finally, let $xn_i$ be an XML fragment, i.e., a set of XML nodes, and $xp_i$ denote an XPath expression.

**Table 9. Translation of the SPARQL Update Operations in XQuery**

| SPARQL | | Translated XQuery Expressions |
|---|---|---|
| **SPARQL Update Operation** | **Syntax Template** [1] | |
| DELETE DATA | **Delete data{**<br>  *tr*<br>  **}** | **delete nodes collection**("http://dataset...")/$xp_1$<br>**...**<br>**delete nodes collection**("http://dataset...")/$xp_n$ |
| INSERT DATA | **Insert data{**<br>  *tr*<br>  **}** | **let** $n_1$ := $xn_1$<br>**...**<br>**let** $n_n$ := $xn_n$<br>**let** $data_1$ := ($n_k$, $n_{m}$,...)<br>**let** $insert\_location_1$ := **collection**(http://dataset..)/$xp_1$<br>**...**<br>**let** $insert\_location_p$ := **collection**(http://dataset..)/$xp_p$<br>**let** $data_p$ := ($n_j$, $n_v$,...)<br>**return(**<br>    **insert nodes** $data_1$ **into** $insert\_location_1$,<br>    **...**<br>    **insert nodes** $data_p$ **into** $insert\_location_p$<br>**)** |
| DELETE / INSERT | a)<br>**Delete {**<br>  *trp*<br>**}Where{**<br>  *gp*<br>**}**<br><br>b)<br>**Insert{**<br>  *trp*<br>**}Where{**<br>  *gp*<br>**}**<br><br>c)<br>**Delete {**<br>  *trp*<br>**}Insert{**<br>  *trp*<br>**}Where{**<br>  *gp*<br>**}** | a)<br>**let** $where\_gp$ := $xE_W$<br>**let** $delete\_gp$ := $xE_D$ ($where\_gp$)<br>**return delete nodes** $delete\_gp$<br><br>b)<br>**let** $where\_gp$ := $xE_W$<br>**let** $insert\_location_1$ := $xp_1$<br>**for** $it_1$ **in** $insert\_location_1$<br>  $xE_I$($where\_gp$, $it_1$)<br>**return insert nodes into** $it_1$<br>**...**<br>**let** $where\_gp$ := $xE_W$<br>**let** $insert\_location_n$ := $xp_n$<br>**for** $it_n$ **in** $insert\_location_n$<br>  $xE_I$($where\_gp$, $it_n$)<br>**return insert nodes into** $it_n$<br><br>c)<br>Translate Delete Where same as a),<br>then translate Insert Where same as b) |

## 12  SPARQL TO XQUERY TRANSLATION — A STEP-BY-STEP EXAMPLE

In this section, we present in detail a SPARQL to XQuery translation example. As an input, consider the query presented below in natural language and in SPARQL syntax, expressed over the Persons ontology (see Table 4 and Table 5 for details). The given SPARQL query contains several SPARQL features like: (a) SPARQL operators (e.g., AND, OPTIONAL, FILTER); (b) Solution sequence modifiers (e.g., Order by Asc/Desc, Limit, Offset); (c) Schema Triples (e.g., rdfs:subClassOf); (d) Built-in functions (e.g., regex).

In the rest of this section we present the translation process, phase by phase: the determination of the variable types (Section 12.1), the Schema Triple processing (Section 12.2), the variable bindings (Section 12.3), the graph pattern, sequence modifier and query form translation (Section 12.4), the XQuery rewriting (Section 12.5) and the resulting XQuery query (Section 12.6).

---

**Natural Language Query**

---

*"For the instances of the Person_Type subclasses, return their SSN code, their last name(s) and their e-mail(s) for the ones that their first name is "John", their last name starts with "B" and are older than 25 years old. The (existence of) e-mail is optional. The query must return at most 30 result items ordered by the last name value in ascending order and by the SSN value in descending order and skipping the first 5 items."*

---

**SPARQL Query**

---

**PREFIX** ns:  <http://example.com/ns#>
**PREFIX** rdfs: <http://www.w3.org/2000/01/rdf-schema#>
**PREFIX** rdf:  <http://www.w3.org/1999/02/22-rdf-syntax-ns#>

**SELECT**  ?SSN ?lname ?email
**WHERE**{ { ?studCl      rdfs:subClassOf          ns:Person_Type.
               ?stud        rdf:type                  ?studCl .
               ?stud      ns:SSN__xs_integer        ?SSN .
               ?stud      ns:FirstName__xs_string    "John" .
               ?stud      ns:LastName__xs_string     ?lname .
               ?stud      ns:Age__validAgeType       ?age.
               **FILTER** ( regex( ?lname, "^B") && ?age>25)   }
         **OPTIONAL** { ?stud   ns:Email__xs_integer      ?email }
}**ORDER BY ASC** ( ?lname ) **DESC**( ?SSN )
**LIMIT** 30  **OFFSET** 5

---

### 12.1  Variable Types

In this section we outline the *variable type determination* phase for the SPARQL query defined above. Consider the following sequence of triple patterns:

| | | |
|---|---|---|
| ?studCl | rdfs:subClassOf | ns:Person_Type |
| ?stud | rdf:type | ?studCl |
| ?stud | ns:SSN__xs_integer | ?SSN |
| ?stud | ns:FirstName__xs_string | "John" |
| ?stud | ns:LastName__xs_string | ?lname |
| ?stud | ns:Age__validAgeType | ?age |
| ?stud | ns:Email__xs_integer | ?email |

Since the Schema Triple patterns are pruned from the determination of the variable types the following three triple patterns are resulted:

$t_1$ = ⟨?stud ns:SSN__xs_integer ?SSN⟩, $t_2$ = ⟨?stud ns:FirstName__xs_string "John"⟩,

$t_3$ = ⟨?stud ns:LastName__xs_string ?lname⟩, $t_4$ = ⟨?stud ns:Age__validAgeType ?age⟩ and

$t_5$ = ⟨?stud ns:Email__xs_integer ?email⟩.

Initially, all the variables are initialized as *Unknown Predicate Variable Type* (*UPVT*); as a consequence, the following hold:

$T_{stud}$ = *UVT*, $T_{age}$ = *UVT*, $T_{SSN}$ = *UVT*, $T_{lname}$ = *UVT*, and $T_{email}$ = *UVT*.

Using the variable type determination rules presented in Section 6.2.1, the following hold:

For $t_1$ = ⟨ ?stud ns:SSN__xs_integer ?SSN ⟩ hold:

$T_{stud}$ = *CIVT*        (*Rule 1*)

$T_{SSN}$ = *LVT*        (*Rule 4*)

For $t_2$ = ⟨ ?stud ns:FirstName__xs_string "John" ⟩ it holds that:

$T_{stud}$ = *CIVT*        (*no change – Rule 1*)

For $t_3$ = ⟨ ?stud ns:LastName__xs_string ?lname ⟩ hold:

$T_{stud}$ = *CIVT*        (*no change – Rule 1*)

$T_{lname}$ = *LVT*        (*Rule 4*)

For $t_4$ = ⟨ ?stud ns:Age__validAgeType ?age ⟩ hold:

$T_{stud}$ = *CIVT*        (*no change – Rule 1*)

$T_{age}$ = *LVT*        (*Rule 4*)

For $t_5$ = ⟨ ?stud ns:Email__xs_integer ?email ⟩ hold:

$T_{stud}$ = *CIVT*        (*no change – Rule 1*)

$T_{email}$ = *LVT*        (*Rule 4*)

Finally, the following hold:

$T_{stud}$ = *CIVT*, $T_{age}$ = *LVT*, $T_{SSN}$ = *LVT*, $T_{lname}$ = *LVT* and $T_{email}$ = *LVT*.

## 12.2    Schema Triples

In this section we outline the *Schema Triple processing* phase. From the triple pattern sequence, we isolate and process the following Schema Triples:

$st_1$ = ⟨ ?studCl rdfs:subClassOf ns:Person_Type ⟩  and $st_2$ = ⟨ ?stud rdf:type ?studCl ⟩.

In the first step of processing, the variables of the Schema Triples are bound to ontology constructs, resulting in:

?studCl = ⟨ Student_Type ⟩ (from the processing of *st$_1$*) and ?stud = ⟨ Student_Type ⟩ (from the processing of *st$_2$*).

In the next processing step, for each variable we determine the XPath Sets (based on the mappings specified in *Example 13*), resulting in: $X_{stud}$ = { /Persons/Student }.

These bindings are going to be the initial bindings for the variable binding phase.

## 12.3 Variable Bindings

In this section we present the variable binding phase. The Schema Triple patterns are not taken into account, thus, the SPARQL query consists of the following two Basic Graph Patterns. $BGP_1 = \langle$ ?studCl rdfs:subClassOf ns:Person_Type . ?stud rdf:type ?studCl . ?stud ns:SSN__xs_integer ?SSN . ?stud ns:FirstName__xs_string "John" . ?stud ns:Last-Name__xs_string ?lname . ?stud ns:Age__validAgeType ?age. FILTER ( regex( ?lname, "^B") && ?age>25) $\rangle$ and $BGP_2 = \langle$ ?stud ns:Email__xs_integer ?email $\rangle$. For each of these $BGPs$, the *variable binding algorithm* (Section 7.1) is exploited.

Thus, for $BGP_1$ hold the following:

From the *determination of the variable types* phase (Section 12.1) the following variable types have been determined:

$T_{stud} = CIVT$, $T_{age} = LVT$, $T_{SSN} = LVT$ and $T_{lname} = LVT$.

From the Schema Triple processing (Section 12.2) the following initial bindings have been determined:

$\mathbf{X}_{stud}^{Sch} = \mathbf{X}_{stud}^0 = \{$ /Persons/Student$\}$, $\mathbf{X}_{age}^0 = \ominus$, $\mathbf{X}_{SSN}^0 = \ominus$ & $\mathbf{X}_{lname}^0 = \ominus$.

### *1st Iteration*

For $t_1 = \langle$ ?stud ns:SSN__xs_integer ?SSN $\rangle$ (Type 2) hold the following:

$\mathbf{X}_{stud}^1 = \mathbf{X}_{stud}^0 \sqcap \mathbf{X}_{SSN\_\_xs\_integerD} \ll \mathbf{X}_{SSN}^0 = \{$ /Persons/Student$\} \sqcap \{$ /Persons/Person, /Persons/Student$\} \ll \ominus$
$= \{$ /Persons/Student $\}$ and $\mathbf{X}_{SSN}^1 = Non\ Determinable$

For $t_2 = \langle$ ?stud ns:FirstName__xs_string "John" $\rangle$ (Type 3) it holds that:

$\mathbf{X}_{stud}^1 = \mathbf{X}_{stud}^0 \sqcap \mathbf{X}_{LastName\_\_xs\_stringD} = \{$ /Persons/Student$\} \sqcap \{$ /Persons/Person, /Persons/Student$\}$
$= \{$ /Persons/Student$\}$

For $t_3 = \langle$ ?stud ns:LastName__xs_string ?lname $\rangle$ (Type 2) hold the following:

$\mathbf{X}_{stud}^1 = \mathbf{X}_{stud}^0 \sqcap \mathbf{X}_{LastName\_\_xs\_stringD} \ll \mathbf{X}_{lname}^0 = \{$ /Persons/Student$\} \sqcap \{$ /Persons/Person, /Persons/Student$\} \ll \ominus$
$= \{$ /Persons/Student$\}$ and $\mathbf{X}_{lname}^1 = Non\ Determinable$.

For $t_4 = \langle$ ?stud ns:Age__xs_integer ?age $\rangle$ (Type 2) hold the following:

$\mathbf{X}_{stud}^1 = \mathbf{X}_{stud}^0 \sqcap \mathbf{X}_{Age\_\_xs\_integerD} \ll \mathbf{X}_{age}^0 = \{$ /Persons/Student$\} \sqcap \{$ /Persons/Person, /Persons/Student$\} \ll \ominus$
$= \{$ /Persons/Student $\}$ and $\mathbf{X}_{age}^1 = Non\ Determinable$.

### *2nd Iteration*

*Nothing changes from the second iteration, so the Variable Binding Algorithm terminates.*

### *Finally*

The following final bindings have been determined:

$\mathbf{X}_{stud} = \{$ /Persons/Student $\}$

For $BGP_2$ we have:

From the *determination of the variable types* phase (Section 12.1) we have the following variable types:

$T_{stud} = CIVT$ and $T_{email} = LVT$

We have the following initial bindings:

$\mathbf{X}_{stud}^0 = \{$ /Persons/Student$\}$ and $\mathbf{X}_{email}^0 = \ominus$

***1st Iteration***

For $t_1 = \langle$ ?stud ns:Email__xs_integer ?email $\rangle$ (Type 2):

$\mathbf{X}_{stud}^1 = \mathbf{X}_{stud}^0 \; \overline{\sqcap} \; \mathbf{X}_{Email\_\_xs\_string} \; {}_D \lessdot \mathbf{X}_{email}^0 = \{$ /Persons/Student $\} \; \overline{\sqcap} \; \{$ /Persons/Person, /Persons/Student $\} \lessdot \ominus$

$\qquad = \{$ /Persons/Student $\}$ and $\mathbf{X}_{email}^1 = $ *Non Determinable*

***2nd Iteration***

*Nothing changes from the second iteration, so the Variable Binding Algorithm terminates.*

***Finally***

$\mathbf{X}_{stud} = \{$ /Persons/Student $\}$ and $\mathbf{X}_{email} = $ *Non Determinable*

## 12.4 Building XQuery Expressions

In this section we outline the translation of the SPARQL elements to XQuery expressions. We present the graph pattern translation first (Section 12.4.1), afterwards the translation of the solution sequence modifiers (Section 12.4.2) and finally the translation of the query forms (Section 12.4.3).

### 12.4.1 Graph Pattern Translation

We first translate the *BGPs* defined in the SPARQL query exploiting the *BGP2XQuery* algorithm (Section 8.3.1).

For *BGP₁* hold the following:

<u>*Subject Translation*</u>

*?stud* $\qquad \Rightarrow \qquad$ *for $stud in $doc/Persons/Student*

<u>*Predicate Translation*</u>

*Nothing to be translated, all the predicate parts are constants (i.e., IRIs).*

<u>*Object Translation*</u>

*?SSN* $\qquad \Rightarrow \qquad$ *for $SSN in $stud/@SNN*

*"John"* $\qquad \Rightarrow \qquad$ *for $stud in $doc/Persons/Student* **[ ./FirstName = "John"]**

*?lname* $\qquad \Rightarrow \qquad$ *for $lname in $stud/LastName*

*?age* $\qquad \Rightarrow \qquad$ *let $age := $stud/Age*

<u>*Filter Translation*</u>

**FILTER**( *regex( ?lname, "^B") && ?age>25)* $\Rightarrow \qquad$ *for $lname in $stud/LastName* **[matches( . , "^B" )]**

$\qquad\qquad\qquad\qquad\qquad\qquad\qquad\qquad\qquad$ *let $age := $stud/Age* **[.>25]**

<u>*Return Clause Building*</u>

**SELECT** ⌐?SSN ?lname¬ *?email* $\Rightarrow$*return(<Result> <SSN>{ string($SSN) }</SSN> , <lname>{ string($lname) }</lname>*

$\qquad\qquad\qquad\qquad\qquad\qquad$ *</Result>)*

For *BGP₂* hold the following:

<u>*Subject Translation*</u>

*The XQuery variable ?stud from the translation of the subject of the triple pattern $t_1$ (i.e., left optional operand) is used in BGP$_2$.*

<u>*Predicate Translation*</u>

*Nothing to be translated, all the predicate parts are constants (i.e., IRIs).*

<u>*Object Translation*</u>

*?email ⟹ for $email in $stud/Email*

<u>*Return Clause Building*</u>

**SELECT** *?SSN ?lname ⟦?email⟧ ⟹ return(<Result> <email>{ string($email)}</email> </Result>)*

The *GP2XQuery* algorithm traverses the SPARQL execution tree translating the SPARQL operators that appear in the *BGPs*, that result in the following:

**OPTIONAL** ⟹    *return( if(exists($BGP_2) ) then(*

   *for $bgp2_it in $BGP_2*

   *return ( <Result> <SSN>{ string($SSN) }</SSN> , <lname>{ string($lname) }</lname>, {$bgp2_it/Result} </Result>) )*

   *else(*

   *(<Result><SSN>{ string($SSN) }</SSN> , <lname>{ string($lname) }</lname> </Result>))*

   *)*

*$BGP_2 is XQuery variables in which the BGP$_2$ evaluation results have been assigned, using* Let *XQuery clause.*

*12.4.2    Solution Sequence Modifier Translation*

In this section we outline the translation of the solution sequence modifiers to XQuery expressions that result in the following XQuery expressions for our example query:

**ORDER BY ASC** *( ?lname )* **DESC***( ?SSN ) ⟹ let $Ordered_Results :=(*

   *for $iter in $Results*

   *order by $iter/lname empty least , $iter/SSN descending empty least*

   *return($iter) )*

**LIMIT** *30* **OFFSET** *5        ⟹ return ($Ordered_Results[position( )>5 and position( )<=35])*

*12.4.3    Query Form Translation*

In this section we outline the translation of the query form to XQuery expressions that result in the following XQuery expressions for our example query:

⟦**SELECT**⟧ *?SSN ?lname ?email        ⟹ return ( **<Results>** { $Modified_Results } **</Results>** )*

## 12.5 XQuery Rewriting

In this section we present the rewriting of the XQuery expressions generated by the Basic Graph Patterns translation. The SPARQL query consists of two *BGPs* (i.e., *BGP₁* and *BGP₂*). Regarding *BGP₂*, none of the rewriting rules can be applied on the generated XQuery expressions (i.e., one For and one Return clause). We present below the rewriting of the XQuery expressions generated by the translation of *BGP₁*.

Applying the *Changing For Clauses to Let Rule* (Rewriting Rule 1):

| *Initial XQuery Expressions* | | *Rewritten by Rule 1 XQuery Expressions* |
|---|---|---|

*let $doc := collection("http://www.music.tuc.gr/...")*    ⟹    *let $doc := collection("http://www.music.tuc.gr/...")*

*for $stud in $doc/Persons/Student[./FirstName = "John" ]*          *for $stud in $doc/Persons/Student[./FirstName = "John" ]*

*for $SSN in $stud/@SNN*          **let** *$SSN := $stud/@SNN*

*for $lname in $stud/LastName[matches( . , "^B" )]*          *for $lname in $stud/LastName[matches( . , "^B" )]*

*let $age := $stud/Age[.>25]*          *let $age := $stud/Age[.>25]*

*return(<Result><SSN>{ string($SSN) }</SSN> , ... </Result>)*          *return(<Result><SSN>{ string($SSN) }</SSN> , ... </Result>)*

Applying the *Reducing Let Clauses Rule* (Rewriting Rule 2):

| *Rewritten by Rule 1 XQuery Expressions* | | *Rewritten by Rule 2 XQuery Expressions* |
|---|---|---|

*let $doc := collection("http://www.music.tuc.gr/...")*    ⟹    ~~let $doc := collection("http://www.music.tuc.gr/...")~~

*for $stud in $doc/Persons/Student[./FirstName = "John" ]*          *for $stud in* **collection("http://www.music.tuc.gr/...")**

                            */Persons/Student[./FirstName = "John" ]* **[Age[.>25]]**

*let $SSN := $stud/@SNN*          ~~let $SSN := $stud/@SNN~~

*for $lname in $stud/LastName[matches( . , "^B" )]*          *for $lname in $stud/LastName[matches( . , "^B" )]*

*let $age := $stud/Age[.>25]*          ~~let $age := $stud/~~

*return(<Result><SSN>{ string($SSN) }</SSN> , ... </Result>)*          *return(<Result><SSN>{string(***$stud/@SNN***)}</SSN>, ... </Result>)*

Applying the *Unnesting For Clauses Rule* (Rewriting Rule 3):

         *No effect.*

Rewritten XQuery expressions:

         *for $stud in collection("http://www.music.tuc.gr/...") /Persons/Student[./FirstName = "John" ][ Age[.>25]]*

         *for $lname in $stud/LastName[matches( . , "^B" )]*

         *return(<Result><SSN>{string($stud/@SNN)}</SSN>, ... </Result>)*

## 12.6 Resulting XQuery Query

The resulting XQuery query is presented below.

**Translated XQuery Query**

```
let $doc := collection("http://www.music.tuc.gr/...")
let $Modified_Results :=(
    let $Results :=(
        for $stud  in    collection("http://www.music.tuc.gr/...")/Persons/Student[./FirstName = "John" ][ Age[.>25]]
        for $lname in  $stud/LastName[matches( . , "^B" )]
        let $BGP_2 :=(
            for $email in  $stud/Email
            return( <Result> <email>{ string($email) }</email> </Result> )
        )
        return(
            if( exists($BGP_2) ) then(
                for $bgp2_it in $BGP_2
                return (<Result><SSN>{ string($stud/@SNN) }</SSN> , <lname>{ string($lname) }</lname>, {$bgp2_it/Result} </Result>) )
            else(
                (<Result> <SSN>{ string($stud/@SNN) }</SSN> , <lname>{ string($lname) }</lname> </Result>) )
        )
    )
return ( let $Ordered_Results :=(
                        for $iter in $Results
                        order by $iter/lname empty least , $iter/SSN descending empty least
                        return($iter) )
            return ($Ordered_Results[position( )>5 and position( )<=35]) )
)
return ( <Results>{ $Modified_Results }</Results> )
```

## 13   EXPERIMENTAL EVALUATION

In this section we present the results of the experimental evaluation that we have conducted on the SPARQL2XQuery Framework using both synthetic and real datasets. The objective was to evaluate the efficiency of: (a) schema transformation; (b) mapping generation; (c) query translation; and (d) query evaluation. We have used several query sets attempting to cover almost all the SPARQL syntax variations, features and special cases.

The SPARQL2XQuery Framework has been implemented using Java related technologies (Java 2SE and Jena) on top of the open source, native XML database. The experimental evaluation was performed on an Intel Xeon processor at 2.00 Ghz, with 16 GB RAM, running Linux and Java 1.6. We have used two native XML Databases (and their XQuery engines) denoted as "XML Store Y" and "XML Store Z". In addition, we have used a memory-based XQuery engine denoted as "Memory-based XQuery Engine". For RDF store, we have used the Jena TDB 0.10.1 storage component and the Jena ARQ 2.10.1 SPARQL engine. Finally, for the evaluation of the Xs2OwL component we used two XSLT processors, a freeware XSLT processor denoted as "Freeware XSLT Processor", and the XSLT processor that is integrated in a commercial tool, denoted as "Commercial XSLT Tool ". Note that, in all experiments, the default configurations for all the software have been used.

The rest of this section is structured as follows. We discuss the performance of the schema transformation and mapping generation processes in Section 13.1, we examine the efficiency of the translation process in Section 13.2, we present the query evaluation efficiency in Section 13.3 and we provide an evaluation overview in Section 13.4.

### 13.1 Schema Transformation and Mapping Generation Performance

#### 13.1.1 Schemas

In order to evaluate the SPARQL2XQuery Framework, we have used several international standards from different domains (e.g., Digital Libraries, Cultural Heritage, Multimedia, etc.) that have been expressed in XML Schema. The Persons XML Schema that we have defined in Section 3.2 has also been used. These XML Schemas have been used in order to evaluate the schema transformation and mapping generation processes. The basic characteristics (e.g., number of elements, attributes, etc.) of the XML Schemas used in the evaluation can be found in [120].

#### 13.1.2 Evaluation Results

Here we present the results of the experiment we conducted in order to study the schema transformation and mapping generation performance. Both the schema transformation and the mapping generation processes are off-line processes and are performed once[25] for every XML Schema in the context of first scenario (*Querying XML data based on automatically generated ontologies*). Although these processes are off-line and are performed once for every XML Schema, we can observe from this experiment that we can characterize them as lightweight processes that take negligible time even for very large XML Schemas (e.g., schemas with 4000 XML Schema constructs).

**Table 10. Schema Transformation & Mapping Generation Time** (msec)

| XML Schema | Schema Transformation Time | | Mapping Generation Time |
|---|---|---|---|
| | Freeware XSLT Processor | Commercial XSLT Tool | |
| **Persons** (Section 3.2) | 2.4 | 17.9 | 8.7 |
| **DBLP** [26] | 62.5 | 22.5 | 360.9 |
| **METS** [18] | 58.2 | 270.5 | 388.9 |
| **Text MD** [19] | 7.7 | 45.1 | 14.5 |
| **MPEG-7** [20] | 730.7 | 3500.6 | 1954.2 |
| **SCORM 12** [29] | 132.7 | 415.2 | 421.1 |
| **MARC 21** [22] | 6.3 | 51.4 | 12.5 |
| **MODS** [23] | 191.3 | 594.8 | 482.3 |
| **TEI** [24] | 840.0 | 980.1 | 2208.4 |
| **TEI Lite** [24] | 418.0 | 932.6 | 1288.3 |
| **EAD** [25] | 402.7 | 3305.7 | 1052.0 |
| **VRA Core 4** [26] | 47.3 | 290.0 | 304.3 |
| **VRA Core 4 Strict** [26] | 3.3 | 122.1 | 10.0 |
| **MIX** [27] | 200.0 | 601.3 | 495.5 |
| **MADS** [30] | 50.1 | 393.4 | 345.6 |

In this experiment we have used several international standards that have been expressed in XML Schema. For each of these XML Schemas, we have used the Xs2Owl component in order to automatically transform the XML Schema in OWL ontologies, measuring the time required for this transformation (*Schema Transformation Time*). Then, using the generated

---

[25] The schema transformation and mapping generation processes may be applied more than once on the same XML Schema in case of schema updates.

[26] Note that in our experiments, the DTD that originally describes the DBLP dataset has been expressed in XML Schema syntax. The DBLP DTD as well as the DBLP dataset are available at: *http://dblp.uni-trier.de/xml/*.

Schema ontology and the XML Schema, we measure the time required for the Mapping Generator component of the SPARQL2XQuery Framework to automatically discover and generate the mappings (*Mapping Generation Time*).

Table 10 presents the *Schema Transformation Time* and the *Mapping Generation Time* for each XML Schema. Notice that the schema transformation time is presented for both Freeware XSLT Processor and Commercial XSLT Tool. The schema transformation time mainly depends on: (a) The number of the XML Schema constructs, since this number corresponds to the number of the transformations performed; and (b) The size of the XML Schema file, since it should be parsed. Similarly, the mapping generation time basically depends on: (a) The number of the XML Schema constructs, since this number equals to the number of the generated mappings; and (b) The size of the XML Schema and ontology files, since these files should be parsed.

We can observe from Table 10 that, for both the XSLT processors, the TEI and MPEG-7 require the maximum transformation time (840.0 and 730.7 msec respectively) due to their large number of XML Schema constructs (4279 and 2567 constructs respectively, see [120]). On the other hand, due to the small number of XML Schema constructs, the Persons (13 constructs, [120]) and VRA Core 4 Strict (19 constructs, [120]) require the minimum transformation time (2.4 and 3.3 msec respectively). Finally, as at is expected, the XML Schema file size slightly affects the transformation time. For example, despite the large size (345.3 Kb, [120]) of the SCORM 21 XML Schema file, the transformation time is not analogously high (132.7 msec) due to its small number of XML Schema constructs (i.e., small number of transformations).

In addition, we observe that the TEI and MPEG-7 require the maximum mapping generation time (2208.4 and 1954.2 msec respectively) due to their large number of XML Schema constructs (i.e., number of mappings discovered and generated). On the other hand, the Persons and VRA Core 4 Strict require the minimum mapping generation time (8.7 and 10.0 msec respectively).

## 13.2 Translation Efficiency

In this section we present the experimental results related to the efficiency of the SPARQL to XQuery translation process. To evaluate the efficiency of the translation process, we measured the translation time required by the SPARQL2XQuery Framework to translate a SPARQL query to an XQuery query. Below, we present three experiments. In the first experiment (Section 13.2.1.1), we have generated several SPARQL queries by modifying their graph pattern size and type. In the second experiment (Section 13.2.1.2), we have varied in the previously generated queries the number of mappings between the ontology and the XML Schema. Finally, in the third experiment we have employed three SPARQL query sets attempting to cover all the SPARQL grammar variations (Section 13.2.2).

### 13.2.1 Translation Time for different Graph Patterns & Mappings

Here, we examine the efficiency of the query translation process. The translation time mainly depends on two factors:

    (a) The number of the SPARQL variables included in the Graph Pattern, since the SPARQL variable number determines: (i) the number of the XQuery clauses generated throughout the translation; (ii) the number of the required *Variable Binding* phases; and (iii) the number of the required *Variable Type Determination* phases.

    (b) The complexity of the variable binding determination process. In particular, the complexity of the variable binding determination depends on: (i) the number of the XPath Set operations; (ii) the type of the XPath Set operations; and (iii) the size of the operands (i.e., the size of the XPath Sets).

In the first experiment, we have generated several SPARQL queries by modifying the size and the type of their graph patterns. For the SPARQL query generation, we assumed that the queries are expressed on an ontology that has been mapped to an XML Schema. We also assume that the ontology has the properties *ns:Pi* with $1 \leq i \leq 30$ (i.e., *ns:P1*, *ns:P2*,…, *ns:P30*), where *ns* is the namespace of the ontology. In the second experiment, for each of the generated SPARQL queries we have varied the number of the predefined mappings (i.e., the XPath Set sizes) between the ontology and the XML Schema.

Note that the queries generated for these experiments are Select SPARQL queries, containing one *return variable* (*Definition 18*) and their Where clause is a Graph Pattern consisting of sequences of conjunctive triple patterns (i.e., Basic Graph Pattern).

### 13.2.1.1  *Varying the Graph Pattern Type and Size*

In this experiment, we have obtained several (different) SPARQL queries by modifying the type and the size of their graph pattern. To this end, we have varied (a) the number; and (b) the type of the triple patterns included in the graph pattern. The number of triple patterns determines the number of SPARQL variables and, as a consequence, the number of the generated XQuery clauses. The triple pattern type determines: (a) the number of the SPARQL variables; (b) the number of the XPath Set operations; and (c) the type of the XPath Set operations.

We have defined four types of graph patterns ($GP_1$, $GP_2$, $GP_3$ and $GP_4$) by modifying the types of the included triple patterns: **(a)** $GP_1$=$?x_1$ *ns:P1* $?y_1$ . $?x_2$ *ns:P2* $?y_2$ .$\cdots$ .$?x_n$ *ns:Pn* $?y_n$ **(b)** $GP_2$=$?x_1$ *ns:P1 "abc"* . $?x_2$ *ns:P2 "abc"* . $\cdots$ .$?x_n$ *ns:Pn "abc"* **(c)** $GP_3$=$?x_1$ $?y_1$ $?z_1$ . $?x_2$ $?y_2$ $?z_2$ . $\cdots$ .$?x_n$ $?y_n$ $?z_n$ and **(d)** $GP_4$=$?x_1$ $?y_1$ *"abc"* . $?x_2$ $?y_2$ *"abc"* . $\cdots$ .$?x_n$ $?y_n$ *"abc"*, where *n* is the number of triple patterns. Table 11 presents the basic characteristics of the SPARQL to XQuery translation for the previous graph pattern types. The last column refers to the XPath Set operations occurring in the variable binding phase.

**Table 11. Translation Characteristics over the Number of Tripple Patterns** (*n*)

| Graph Pattern Type | Characteristics w.r.t. Number of Triple Patterns (*n*) | | |
|---|---|---|---|
| | SPARQL Variables | Generated XQuery Clauses | XPath Set Operations |
| $GP_1$ | $2n$ | *2n* For/Let, 1 Where and 1 Return | $2n$ ($\overline{\sqcap}$) and $n$ ($\prec$) |
| $GP_2$ | *n* | *n* For/Let, 1 Where and 1 Return | $n$ ($\overline{\sqcap}$) |
| $GP_3$ | $3n$ | *3n* For/Let, 1 Where and 1 Return | $2n$ ($\overline{\sqcap}$), $n$ ($\succ$) and $n$ ($\prec$) |
| $GP_4$ | $2n$ | *2n* For/Let, 1 Where and 1 Return | $2n$ ($\overline{\sqcap}$) and $n$ ($\succ$) |

**Table 12. Query Translation Time & SPARQL Parsing Time vs. Graph Pattern Type and Size**

| Query Translation Time [*SPARQL Parsing Time*] | | | | | | | |
|---|---|---|---|---|---|---|---|
| Graph Pattern Type | Number of Triple Patterns (*n*) | | | | | | |
| | 1 | 3 | 7 | 10 | 15 | 20 | 30 |
| $GP_1$ | 2.09 [*0.14*] | 2.13 [*0.15*] | 2.17 [*0.17*] | 2.20 [*0.66*] | 2.37 [*0.69*] | 2.91 [*0.70*] | 3.93 [*0.72*] |
| $GP_2$ | 2.07 [*0.38*] | 2.07 [*0.37*] | 2.11 [*0.38*] | 2.13 [*0.39*] | 2.29 [*0.42*] | 2.79 [*0.46*] | 3.81 [*0.65*] |
| $GP_3$ | 3.22 [*0.22*] | 3.26 [*0.24*] | 3.28 [*0.29*] | 3.39 [*0.32*] | 3.74 [*0.37*] | 3.89 [*0.41*] | 4.25 [*0.46*] |
| $GP_4$ | 3.21 [*0.21*] | 3.26 [*0.24*] | 3.29 [*0.28*] | 3.35 [*0.30*] | 3.64 [*0.31*] | 3.76 [*0.34*] | 4.04 [*0.40*] |
| Average | 2.65 [*0.24*] | 2.68 [*0.25*] | 2.71 [*0.28*] | 2.76 [*0.42*] | 3.01 [*0.45*] | 3.34 [*0.48*] | 4.01 [*0.56*] |

For each of the above graph pattern types ($GP_1$, $GP_2$, $GP_3$ and $GP_4$), we have constructed graph patterns containing *n* triple patterns with $n = 1, 3, 7, 10, 15, 20, 30$. Finally, for each ontology property *ns:Pi*, we have assumed the mapping *ns:Pi* $\equiv$ {/a/b/i}. The translation time required by the SPARQL2XQuery Framework for the SPARQL to XQuery translation of each query is presented in Table 12. The *SPARQL parsing time* is also presented in Table 12. As *SPARQL Parsing Time* we refer to the time required by an SPARQL query engine to parse the SPARQL query and build the query object. Note that, in order to translate the SPARQL queries, a parsing phase using a SPARQL engine is required. As a result, in Table 12 the translation time *a [b]* means that the total translation time is *a* msec and it includes the SPARQL parsing time, which is *b* msec.

We can observe from Table 12 that the *GP₂* type has achieved the lowest translation time. This is due to the fact that *GP₂* contains only one variable. On the other hand, the *GP₃* type has taken the maximum translation time, since the *GP₃* contains the maximum number of variables in the triples and as a result, a large number of variable binding and determination of variable type phases are required.

In particular, the single variable included in the triple patterns of the *GP₂* graph patterns has resulted in the generation of a small number of XQuery clauses. For a query with *n* triple patterns, *n+2* XQuery clauses (*n* For/Let, one Where and one Return) have been generated (Table 11). Accordingly, the three variables included in the triple patterns of the *GP₃* graph patterns have resulted in a large number (i.e., *3n+2*) of XQuery clauses.

Finally, the triple patterns of the *GP₁* and *GP₄* graph patterns include two variables. Throughout the translation, *2n+2* XQuery clauses have been generated. Despite the same number of generated XQuery clauses, variable binding and variable type determination phases, the *GP₁* type has achieved lower translation time than the *GP₄* type. This is explained as follows: The variable binding determination process for the *GP₄* graph patterns is of higher complexity than that of *GP₁*, since the predefined mappings for the *Pi* properties in the *GP₁* graph patterns reduce the number of possible bindings, and, therefore, the complexity of determining the variable bindings decreases (see Section 7 for details).

*13.2.1.2    Varying the Number of Mappings*

In this experiment we have used the SPARQL queries generated in the previous experiment. In addition, we have varied here the number of the predefined mappings between the ontology and the XML Schema; In particular, we have modified the number of the mappings (i.e., the size of XPath Sets) for all the ontology properties *Pi*, and, therefore we have modified the complexity of the variable binding phase.

In this experiment, for each property *Pi* we have assumed the mapping $Pi \equiv \{/a/b/c\_i\}$, which contains one XPath expression. We have then modified the number of the XPath expressions that correspond to each *Pi* mapping. Hence, for each property *Pi*, we have the mapping $Pi \equiv \{/a/b/i\_1, /a/b/i\_2,..., /a/b/i\_k \}$, with *k*=1, 2, 3, 5 being the number of XPath expressions for each *Pi* mapping.

Figure 8 presents the query translation time with a varying number of XPath expressions per mapping. Each diagram of Figure 8 corresponds to a specific number (*n*) of triple patterns and depicts all the graph pattern types (*GP₁*, *GP₂*, *GP₃* and *GP₄*), while varying the number of XPath expressions from 1 to 5.

As it is expected, the number of XPath expressions per mapping had no effect in the query translation time for a small number of triple patterns (Figure 8 *(a)* and *(b)*), because of the very low translation time required and the small number of operations involved. In particular, for *n*=1 and *n*=3 triple patterns the translation time for all the graph pattern types remains stable as the number of XPath expressions increases. For graph patterns containing *n*=7 triple patterns (Figure 8 *(c)*) we observe only a slight increase in the query translation time as the number of XPath expressions increases. For *n*>7, as the number of XPath expressions increases, the query translation time grows linearly for all the graph pattern types. The former is explained as follows: Increasing the number of XPath expressions results in the (analogous) increase of the iterations for parsing and processing the XPath Sets.

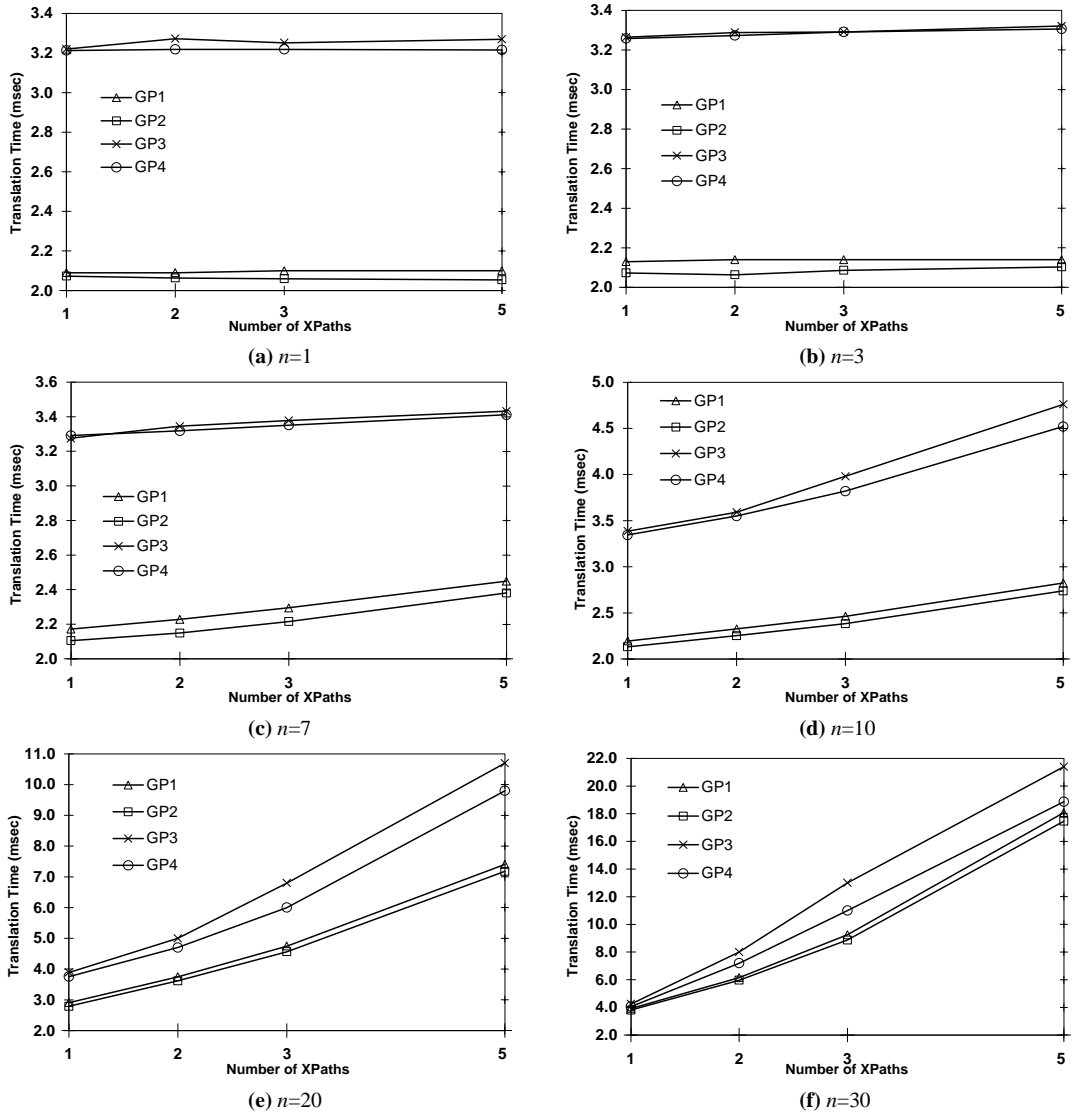

**Figure 8: Query Translation Time vs. Number of Mappings (for *n*=1, 2, 3, 7, 10, 20 and 30 triple patterns)**

*13.2.2    Translation Time for the Persons, DBLP & Berlin Query Sets*

*13.2.2.1    Query Sets*

In this section we present the three query sets that have been exploited in our experiments. The *first query set* comprises the 12 SPARQL queries of the *Berlin SPARQL Benchmark* [117]. An overview of the SPARQL features that are used by the Berlin query set can be found in [120]. The *second query set* (Persons Queries) contains 15 SPARQL queries based on the Persons ontology (Table 4 and Table 5). The *third query set* (DBLP Queries) contains 5 SPARQL queries based on the DBLP ontology. The last two query sets have been used for evaluating our system in terms of: (a) the translation time; and (b) the query evaluation time.

The second and the third SPARQL query sets attempt to cover almost all the SPARQL grammar variations, features and special cases, with varying SPARQL query types (e.g., Select, Ask, etc.), graph patterns with different sizes and complexity. These queries use all the SPARQL algebra operators (e.g., Optional, Union, Filter, etc.), exploit combinations of the solution sequence modifiers (e.g., Limit, Offset, Order by, etc.) and contain several other features (e.g., Built-in functions, Schema triples, complex Filter conditions, etc.). In our attempt to cover almost all the possible SPARQL syntax variations and special cases, we have also considered the existing SPARQL Benchmarks (*Berlin SPARQL Benchmark*[27] [117], *SP²Bench* [118], *W3C SPARQL Implementation Coverage Report*[28] and *W3C DAWG Test cases*[29]) throughout the query set specification.

All the SPARQL queries, the translated XQuery queries, as well as an analysis of their characteristics and features can be found in [120].

### *13.2.2.2    Evaluation Results*

In this experiment we have evaluated the efficiency of the translation process by exploiting several different SPARQL queries. We have utilized three SPARQL query sets, attempting to cover almost all the SPARQL syntax variations, features and special cases. For each query, we have measured the translation time required by the SPARQL2XQuery Framework to translate the SPARQL query in XQuery expressions. The query translation time and the SPARQL parsing time as well as the average parsing and translation time for each query set are presented in Table 13.

Table 13 (a) presents the translation times for the 15 queries of the Persons query set. We can observe from [120] that the queries of the Persons query set have in average 4 triple patterns per query and 1–2 XPaths per mapping. In addition, some of these queries contain one or more solution sequence modifiers and Schema Triples. The translation of the solution sequence modifiers and the Schema Triples has made the translation time for the queries of the Persons query set slightly higher than the translation time of the queries of the previous experiment, since the later have neither solution sequence modifiers nor Schema Triples.

The translation time for the queries of the DBLP query set are presented in Table 13 (b). These queries have some characteristics similar to the ones of the Persons query set (i.e., in average 4 triple patterns per query and 1–2 XPaths per mapping). However, the DBLP queries are more complex in order to encapsulate most of the SPARQL features in five queries, thus resulting in slightly higher translation times compared to the ones of the Persons query set.

Finally, Table 13 (c) presents the translation times for the Berlin query set. This query set is the most complex, with an average of 8 triple patterns per query, 1–4 XPaths per mapping, several solution sequence modifiers per query and several OPTIONAL and FILTER operators. The highest translation times occur in Queries 7 and 8 with 14 and 10 triple patterns respectively, 4 OPTIONAL operators, FILTERs and solution sequence modifiers.

---

**Table 13. Query Translation & SPARQL Parsing Time** (msec) **for (a)** Person, **(b)** DBLP **and (c)** Berlin **Query Sets**

| (a) Person **Query Set** | | | (b) DBLP **Query Set** | | | (c) Berlin **Query Set** | | |
|---|---|---|---|---|---|---|---|---|
| **Persons Query** | **Translation Time** | **SPARQL Parsing Time** | **DBLP Query** | **Translation Time** | **SPARQL Parsing Time** | **Berlin Query** | **Translation Time** | **SPARQL Parsing Time** |
| **Q₁** | 3.35 | 0.90 | **Q₁** | 5.73 | 1.2 | **Q₁** | 4.04 | 1.29 |
| **Q₂** | 3.35 | 0.80 | **Q₂** | 4.19 | 1.4 | **Q₂** | 13.82 | 0.90 |
| **Q₃** | 3.31 | 0.97 | **Q₃** | 7.70 | 1.2 | **Q₃** | 10.54 | 0.88 |
| **Q₄** | 3.32 | 0.74 | **Q₄** | 7.62 | 1.0 | **Q₄** | 7.26 | 0.82 |
| **Q₅** | 3.34 | 0.62 | **Q₅** | 3.89 | 0.6 | **Q₅** | 3.85 | 0.99 |
| **Q₆** | 3.30 | 0.50 | **Avg.** | **5.83** | **1.1** | **Q₆** | 3.61 | 0.50 |
| **Q₇** | 3.32 | 0.87 | | | | **Q₇** | 16.11 | 0.79 |
| **Q₈** | 6.23 | 0.49 | | | | **Q₈** | 19.02 | 0.71 |
| **Q₉** | 6.46 | 0.68 | | | | **Q₉** | 3.55 | 0.28 |
| **Q₁₀** | 3.26 | 0.34 | | | | **Q₁₀** | 3.70 | 0.51 |
| **Q₁₁** | 3.30 | 0.39 | | | | **Q₁₁** | 6.63 | 0.29 |
| **Q₁₂** | 3.29 | 0.39 | | | | **Q₁₂** | 3.72 | 0.48 |
| **Q₁₃** | 3.28 | 0.50 | | | | **Avg.** | **7.99** | **0.70** |
| **Q₁₄** | 3.26 | 0.32 | | | | | | |
| **Q₁₅** | 3.26 | 0.29 | | | | | | |
| **Avg.** | **3.71** | **0.59** | | | | | | |

## 13.3 Query Evaluation Efficiency

In this section we present the experimental results that refer to the efficiency of evaluating the XQuery expressions generated by the SPARQL2XQuery Framework. In Section 13.3.1 we outline the datasets, queries and metrics that are used in our evaluation scenario. In the first part of this experiment (Section 13.3.2) we have employed the synthetic Persons XML dataset and the Persons query set. In the second part (Section 13.3.3) we have utilized the real DBLP XML dataset and the corresponding query set.

### 13.3.1    Methodology

**Datasets.** In order to evaluate the SPARQL2XQuery Framework in term of query evaluation efficiency, we have used both real and synthetic datasets.

The real dataset we have employed is the XML DBLP dataset[26]. The characteristics of the DBLP dataset have been presented in [118]. The size of the DBLP dataset is 833Mb. First we have manually expressed the DTD that describes the DBLP dataset in XML Schema syntax. Then, the XML Schema has been transformed to an OWL ontology using the XS2OWL component. The DBLP XML Schema and the ontology generated by XS2OWL are available in [120].

Our synthetic dataset is structured according to the Persons XML Schema (Figure 3). The SPARQL queries expressed on it are based on the Persons OWL ontology generated for this XML Schema by the XS2OWL component (Table 4 and Table 5). For the generation of the synthetic XML dataset that follows the Persons XML Schema, we have implemented a data generator that takes as input a factor $N$, which is the number of the records to be generated (details regarding the generator implementation can be found in [120]). Finally, all the Persons XML datasets have been transformed in RDF format, in order to be able to perform a native evaluation of the SPARQL queries on them.

Table 14 summarizes the basic features of the Persons XML datasets, including the size in Kilobytes, the approximate number of XML nodes, etc. We have generated 10 datasets ($DT_1$ to $DT_{10}$), varying the $N$ factor from $10^2$ to $5 \cdot 10^6$. In addition, Table 14 presents the characteristics of the corresponding RDF datasets (i.e., number of triples and size in Kilobytes) that have been generated from the XML dataset transformation.

**Table 14. Characteristics of the Persons XML Datasets $DT_1$ to $DT_{10}$ and the Corresponding RDF Datasets**

| Dataset Name | XML Dataset Characteristics | | | Corresponding RDF Dataset Characteristics | |
|---|---|---|---|---|---|
| | $N$ | XML Nodes | Size (Kb) | Triples | Size (Kb) |
| $DT_1$ | $10^2$ | 1450 | 20 | $6 \cdot 10^2$ | 40 |
| $DT_2$ | $5 \cdot 10^2$ | 7250 | $10^2$ | $3 \cdot 10^3$ | $2 \cdot 10^2$ |
| $DT_3$ | $10^3$ | $145 \cdot 10^2$ | $2 \cdot 10^2$ | $6 \cdot 10^3$ | $4 \cdot 10^2$ |
| $DT_4$ | $5 \cdot 10^3$ | $725 \cdot 10^2$ | $10^3$ | $3 \cdot 10^4$ | $2 \cdot 10^3$ |
| $DT_5$ | $10^4$ | $145 \cdot 10^3$ | $2 \cdot 10^3$ | $6 \cdot 10^4$ | $4 \cdot 10^3$ |
| $DT_6$ | $5 \cdot 10^4$ | $725 \cdot 10^3$ | $10^4$ | $3 \cdot 10^5$ | $2 \cdot 10^4$ |
| $DT_7$ | $10^5$ | $145 \cdot 10^4$ | $2 \cdot 10^4$ | $6 \cdot 10^5$ | $4 \cdot 10^4$ |
| $DT_8$ | $5 \cdot 10^5$ | $725 \cdot 10^4$ | $10^5$ | $3 \cdot 10^6$ | $2 \cdot 10^5$ |
| $DT_9$ | $10^6$ | $145 \cdot 10^5$ | $2 \cdot 10^5$ | $6 \cdot 10^6$ | $4 \cdot 10^5$ |
| $DT_{10}$ | $5 \cdot 10^6$ | $725 \cdot 10^5$ | $10^6$ | $3 \cdot 10^7$ | $2 \cdot 10^6$ |

**Queries.** In our evaluation scenario, every SPARQL query $Q_s$ of the Persons and DBLP query sets (Section 13.2.2.1), has been automatically translated by the SPARQL2XQuery Framework to the XQuery query $Q_{Xa}$. Moreover, $Q_S$ has been independently manually translated by an external expert to the XQuery $Q_{Xm}$. The $Q_{Xm}$ queries have been expressed considering the XML Schema semantics and after applying techniques aiming to provide efficient XQuery queries. Finally, the rewriting rules defined in Section 10 have been applied on the automatically generated XQuery queries ($Q_{Xa}$), to obtain the automatically rewritten XQuery queries $Q_{Xa-Rw}$.

**Evaluation Metrics.** In order to study the efficiency of the XQuery queries generated by the SPARQL2XQuery Framework, we have measured and compared the query evaluation times for (a) the original SPARQL queries, natively executed using a SPARQL engine; (b) the automatically generated ($Q_{Xa}$) XQuery queries; (c) the automatically rewritten ($Q_{Xa-Rw}$) XQuery queries; and (d) the manually translated ($Q_{Xm}$) XQuery queries. Note that the XQuery evaluation times heavily rely on the underling XML data management system (e.g., storage, indexing, query engine, query optimizer, configuration, etc.).

### 13.3.2 Synthetic Dataset

In this experiment we study the efficiency of the XQuery queries generated by the SPARQL2XQuery Framework using synthetic datasets (Table 14). We have measured and compared the query evaluation times of the automatically generated, rewritten and manually translated XQuery queries.

In the rest of this section, we analyze the evaluation times for each query (Section 13.3.2.1), we vary the dataset size in order to examine the query evaluation efficiency over the dataset size (Section 13.3.2.3) and we compare the query evaluation time with the query translation time (Section 13.3.2.3).

### 13.3.2.1 Query Evaluation Time Analysis

We have used the synthetic Persons dataset $DT_8$ (Section 13.3.1) and the Persons query set (Section 13.2.2.1). The $DT_8$ dataset comprises $5 \cdot 10^5$ records of persons and students (250,000 persons and 250,000 students), is of size $10^5$ Kb and has approximately $725 \cdot 10^4$ XML nodes.

Table 15 summarizes the results of the comparison of the execution of the SPARQL as well as the automatically generated, rewritten and manually translated XQuery queries. In particular, for each query, Table 15 contains the evaluation times for (a) the SPARQL queries (denoted as SPARQL); (b) the manually translated XQuery queries (denoted as Manual); (c) the automatically rewritten XQuery queries (denoted as Auto-Rw); and (d) the automatically generated (without rewriting)

XQuery queries (denoted as Auto). In addition, Table 15 presents the improvement of the rewritten compared to the automatically generated queries (denoted as Auto-Rw vs. Auto) as well as the comparison between the automatically rewritten and manually translated XQuery queries (denoted as Auto-Rw vs. Manual). The measuring unit for the evaluation time is second (sec).

**Table 15. Query Evaluation Time over the Persons *DTs* Dataset (XML Store Y)**

| | **Query Evaluation Time** (sec) | | | | | |
|---|---|---|---|---|---|---|
| **Query** | **SPARQL** $(Q_S)$ | **Manual** $(Q_{Xm})$ | **Auto-Rw** $(Q_{Xa\text{-}Rw})$ | **Auto** $(Q_{Xa})$ | **Auto-Rw vs. Auto** | **Auto-Rw vs. Manual** |
| **Q₁** | 1.66 | 5.95 | 4.30 | 6.78 | 57.7 % | 27.7 % |
| **Q₂** | 1.69 | 5.96 | 4.28 | 6.76 | 57.8 % | 28.1 % |
| **Q₃** | 1.53 | 0.41 | 0.42 | 0.45 | 7.6 % | -1.0 % |
| **Q₄** | 2.78 | 10.79 | 11.00 | 11.08 | 0.7 % | -1.9 % |
| **Q₅** | 10.83 | 55.70 | 55.77 | 63.97 | 14.7 % | -0.1 % |
| **Q₆** | 1.55 | 6.55 | 6.49 | 6.89 | 6.1 % | 0.9 % |
| **Q₇** | 1.36 | 0.91 | 0.92 | 0.93 | 1.2 % | -0.2 % |
| **Q₈** | 6.03 | 12.93 | 13.09 | 13.11 | 0.2 % | -1.3 % |
| **Q₉** | 5.34 | 3.21 | 3.22 | 5.76 | 79.1 % | -0.3 % |
| **Q₁₀** | 0.00 | 6.63 | 5.74 | 6.91 | 20.4 % | 13.4 % |
| **Q₁₁** | 21.74 | 14.89 | 15.07 | 16.47 | 9.3 % | -1.2 % |
| **Q₁₂** | 2.44 | 15.47 | 15.49 | 15.74 | 1.6 % | -0.1 % |
| **Q₁₃** | 0.00 | 0.23 | 0.24 | 0.25 | 5.5 % | 2.1 % |
| **Q₁₄** | 1.37 | 3.69 | 3.61 | 3.80 | 5.2 % | 2.2 % |
| **Q₁₅** | 2.74 | 9.14 | 15.69 | 15.88 | 1.2 % | -71.7 % |
| **Average** | **4.07** | **10.17** | **10.36** | **11.65** | **12.5 %** | **-1.9 %** |

**Automatically Rewritten vs. Automatically Generated (Auto-Rw vs. Auto).** We can observe from Table 15 that for almost all the queries the evaluation times for the rewritten queries presented a notable performance improvement compared to the automatically generated ones. The average reduction in the evaluation time for the rewritten queries was 12.5%, and the maximum 79.1%.

For five ($Q_4$, $Q_7$, $Q_8$, $Q_{12}$ and $Q_{15}$) out of fifteen queries, the query evaluation time was almost the same for the rewritten and the automatically generated queries (with a time decrease between 0.2% and 1.6 %). For seven queries ($Q_3$, $Q_5$, $Q_6$, $Q_{10}$, $Q_{11}$, $Q_{13}$ and $Q_{14}$), the rewritten queries have presented a slight improvement with an evaluation time decrease between 5.2% and 20.4% compared to the automatically generated ones. Finally, three queries ($Q_1$, $Q_2$, and $Q_9$) have presented a significant performance improvement with a time decrease between 57.7% and 79.1%. In more detail:

– For the queries $Q_4$, $Q_8$, $Q_{13}$ and $Q_{15}$, the only difference between the rewritten the automatically generated queries, is that the rewritten have one Let XQuery clause less. In particular, in the rewritten queries the Let clause that is used to assign the XML data on which the query is evaluated (i.e., *let $doc := collection(…)* ), has been removed. The XML data declaration (i.e., *collection(…)*) is directly used instead of the *$doc* XQuery variable. Hence, it is expected that the evaluation of these queries has not shown any significant efficiency improvement.

– For the queries $Q_1$, $Q_2$, $Q_3$, $Q_6$, $Q_7$ and $Q_{10}$ the rewriting rule *Rule 3* (*Unnesting For Clauses*) has been applied, which removes one For clause in each query, thus resulting in a less nested For clause. For the queries $Q_3$, $Q_6$ and $Q_7$, the improvement of the rewritten queries is not significant (1.2% to 7.6 %), since the outer loops (i.e., outer For clauses) are restricted with conditions (i.e., predicates over the XPath of the For clause) resulting into very few inner loops. For the queries $Q_1$ and $Q_2$, though, which have no restrictions in the For clauses, the improvement is significant (57%). Finally, we expected that the same should hold for the query $Q_{10}$; however, its improvement was not as significant as expected (20% improvement). This may happen, because this query returns only the first 100 of the results, thus, the query engine possibly selects an efficient execution plan (although the query optimizer has been turned-off).

–  For the queries $Q_{13}$ and $Q_{14}$ the rewriting rule *Rule 2* (*Reducing Let Clauses*) has been applied, which removes one Let clause from each query. As is expected, this rewriting resulted in a slight improvement of 5%.

–  Finally, for the queries $Q_5$ and $Q_9$, the rewriting rule *Rule 1* (*Changing For Clauses to Let*) has been initially applied. *Rule 1* considers the exact cardinality of one in the SSN attribute and the Age element. As a result, two For clauses in each query are transformed to Let clauses. Then, the *Rule 2* has been applied on the queries.

   From query $Q_5$, the two Let clauses that resulted from the application of *Rule 1* on the For clause have been removed. In addition, from query $Q_9$ four Let clauses have been removed (two of them have resulted from the application of *Rule 1* on the For clause). Compared to the initial queries, the query $Q_5$ has two For clauses less and the query $Q_9$ has two For and two Let clauses less. These rewritings have resulted in a considerable improvement of 14.7% and 79.1% for the queries $Q_5$ and $Q_9$ respectively.

**Automatically Rewritten vs. Manually Translated (Auto-Rw vs. Manual).** The evaluation times of the automatically rewritten queries are very close to the ones of the manually translated queries as shown in Table 15, with an average increase of 1.9%.

For three ($Q_1$, $Q_2$ and $Q_{10}$) out of the fifteen queries, the rewritten queries have considerably outperformed the manually translated ones, with an evaluation time decrease between 13.4% and 28.1%. In addition, in other cases ($Q_6$, $Q_9$, $Q_{13}$ and $Q_{14}$), the rewritten queries have shown a slight improvement (with an evaluation time decrease between 0.9% and 13.4%) compared to the manually translated ones. For the remaining queries, (with the exception of query $Q_{15}$), the evaluation time of the rewritten and the manually translated queries was almost the same. For query $Q_{15}$ the manual translation has shown a significant evaluation time increase (71.7%). In more detail:

–  For the queries $Q_1$, $Q_2$ and $Q_{10}$, the rewritten queries have one For clause less compared to the manually translated ones. The use of the rewriting rule *Rule 3* has resulted to unnested For clauses in the rewritten queries. The resulting For clauses have presented an evaluation time improvement of 13.4% to 28.1% in the rewritten queries compared to the manually translated ones.

–  For the remaining queries (except $Q_{15}$), the performance of the rewritten queries is almost the same with the manually translated ones. The only reason for delays in few rewritten queries is the use of several "special" markup tags (e.g., <Result>, <Results>, etc.) which are exploited to structure the query results. These markup tags have resulted in a larger size of the results, hence a slight delay in evaluation time has been observed.

–  Finally, for $Q_{15}$, the manually translated query takes into account the cardinality of the elements FirstName and Last-Name, which have been defined in the XML Schema to be more than one. In that case, there is no need to check if the $FirstName and $LastName XQuery variables were bound to some values during the construction of the RDF graph. This is done in the automatically generated queries (rewritten and not-rewritten) by using the fn:exists( ) XQuery built-in function. This query is the only case with a considerable difference in the evaluation time of the manually translated query compared to the rewritten one. However, it is obvious that simple rewriting rules similar to *Rule 1* can be defined in order to exploit the XML Schema cardinality in several cases. For example, during the translation of Construct SPARQL queries, the cardinality value of more than one for elements or attributes can be considered by a rewriting rule, in order to avoid the unnecessary check if some values exist for these elements or attributes.

*13.3.2.2   Varying the Size of the Dataset*

In order to study the query evaluation efficiency over the dataset size, we have used the 10 synthetic Persons XML datasets. We first present an overview of the effect of the dataset size on the evaluation time. In the following figures, we present the results obtained using different XQuery engines. In particularly, Figure 9 corresponds to XML Store Y, Figure 10 corresponds to XML Store Z, and Figure 11 corresponds to Memory-based XQuery Engine. The figures show the query evaluation times for all the queries on three datasets ($DT_1$, $DT_8$ and $DT_{10}$). Each of the diagrams corresponds to one dataset. We can observe that in all cases the automatically rewritten queries outperform the automatically generated ones. In addition, the improvement of the automatically rewritten XQueries against the automatically generated XQueries does not show significant variations (is almost constant) over the dataset size.

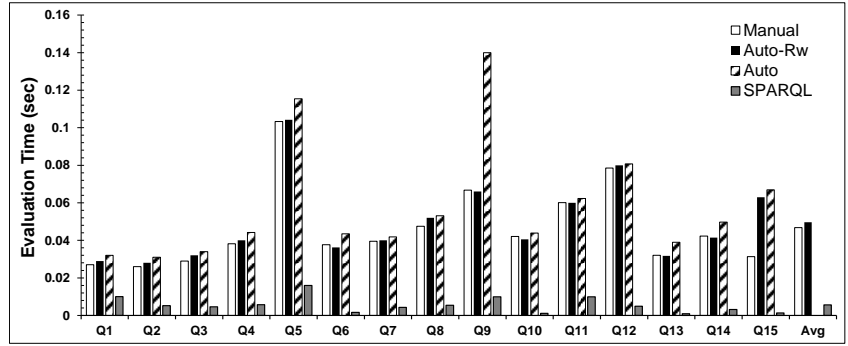

**(a)** $DT_1$ Persons Dataset (XML Store Y)

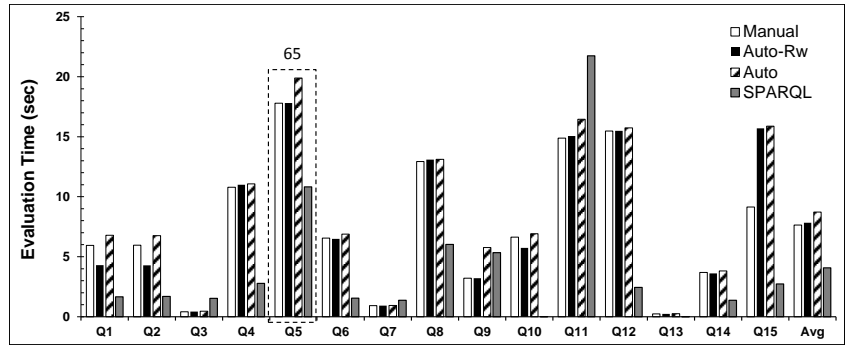

**(b)** $DT_8$ Persons Dataset (XML Store Y)

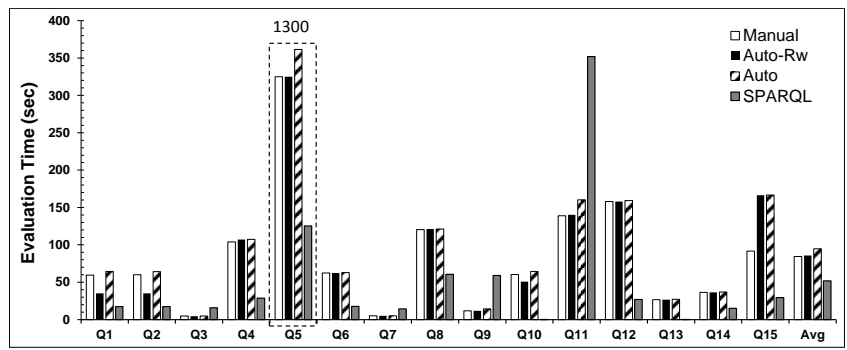

**(c)** $DT_{10}$ Persons Dataset (XML Store Y)

**Figure 9: Query Evaluation Time over the Persons Datasets $DT_1$, $DT_8$ and $DT_{10}$ (XML Store Y)**

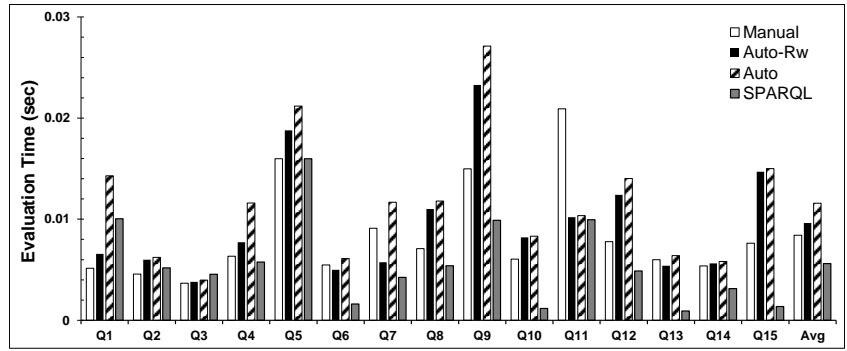

**(a)** *DT₁* Persons Dataset (XML Store Z)

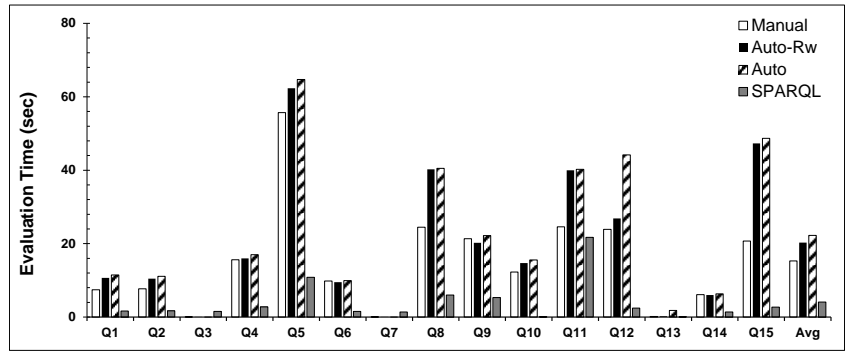

**(b)** *DT₈* Persons Dataset (XML Store Z)

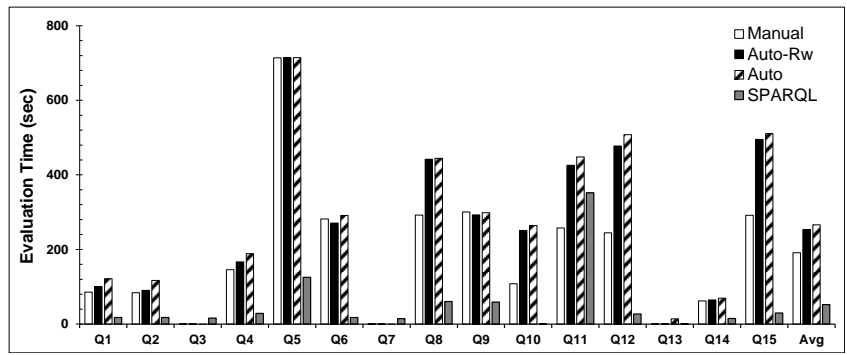

**(c)** *DT₁₀* Persons Dataset (XML Store Z)

**Figure 10: Query Evaluation Time over the Persons Datasets *DT₁*, *DT₈* and *DT₁₀* (XML Store Z)**

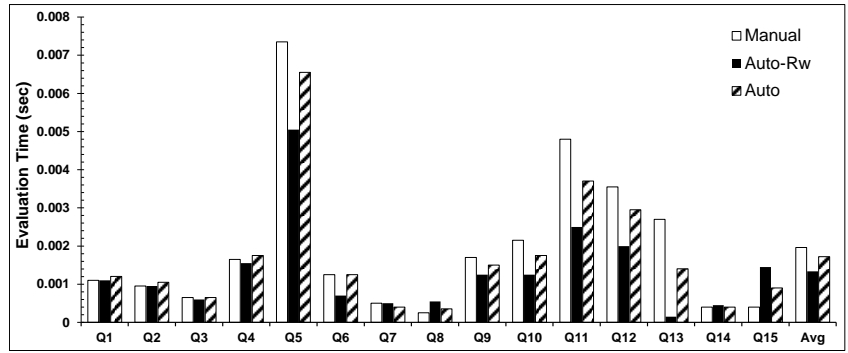

**(a)** *DT₁* Persons Dataset (Memory-based XQuery Engine)

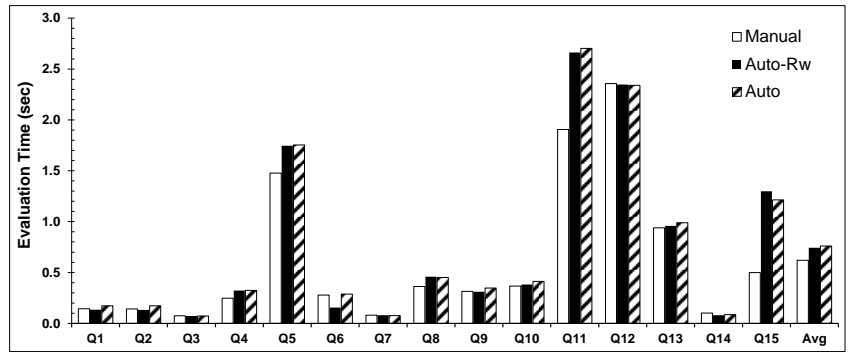

**(b)** *DT₈* Persons Dataset (Memory-based XQuery Engine)

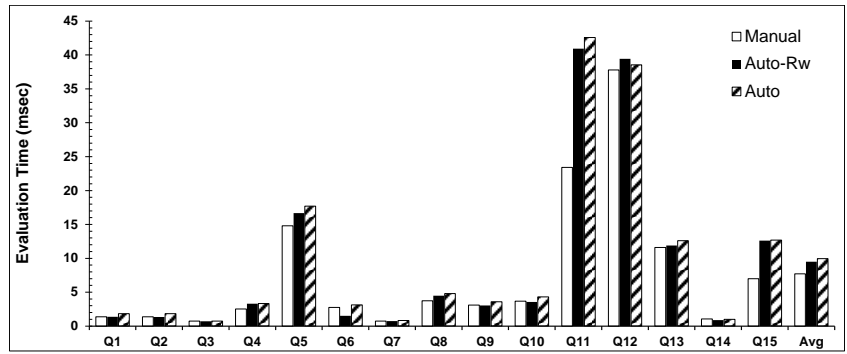

**(c)** *DT₁₀* Persons Dataset (Memory-based XQuery Engine)

**Figure 11: Query Evaluation Time over the Persons Datasets *DT₁*, *DT₈* and *DT₁₀* (Memory-based XQuery Engine)**

Figure 12 provides a thorough look at the query evaluation time over the dataset size. Particularly, Figure 12 presents the evaluation times (in logarithmic scale) for (a) the manually translated; (b) the automatically generated; and (c) the automatically rewritten XQuery queries over the 10 datasets. Each of the first 15 diagrams corresponds to one query (e.g., Figure 12 *(a)* corresponds to $Q_1$, Figure 12 *(b)* corresponds to $Q_2$, etc.) and the last diagram (Figure 12 *(q)*) corresponds to the average evaluation times for all the queries (Queries 1−15).

We observe that the evaluation times for both the manually and automatically rewritten queries have almost similar performance over the dataset size. As the dataset size increases, the evaluation times increase in a sublinear manner for the specific query set. For some of the queries, the increase is less sharp than for others (e.g., Queries 3, 7, 9); this is due to the high selectivity (i.e., small result set) of these queries. However, for all the queries the increase is sharper for datasets larger than $10^5$ records. Finally, with the exception of the queries 7, 10, 11 and 12 where the evaluation times are almost equal from the smallest dataset to the largest, as the dataset size increases, the difference between the evaluation times decreases, with most of the queries having almost equal evaluation times for the larger datasets ($DT_7$ to $DT_{10}$).

The average evaluation times (Figure 12 $(q)$) increase very fast with a sharper increase for datasets larger than $10^5$ records. In addition, as the dataset size increases, the difference between the evaluation times decreases.

In more detail, for the smallest dataset ($10^2$ records), the average evaluation time for the automatically generated and rewritten queries has a 6.1% overhead compared to that of the manually translated ones. In addition, the rewritten queries have shown an evaluation time decrease of 18% compared to the automatically generated ones.

Regarding the $DT_7$ dataset ($10^5$ records), the automatically rewritten queries have a 4.1% overhead compared to the manually translated ones. In addition, the rewritten queries have shown an evaluation time decrease of 11.8% compared to the automatically generated ones.

Finally, for the largest dataset ($5 \cdot 10^6$ records), the automatically rewritten queries have a 1.0% overhead compared to the manually translated ones. In addition, the rewritten queries have shown an evaluation time decrease of 10.8% compared to the automatically generated ones.

The results show that even without extensive optimization, a noticeable performance improvement can be achieved. The query evaluation time decreases in average by 13% compared to the not-rewritten ones, with a maximum decrease 83% in some cases. Even the automatically generated queries have reasonable performance and scale rather well for sizes up to $725 \cdot 10^5$ XML nodes.

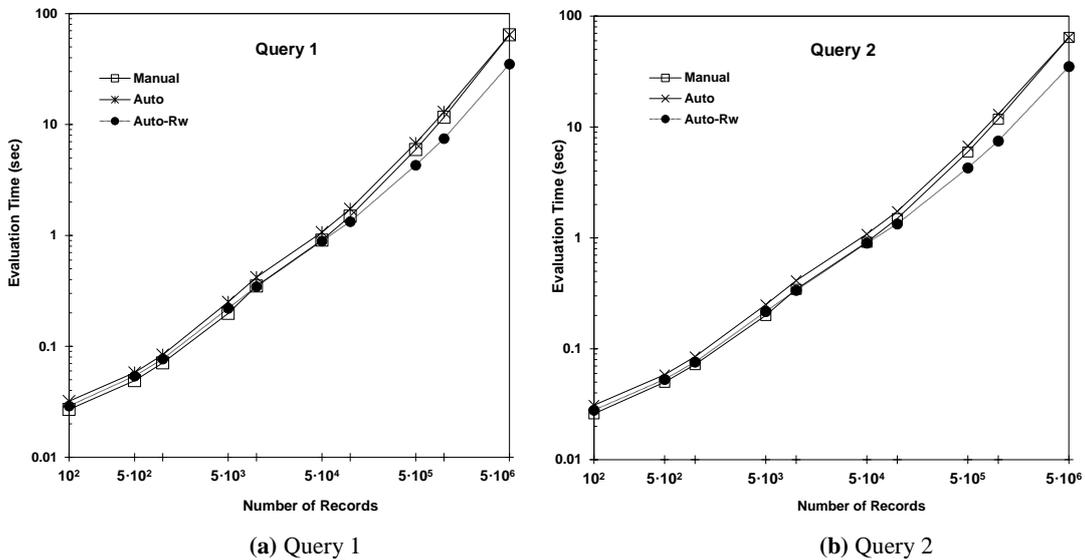

**(a)** Query 1        **(b)** Query 2

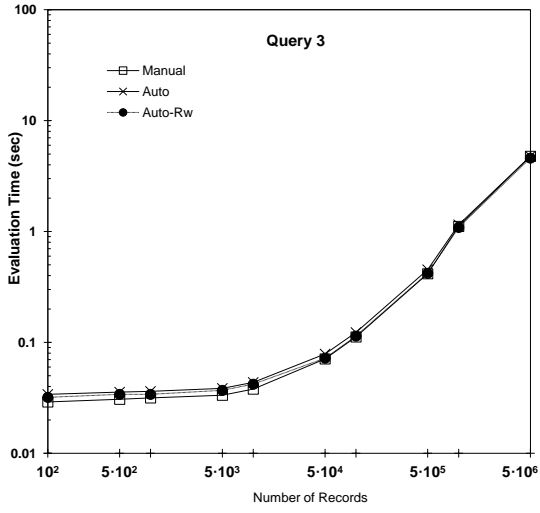

**(c)** Query 3

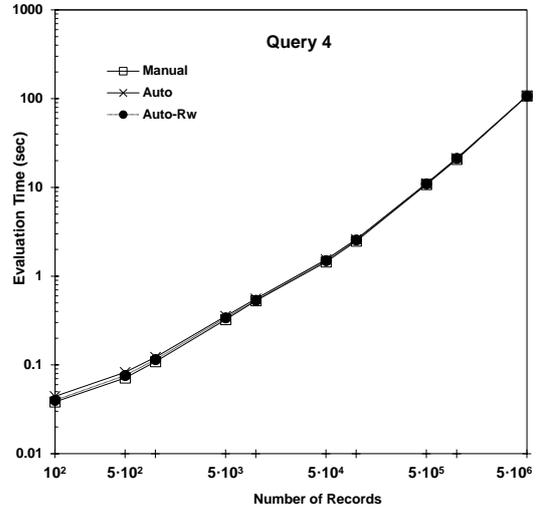

**(d)** Query 4

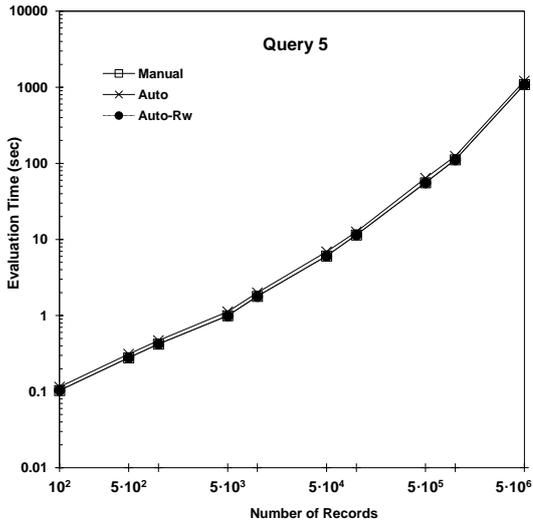

**(e)** Query 5

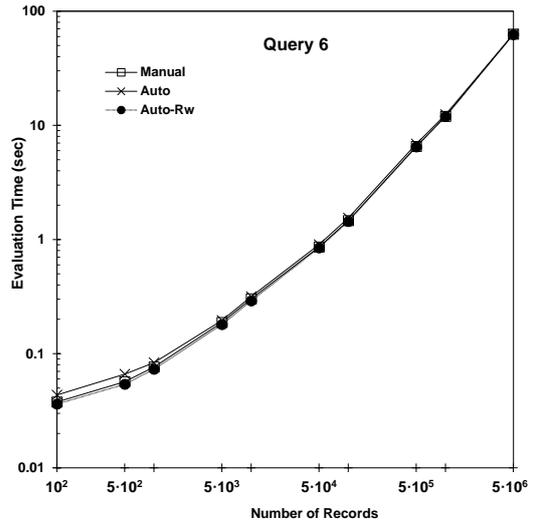

**(f)** Query 6

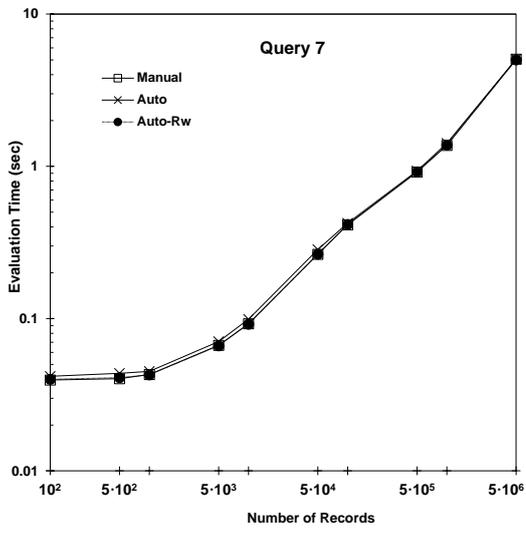

**(g)** Query 7

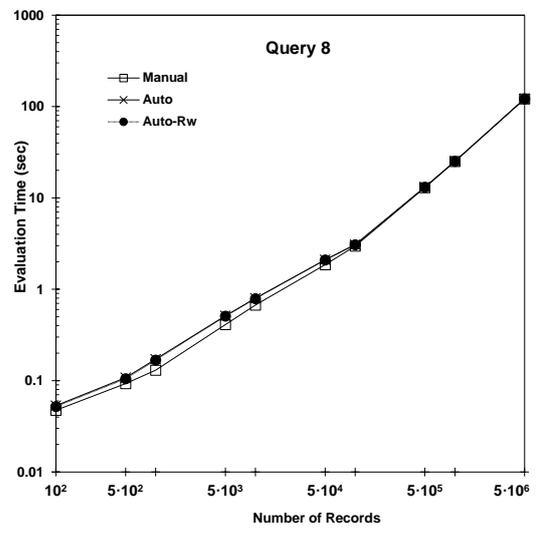

**(h)** Query 8

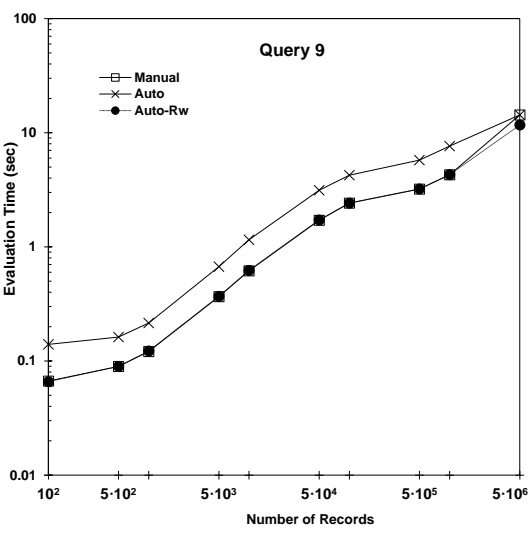

**(i)** Query 9

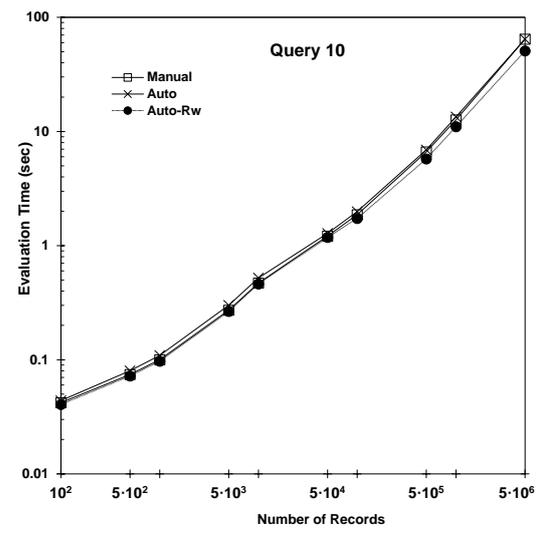

**(k)** Query 10

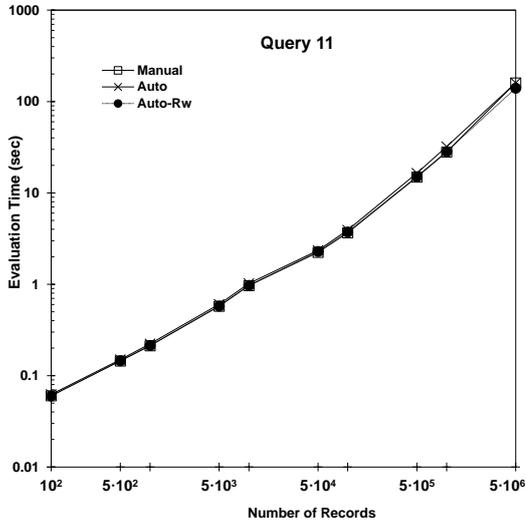

**(l)** Query 11

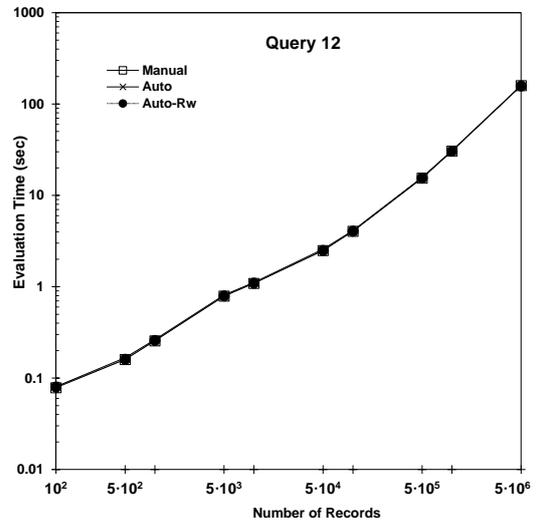

**(m)** Query 12

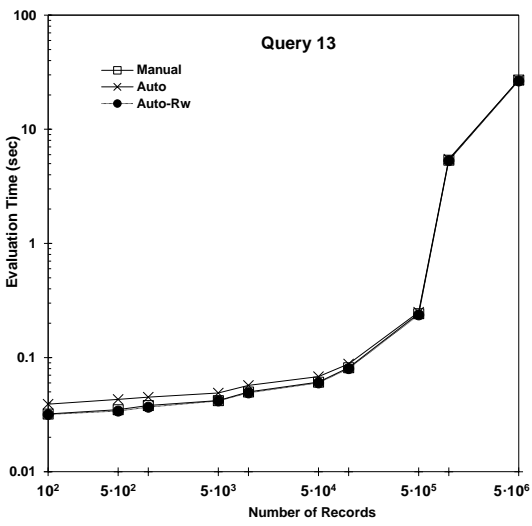

**(n)** Query 13

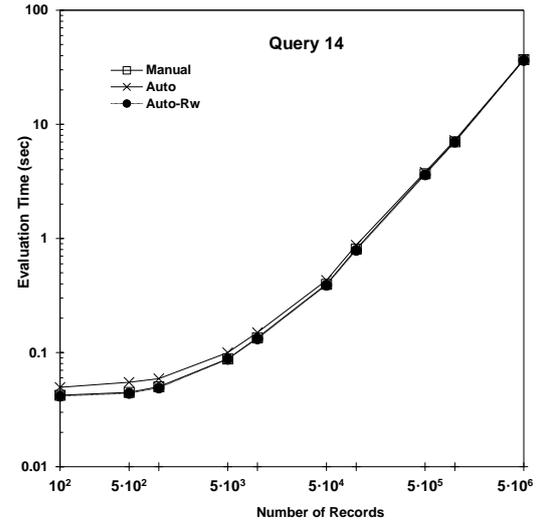

**(o)** Query 14

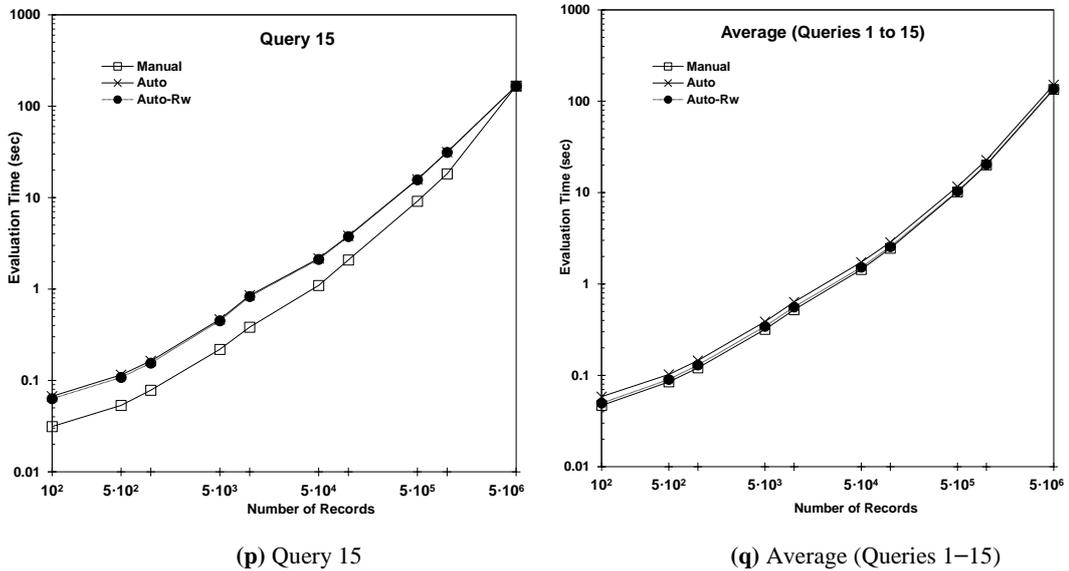

**(p)** Query 15         **(q)** Average (Queries 1−15)

**Figure 12: Query Evaluation Times vs. Dataset Size (Persons Datasets $DT_1$-$DT_{10}$) (XML Store Y)**

### 13.3.2.3 *Query Evaluation Time vs. Query Translation Time*

We present here a comparison of the query evaluation time with the query translation time. We have compared the query translation time and the query evaluation time to the *total time* which is the sum of the two (the diagrams are available in [120]).

We observe that the query translation takes negligible time in comparison to the query evaluation time even for the smallest dataset (i.e., the lowest evaluation times). In particular, for the dataset $DT_1$, the lowest ratio of translation time to total time (equal to 2.8%) occurs in query $Q_5$, while the highest ratio of translation time to total time (equal to 10.5%) occurs in query $Q_8$. Finally, the average ratio of translation time to total time is equal to 5.9%. Regarding the dataset $DT_8$, the lowest ratio of the translation time to the total time (equal to 0.01%) occurs in query $Q_5$, while the highest ratio of the translation time to the total time (equal to 1.3%) occurs in query $Q_{13}$. Finally, the average ratio of the translation time to the total time is equal to 0.04%.

### 13.3.3 *Real Dataset*

In this experiment we have studied the efficiency of the automatically generated XQuery queries using a real dataset. We have utilized the real DBLP dataset, as well as the DBLP query set, including 5 queries (Section 13.2.2.1). In an analogous manner with the previous experiment, we have measured and compared the query evaluation times for the automatically generated, rewritten and manually translated XQuery queries. Table 16 summarizes the experimental results.

**Table 16.** Query Evaluation Time for the DBLP Dataset (XML Store Y)

| Query-Evaluation-Time-header | | | | | | |
|---|---|---|---|---|---|---|
| **Query** | **SPARQL** ($Q_S$) | **Manual** ($Q_{Xm}$) | **Auto-Rw** ($Q_{Xa\text{-}Rw}$) | **Auto** ($Q_{Xa}$) | **Auto-Rw vs. Auto** | **Auto-Rw vs. Manual** |
| $Q_1$ | 2.88 | 40.14 | 40.12 | 44.56 | 10.0 % | 0.1 % |
| $Q_2$ | 0.07 | 0.19 | 0.19 | 0.21 | 11.2 % | 0.5 % |
| $Q_3$ | 0.06 | 16.61 | 16.63 | 18.72 | 11.2 % | -0.1 % |
| $Q_4$ | 14.24 | 20.52 | 20.82 | 29.57 | 29.6 % | -1.5 % |
| $Q_5$ | 0.26 | 9.73 | 10.69 | 11.51 | 7.1 % | -9.9 % |
| **Average** | **3.50** | **17.44** | **17.69** | **20.92** | **13.8 %** | **-2.2 %** |

**Automatically Rewritten vs. Automatically Generated (Auto-Rw vs. Auto) Queries.** Table 16 shows that the evaluation times for the rewritten queries have presented a significant performance improvement compared to the automatically generated ones, with an average evaluation time decrease of 13.8%. In more detail:

- For the queries $Q_1$, $Q_3$ and $Q_5$, the rewriting rule *Rule 1* (*Changing For Clauses to Let*) has been firstly applied. *Rule 1* exploits the exact cardinality for the Title and Year elements. As a result, two For clauses for $Q_1$ and one For clause for $Q_3$ and $Q_5$ have been transformed to Let clauses. Afterwards, *Rule 2* (*Reducing Let Clauses*) has been applied and has removed the Let clauses generated from *Rule 1*. Compared to the initial queries, the query $Q_1$ has two For clauses less and the queries $Q_3$ and $Q_5$ have one For clause less. The above rewritings have resulted in an improvement of 10.0%, 11.2% and 7.1% for the queries $Q_1$, $Q_3$ and $Q_5$, respectively.

- For query $Q_2$, the rewriting rule *Rule 2* has been applied and has removed one Let clause, resulting in an improvement of 11.2%.

- Finally, for query $Q_4$ the rewriting rule *Rule 3* (*Unnesting For Clauses*) has been applied and has removed two For clauses, resulting in an improvement of 29.6%.

**Automatically Rewritten vs. Manually Translated (Auto-Rw vs. Manual) Queries.** We can see from Table 16 that the evaluation times of the automatically rewritten queries are almost similar to the manually translated queries, with an average evaluation time increase of 2.2%. In more detail:

- For all the queries, with the exception of $Q_5$, the evaluation time for the rewritten queries is almost to the same with the manually translated ones. The only delay reason in the rewritten queries is the use of several "special" markup tags (e.g., *<Result>*, *<Results>*, etc.) which are exploited to structure the query results. These markup tags have resulted in a larger size of the results, hence a slight delay in evaluation time has been observed.

- For $Q_5$, the manually translated query has taken into account the cardinality of the elements Author and Title, which have been defined in XML Schema to be more than one. Thus, there was no need to check the existence of these values, as was done in the automatically generated query using the fn:exists( ) XQuery function.

Finally, we can observe from Table 16 that the query evaluation performance for the DBLP dataset is similar with that of the synthetic dataset of the same size.

In the following figure, we present the results obtained using different XQuery engines. In particularly, Figure 13(a) corresponds to XML Store Y, Figure 13(b) corresponds to XML Store Z, and Figure 13(c) corresponds to Memory-based XQuery Engine. The figures show the query evaluation times for all the queries over the DBLP dataset.

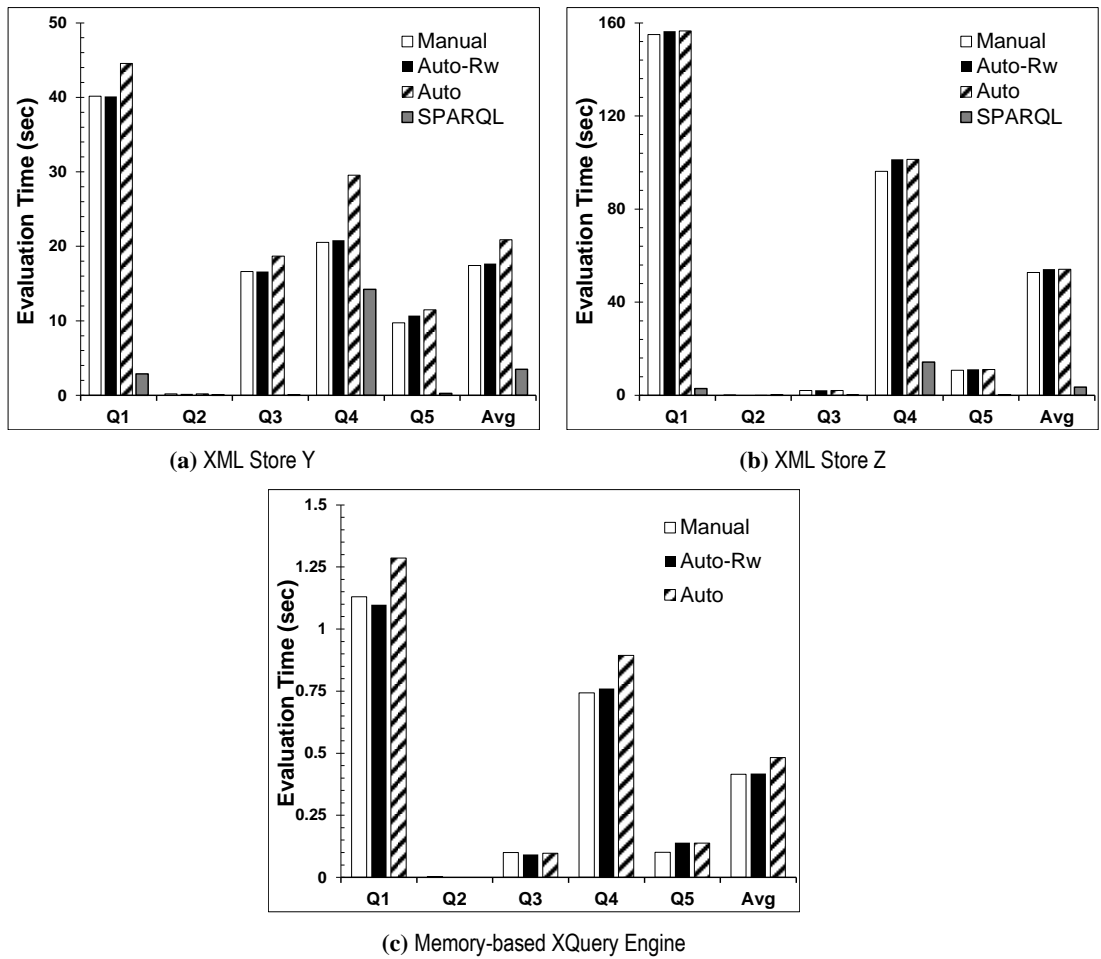

**Figure 13: Query Evaluation Time over the DBLP Dataset (Using different XQuery Engines)**

## 13.4 Evaluation Result Overview

**Schema Transformation and Mapping Generation.** Although both the schema transformation and mapping generation processes are off-line processes, we wanted to have an indication of their performance. To this end, we have used several international XML Schema standards and have measured the time required for schema transformation and for mapping discovery and generation. We observed that both processes took negligible time even for very large XML Schemas.

**Translation Efficiency.** In order to demonstrate the efficiency of the SPARQL to XQuery translation process we measured the translation time required by the SPARQL2XQuery Framework. In the first experiment, we generated several SPARQL queries by modifying their graph pattern size and type. In the second experiment, for the queries generated in the first experiment, we modified the number of the predefined mappings. Finally, in the third experiment, we have used three SPARQL query sets attempting to cover almost all the SPARQL grammar variations. The query sets used are the *Berlin SPARQL Benchmark* query set, a query set over the DBLP schema and a set over the Persons schema.

**Query Evaluation Efficiency.** Regarding the efficiency of the generated XQuery expressions, we have defined a small set of simple rewriting rules aiming to provide more efficient XQuery expressions. We have applied these rules on the automatically generated XQuery expressions. Then, we have compared the evaluation time of the automatically generated, rewritten and manually translated XQuery expressions. In the first set of experiments, a synthetic dataset and a set of 15 queries have been used. We have modified the dataset size and we have measured the query evaluation time for the automatically generated, rewritten and manually translated XQuery queries. In the second set of experiments, the real DBLP dataset has been utilized for demonstrating the query evaluation efficiency.

The results are similar for both the real and synthetic datasets. In particular, for the largest synthetic dataset ($5 \cdot 10^6$ records) the rewritten queries have presented an evaluation time decrease of 10.8% compared to the not-rewritten ones. In general, the rewriting rules have resulted in significant performance improvement, with an average evaluation time decrease of 13%, reaching 83% in some cases. Moreover, the average evaluation time for the automatically generated and rewritten queries has 1.0% overhead compared to the manually specified ones. Finally, the query evaluation times have been compared to the query translation times. The conclusion was that the query translation takes negligible time in comparison to the evaluation time, even for very small datasets.

## 14 CONCLUSIONS & FUTURE WORK

The *Web of Data* (*WoD*) is an open environment comprised of hundreds of large interlinked, user contributed datasets. The *WoD* is founded on technologies and standards developed by the *Semantic Web* (*SW*) community (e.g., OWL, RDF/S, SPARQL, etc.) for Web information representation and management. On the other hand, in the current Web infrastructure the XML/XML Schema are the dominant standards for information exchange, and for the representation of semi-structured information. In addition, many international standards (e.g., *Dublin Core*, *MPEG-7*, *METS*, *TEI*, *IEEE LOM*, etc.) have been expressed in XML Schema. The aforementioned have led to an increasing emphasis on XML data.

In the WoD users should not interact with different data models and languages for developing their applications or expressing their queries. In addition, it is unrealistic to expect that all the legacy data (e.g., Relational, XML, etc.) will be converted to RDF data. Thus, it is crucial to provide interoperability mechanisms that allow the WoD users to transparently access external heterogeneous data sources from their own working environment. Finally, in the Linked Data era, offering SPARQL endpoints (i.e., SPARQL-based search services) over legacy data has become a major research challenge. However, despite the significant body of related work on relational data, to the best of our knowledge there is no work addressing neither the SPARQL to XQuery translation problem nor offering SPARQL endpoints over XML data. In the most recent research approaches, a combination of SW (SPARQL) and XML (XQuery, XPath and XSLT) technologies is exploited in order to transform XML data to RDF and vice versa.

In this paper we have proposed the SPARQL2XQuery Framework, which bridges the heterogeneity gap and creates an interoperable environment between the SW and XML worlds. The SPARQL2XQuery Framework comprises the key component for several WoD applications, allowing the establishment of SPARQL endpoints over XML data, as well as a fundamental component of ontology-based integration frameworks involving XML sources.

The SPARQL2XQuery Framework allows arbitrary SPARQL queries posed over ontologies to be automatically translated to XQuery expressions which are evaluated over XML data with respect to a set of predefined mappings. To this end, our Framework allows both manual and automatic mapping specification between ontologies and XML Schemas. Finally, the query results are returned either in RDF or in SPARQL Query Result XML Format. Thus, the WoD users are no longer required to interact with more than one models or query languages.

In more detail, we have introduced a mapping model for the expression of OWL–RDF/S to XML Schema mappings, as well as a method for SPARQL to XQuery translation both provided by the SPARQL2XQuery Framework. To the best of our knowledge, this is the first work addressing these issues. Moreover, we have presented the XS2OWL component, which allows transforming XML Schemas to OWL ontologies, exploiting the latest versions of the standards (XML Schema 1.1. and OWL 2). As far as we know, this is the first work that fully captures the XML Schema semantics and supports the XML Schema 1.1 constructs. The XS2OWL component has been integrated in the SPARQL2XQuery framework in order to provide automatic mapping generation and maintenance.

A thorough experimental evaluation of the SPARQL2XQuery framework has been conducted and presented, in order to demonstrate the efficiency of (a) schema transformation; (b) mapping generation; (c) query translation; and (d) query evaluation.

We have also discussed, in this paper, the major technical and theoretical challenges we have faced throughout the development of the SPARQL2XQuery Framework. The major difficulties have arisen from the different data models and semantics adopted by the SW and XML worlds. In summary, we had to overcome several heterogeneity issues like *Directed graphs* vs. *Tree structures*, *Three-valued logic* vs. *Two-valued logic*, *Graph patterns* vs. *Iterative procedures*, etc. We have also discussed issues involved in the translation process that are related to the SPARQL semantics like *Well Designed* vs. *Non-Well Designed Graph Patterns*, *Safe* vs. *non-Safe Filter Expressions*, etc.

Our current work can be categorized in three parts; (1) performance issues, (2) ontology-based integration and (3) SPARQL 1.1 features.

*Performance Issues.* We study several performance issues, including the specification of sophisticated XQuery rewriting rules that exploit the XML Schema semantics, as well as the adoption of XQuery query optimization techniques aiming to provide more efficient XQuery expressions.

*Ontology-based Integration.* The SPARQL2XQuery Framework is going to be part of an ontology-based mediator [40][41][42][43] that we are developing now and is going to provide semantic interoperability and integration between distributed heterogeneous sources using the standard SW and XML technologies.

*SPARQL 1.1 Features.* We examine the support of the new SPARQL features (e.g., nested queries, aggregate functions, etc.) that are going to be introduced by the upcoming SPARQL 1.1 standard [14]. A remarkable extension of the upcoming SPARQL 1.1 is related to the support of update operations (i.e., update, insert, and remove) [16].

## ACKNOWLEDGMENTS

We would like to thank *Yiannis Karagiorgos*, *Giorgos Giannopoulos*, *Konstantinos Makris*, *Dimitris Sacharidis*, and *Theodore Dalamagas*, for many helpful comments on earlier versions of this article.